%% file: thesis.tex
\documentclass[PhD]{iitmdiss}
\usepackage[toc,page]{appendix}
\usepackage[dvips]{graphicx}
\usepackage{amssymb}
\usepackage[hypertex]{hyperref} % hyperlinks for references.
\usepackage{amsmath} % easier math formulae, align, subequations \ldots
\newcommand{\beqar}{\begin{eqnarray}}
\newcommand{\eeqar}{\end{eqnarray}}
\newcommand{\beq}{\begin{equation}}
\newcommand{\eeq}{\end{equation}}
\newcommand{\ket}[1]{\left|{#1}\right\rangle}
\newcommand{\bra}[1]{\left\langle{#1}\right|}
\newcommand{\modu}[1]{\left|{#1}\right|}
\newcommand{\expect}[3]{\left\langle{#1}\right|{#2}\left|{#3}\right\rangle}
\newcommand{\aver}[1]{\left\langle{#1}\right\rangle}
\newcommand{\added}[2]{\left|{#1},{#2}\right\rangle}
\newcommand{\inner}[2]{\left\langle{#1}|{#2}\right\rangle}

\begin{document}

%%%%%%%%%%%%%%%%%%%%%%%%%%%%%%%%%%%%%%%%%%%%%%%%%%%%%%%%%%%%%%%%%%%%%%
% Title page

\title{Non-classical effects in wave packet dynamics}

\author{C. Sudheesh}

\date{December 2005}
\department{PHYSICS}

%\nocite{*}
\maketitle

%%%%%%%%%%%%%%%%%%%%%%%%%%%%%%%%%%%%%%%%%%%%%%%%%%%%%%%%%%%%%%%%%%%%%%
% Certificate
\certificate

\vspace*{0.5in}

\noindent This is to certify that the thesis titled 
{\bf Non-classical effects in wave packet dynamics}, 
submitted by {\bf C. Sudheesh}  
to the Indian Institute of Technology Madras for
the award of the degree of {\bf Doctor of Philosophy}, is a bona fide
record of the research work done by him under our supervision.  The
contents of this thesis, in full or in part, have not been submitted
to any other Institute or University for the award of any degree or
diploma.

\vspace*{1in}

%\hspace*{-0.25in}
%\parbox{2.5in}{
%\noindent {\bf S.~Lakshmi~Bala } \\[-8pt]
%\noindent Research Guide \\ 
%\noindent Associate Professor \\[-8pt]
%\noindent Dept. of Physics\\ [-8pt]
%\noindent IIT Madras \\
%} 
\hspace*{4.0in} 
\parbox{2.5in}{

\noindent {\bf S.~Lakshmi~Bala } \\[-8pt]
\noindent Research Guide \\ 
\noindent Associate Professor \\[-8pt]
\noindent Dept. of Physics\\ [-8pt]
\noindent IIT Madras \\[16pt]

\noindent {\bf V.~Balakrishnan} \\ [-8pt]
\noindent Research Guide \\ 
\noindent Professor \\[-8pt]
\noindent Dept. of  Physics\\[-8pt]
\noindent IIT Madras \\
}  

\vspace*{0.25in}
\noindent Place: Chennai 600 036, India\\
Date: 26 December 2005 

\acknowledgments
I take this opportunity to express my sincere gratitude
to my principal supervisor, Dr. S. Lakshmi Bala, although
a few words cannot do justice to the pivotal role she has 
had in my life as a graduate student.
Her commitment, hard work and  attention to detail
have set an example I hope to match some day.
I have benefited immensely from her encouraging support
and constructive criticism.
I am greatly indebted to my second
supervisor, Dr. V. Balakrishnan, for insightful
comments and valuable discussions throughout the course of 
my doctoral research. His dedication, excellence in 
teaching, and exemplary academic standards 
have always inspired me to strive hard for perfection.

Special words of appreciation go to Dr. Suresh Govindarajan
and Dr. M. V. Satyanarayana for their 
constant encouragement and useful  discussions.
I would also like to thank the other members of my  
doctoral committee, 
Dr. S. G. Kamath and  Dr. K. Mangala Sunder, 
for their efforts in reviewing and commenting on this work.
My thanks also go to
Dr. Neelima Gupte, Dr. Arul Lakshminarayan, 
Dr. M. S. Sriram and
Dr. Radha Balakrishnan for their help and discussions.

It is a pleasure to thank 
Dr. A. Subrahmanyam, Head of the Department
of Physics, and Dr. V. R. K. Murthy, former 
Head of the Department of Physics, for their 
kind help and advice 
whenever I needed these. 
I am thankful to Dr. V. G Idichandy, Dean (Students), 
for taking necessary steps
when I had health-related problems.

I would like to thank my parents and my uncles, 
Mr. P. Asokan and Mr. P. Gangadharan, for love, support and
encouragement.

I am grateful to Aji, Shajahan and Vinu 
for providing useful reference material.
I would like to extend my thanks to all my 
friends and colleagues for valuable
discussions and timely help --- in particular, 
Ranjith, Hari Varma, Subeesh,
Raghu, Sree Ranjani, Madhu, Zahera, Tanima, Meenakshi,
Basheer, Pramod, Sandeep and Sanoop. Special thanks go 
to Sreejith, whom 
I am fortunate to have as a friend, and 
who has always been with me in my ups and downs 
since my college
days. 

I am grateful to IIT Madras for financial support in the form of 
a Research Assistantship  
during the course of my doctoral research.

The work presented here was supported in part by the 
Department of Science and Technology, 
India, under Project No. SP/S2/K-14/2000.

\abstract

\noindent {\bf Keywords}: 
\hspace*{0.5em} \parbox[t]{4.4in}{Wave packets; Kerr medium; 
revivals; fractional 
revivals; coherent states; photon-added coherent states; 
expectation values; higher moments; squeezing; Wigner function; 
entropy; entanglement; time series; Lyapunov exponent.}

\vspace*{24pt}
Starting with an initial wave packet 
that corresponds to an ideal coherent state, we show that,  
in the framework of a generic Hamiltonian that
models the propagation of a single-mode 
electromagnetic field in a Kerr-like
medium, distinctive signatures of wave packet 
revivals and fractional revivals are displayed by the time evolution
of the expectation values and cumulants 
of appropriate observables. This enables
selective identification of different fractional revivals. 
Using a class of photon-added coherent states as 
the initial states, we further show that the revival phenomena 
provide clear signatures of the 
extent of the departure from coherence, and of the 
deviation from Poissonian 
photon number statistics, of the
initial state of the radiation field.
A comparison is also made of the manner in which other 
non-classical effects are 
displayed as the wave packet evolves, 
in the cases of an initial photon-added coherent 
state and an initial ideal coherent state, 
respectively. These effects comprise squeezing, 
higher-order squeezing, and 
the non-positivity of the Wigner function. 

Extending these studies to the case of two interacting modes, we 
consider another generic Hamiltonian that models the interaction of
a single-mode electromagnetic field propagating 
in a nonlinear medium, an atom of which is 
modeled by an anharmonic oscillator. The dynamics of
expectation values and higher moments of observables pertaining to the 
field sub-system as well as the total system are studied over a range
of values of a relevant parameter, namely, the ratio of the 
strength of the anharmonicity to that of 
the inter-mode coupling. Analogs of 
revival phenomena are found, and their signatures identified in
quantifiers such as sub-system entropies that 
serve to characterize the degree of entanglement. A time series analysis of
the mean photon number (or, equivalently, 
the mean energy of the field mode) is
carried out, to deduce the embedding dimension of the reconstructed phase
space and the maximal Lyapunov exponent. The corresponding 
dynamical behavior is found to range from 
mere ergodicity (as in quasi-periodicity) to exponential instability 
characterized by a positive maximal
Lyapunov exponent, as the parameter ratio mentioned above is varied. 

Our results shed light on the non-classical effects that occur during
the propagation of a quantum mechanical wave packet 
in a nonlinear medium. Several avenues for
further research are also revealed.

\pagebreak

%%%%%%%%%%%%%%%%%%%%%%%%%%%%%%%%%%%%%%%%%%%%%%%%%%%%%%%%%%%%%%%%%
% Table of contents etc.

\begin{singlespace}
\tableofcontents
\thispagestyle{empty}

%\listoftables
%\addcontentsline{toc}{chapter}{LIST OF TABLES}
\listoffigures
\addcontentsline{toc}{chapter}{LIST OF FIGURES}
\end{singlespace}

%%%%%%%%%%%%%%%%%%%%%%%%%%%%%%%%%%%%%%%%%%%%%%%%%%%%%%%%%%%%%%%%%%%%%%
% Abbreviations
%\abbreviations

%\noindent 

%\textbf{CS} \> Coherent states\\
%\textbf{PACS} \> Photon-added coherent states \\
%\textbf{SVNE} \> Sub-system von Neumann entropy \\
%\textbf{SLE}   \> Sub-system linear entropy \\

\pagebreak

\glossary
\begin{tabbing}
xxxxxxxx \= xxxxxxxxxxxxxxxxxxxxxxxxxxxxxxxxxxxxxxxxxxxxxxxx \kill
$H$ \> Hamiltonian \\ 
$a,a^{\dagger}$ \> ladder operators for a
  single-mode electromagnetic field \\
${\sf N}$ \> photon number operator $a^{\dagger}\,a$\\
$\ket{n}$ \> a Fock state \\
$x, p$\> $(a+ a^{\dagger})/\sqrt{2}\,\,,\,\, (a-a^{\dagger})/i\sqrt{2}$\\
$\chi$ \> characteristic frequency parameter in the Hamiltonian \\
$T_{\rm rev}$ \> revival time \\
$\ket{\alpha}$ \> coherent state (CS) of a single-mode
  electromagnetic field\\
$\alpha$ \> complex number labeling a CS \\
$\nu$ \> $|\alpha|^2$ \\
$\theta$ \> argument of $\alpha \,\,( = \nu^{1/2}\,e^{i\theta}$) \\
$\tau$\> $\sin\,(2\chi t)$ (in Chapter 2) \\
$\delta x$ \> $x - \aver{x}$ (similarly, $\delta p$)\\
$\varDelta x$\> variance of $x$ (similarly, $\varDelta p$)\\
$\beta_{1}^{(x)}$\> square of the skewness in $x$ (similarly, 
$\beta_{1}^{(p)}$) \\ 
$\beta_{2}^{(x)}$ \> kurtosis in $x$ (similarly, 
$\beta_{2}^{(p)}$)\\
$\ket{\alpha,m}$  \> $m$-photon-added coherent state (PACS)\\
$\aver{\sf N}$\> mean photon number in a given state\\
$\aver{\sf N}_{m}$ \> mean photon number in the PACS 
$\ket{\alpha,m}$  \\
$L_m$ \> Laguerre polynomial of order $m$ \\
$L_{m}^{s}$ \> associated Laguerre polynomial\\
$\ket{\phi}$ \> $\sum_{n=0}^{\infty} e^{i n\phi} \ket{n}$ \\
$P(\phi)$ \> phase distribution function \\ 
$A_{\varphi}$ \>  $(a e^{i \varphi} + a^{\dagger} 
e^{-i \varphi})/\sqrt{2}$\\
$B_{\varphi}$ \>  $(a e^{i \varphi} - a^{\dagger} 
e^{-i \varphi})/(i\sqrt{2})$ \\
$q$ \> a non-negative integer \\
$Z_1$ \> $(a^q+a^{\dagger q})/\sqrt{2}$ or 
$(a^q+a^{\dagger q}+b^q+b^{\dagger q})/(2\sqrt{2})$\\
$Z_2$ \> $(a^q-a^{\dagger q})/(i\sqrt{2}$ or 
$(a^q-a^{\dagger q}+b^q-b^{\dagger q})/(2i\sqrt{2})$\\
$F_q({\sf N})$ \> $[a^q\,,\, a^{\dagger q}]$, a polynomial of 
order $(q-1)$ in ${\sf N}$ \\
$D_{q}(t)$\> a measure of $q^{\rm th}$-power amplitude squeezing \\
$D_{q}^{(m)}(t)$\> $D_{q}(t)$ for an initial PACS $\ket{\alpha,m}$ \\
$\beta$ \> a complex number\\
$W(\beta;t)$ \> Wigner function\\  
$\delta$ \> non-classicality indicator $\int\!d^2 \beta\,
|W(\beta;t)| - 1$ \\
$\rho (t)$ \> density matrix  \\
$\rho_{nl}(t)$ \> density matrix element \\
$b,b^{\dagger}$ \> ladder operators for an atomic oscillator modeling
  the nonlinear medium \\ 
${\sf N}_{\rm tot}$ \> total number operator 
$a^{\dagger}\,a + b^{\dagger}\,b$\\
$\gamma$ \> anharmonicity parameter in the two-mode Hamiltonian \\
$g$ \> coupling constant in the two-mode Hamiltonian \\
$\lambda$ \> limiting value of $\gamma/\hbar$ in the classical limit \\ 
$H_{\rm cl}$\> Hamiltonian in the classical limit \\
$N_{\rm cl}$ \> second constant of the motion in the classical system \\ 
$\rho_{k}(t)$ \> reduced density matrix for a sub-system \\
$\lambda_{k}^{(i)}(t)$ \> eigenvalue of the reduced density matrix 
$\rho_{k}(t)$ \\
$S_{k}(t)$\> sub-system von Neumann entropy (SVNE) \\
$\delta_{k}(t)$ \> sub-system linear entropy (SLE) \\  
$\xi\,,\,\eta$ \> quadrature variables for the 
coupled two-mode system \\
$\delta t$ \> time step for time series analysis \\
$\tau$ \> delay time in time series analysis (in Chapter 6) \\
$d_{\rm emb}$ \> embedding dimension (dimensionality of 
reconstructed phase space) \\
$S(f)$ \> power spectrum at frequency $f$ \\
$d_{j}(k)$ \> distance between $j^{\rm th}$ pair of 
nearest-neighbor points at time 
$k\,\delta t$\\
$\lambda_{\rm max}$ \> maximal Lyapunov exponent \\
\end{tabbing}
%%%%%%%%%%%%%%%%%%%%%%%%%%%%%%%%%%%%%%%%%%%%%%%%%%%%%%%%%%%%%%%%%%%%%%
% Notation
%\include{misc/symbols}
\pagebreak
\clearpage

% The main text will follow from this point so set the page numbering
% to arabic from here on.
\pagenumbering{arabic}

%%%%%%%%%%%%%%%%%%%%%%%%%%%%%%%%%%%%%%%%%%%%%%%%%%
% Introduction.
\include{intro}

\include{manifest}

\include{pacs}

\include{squeeze}

\include{twomode}

\include{timeseries}

\include{conclude}

%%%%%%%%%%%%%%%%%%%%%%%%%%%%%%%%%%%%%%%%%%%%%%%%%%%%%%%%%%%%
% Appendices.

\begin{appendices}
\include{appendixA}
\include{appendixB}

\include{appendixC}

\include{appendixD}

\end{appendices}

%%%%%%%%%%%%%%%%%%%%%%%%%%%%%%%%%%%%%%%%%%%%%%%%%%%%%%%%%%%%
% Bibliography.

\begin{singlespace}
 % \bibliography{refs}dd
\bibliography{thesis}
\end{singlespace}

%%%%%%%%%%%%%%%%%%%%%%%%%%%%%%%%%%%%%%%%%%%%%%%%%%%%%%%%%%%%
% List of papers

\listofpapers
~~~~{\large {\bf Papers in refereed journals}}
\begin{enumerate}  
\item C. Sudheesh, S. Lakshmibala and V. Balakrishnan,\\
{\it Manifestations of wave packet revivals 
in the moments of observables}. \\
Phys. Lett. A {\bf 329}, 14-21 (2004).
\item C. Sudheesh, S. Lakshmibala and V. Balakrishnan,\\
{\it Wave packet dynamics of photon-added coherent states}.\\ 
Europhys. Lett. {\bf 71}, 744-750 (2005).
\item C. Sudheesh, S. Lakshmibala and V. Balakrishnan,\\
{\it Squeezing and higher-order squeezing of photon-added coherent 
states propagating in a Kerr-like medium}.\\
J. Opt. B: Quant. Semiclass. Opt. {\bf 7}, S728-S735 (2005).
\item C. Sudheesh, S. Lakshmibala and V. Balakrishnan,\\
{\it Wave packet dynamics of entangled two-mode states}.\\ 
J. Phys. B: At. Mol. Opt. Phys. {\bf 39}, 3345-3359 (2006).
\end{enumerate}  
{\large {\bf Presentations in conferences}}
\begin{enumerate}  
\item 
C. Sudheesh, S. Lakshmibala and
V. Balakrishnan, \\
{\it Dynamical squeezing and higher-order squeezing in wave
packet propagation in a Kerr-like medium}.\\
ICSSUR`05: 9th International 
Conference on Squeezed States and
Uncertainty
Relations, Besan\c con, France, May 2005.
\item
C. Sudheesh, S. Lakshmibala and
V. Balakrishnan,\\
{\it Dynamics of a wave packet governed by a nonlinear quantum
Hamiltonian: Classical correspondence}.\\
II National Conference on Nonlinear Systems and
Dynamics, Aligarh Muslim University, Aligarh, February 2005.
\item
C. Sudheesh, S. Lakshmibala and V. Balakrishnan,\\
{\it Entanglement dynamics in the propagation of a single-mode 
field through a nonlinear medium}.\\
Topical Conference on Atomic, Molecular and Optical Physics,
Indian Association for the Cultivation of Science, Kolkata, December
2005. 
\end{enumerate}  
{\large \bf Submitted for publication}\\
\phantom{xxxxx}C. Sudheesh, S. Lakshmibala and V. Balakrishnan,\\ 
\phantom{xxxx}
{\it Ergodicity properties of quantum expectation values in entangled two-mode states, arXiv:0706.2954 (2007)}.

\newpage
\begin{center}
{\Large{\bf CURRICULUM VITAE}}\\
\end{center}
\vspace{1.0cm}
\noindent
\begin{tabular}{lcl}
{\bf Name} & : &  C. Sudheesh	\\
{\bf Date of Birth} &: &  25.05.1977\\
{\bf Nationality} &: & Indian\\
{\bf Permanent Address}& : & Chethil House, Iringal PO\\
& & Vadakara, Kozhikode 673 521.\\
{\bf Present Address} & : & Postdoctoral Fellow\\
& & Theoretical Physics Division \\
& & Physical Research Laboratory\\
& & Ahmedabad 380 009.\\
{\bf Email} &: & chethil.sudheesh@gmail.com\\\\
\end{tabular}

\noindent
~~{\bf {Academic Record}}:
\begin{itemize}
\item B.Sc (Physics), I Class (1st Rank in College), Govt. Madappally College, University of Calicut, (1997).
\item M.Sc (Physics), Indian Institute of Technology, Madras (2001).
\item Qualified in GATE (Physics), 97.71 percentile (2001).
\item Qualified for Lecturership in the CSIR-UGC National Eligibility Test (NET), (2001).
\end{itemize}
~~{\bf {Conferences attended}}:
\begin{itemize}
\item II National Conference on Nonlinear Systems and
Dynamics, Aligarh Muslim University, Aligarh, February 2005.
\item ICSSUR`05: 9th International 
Conference on Squeezed States and
Uncertainty
Relations, Besan\c con, France, May 2005.
\item III National Conference on Nonlinear Systems and
Dynamics, Ramanujan Institute for Advanced Study in Mathematics, University 
of Madras, Chennai, February 2006.
\end{itemize}
~~{\bf \underline{Schools attended}}:
\begin{itemize}
\item IV School on Physics of Beams, Centre for Advanced Technology, Indore (27 Dec. 1999 to 7 Jan. 2000).
\item SERC School on Quantum Information and Quantum Optics, Physical Research Laboratory, Ahmedabad (February 1-14, 2004).
\item SERC School on Statistical Physics, Tata Institute of Fundamental Research, Mumbai (February 16-28, 2004).
\end{itemize}

\end{document}

%% file: intro.tex
\chapter{Introduction}\label{intro}

The long-time behavior of classical dynamical systems, and the degree of
randomness they exhibit (ranging from mere ergodicity to fully-developed
chaos), have been investigated extensively in the literature. 
They continue to be subjects of abiding interest and current
research. Ergodicity
theorems of wide applicability ensure that, generically, the
dynamics is metrically transitive --- either on the energy surface, in
conservative (Hamiltonian) systems, or on some attractor, in dissipative
systems. While the evolution equations for the dynamical variables
describing the behavior of a classically chaotic system are
necessarily nonlinear, the corresponding quantum dynamics is governed by
the Schr\"odinger equation (or, in the more general case of mixed states, 
by
the Liouville equation), which is inherently linear. It is not
straightforward, therefore, to assess the extent of randomness
in a quantum system, or
the precise manner in which the information content of a quantum system
changes with time.

At least two different approaches have been 
employed in identifying signatures of
the nature of the ergodicity
of the quantum mechanical counterpart of a generic classical system \cite
{haake}. One approach relies on  the
observation that such signatures
are generally   
manifested in the
statistics of the energy levels of the corresponding  quantum system.
If the system
is classically integrable, the quantum levels cluster together, 
and could even cross when a specific parameter in the Hamiltonian 
is varied \cite{berry1}.
A classically chaotic system, on the other hand, has its corresponding   
quantum energy
levels so correlated as to resist such crossings \cite{berry2,berry4}.
Another approach
is based on investigating the dynamics
of the overlap between two quantum states of the same physical                 
system which 
originate from the same initial
state, but with {\it slightly} different values of one of the control  
parameters \cite{peres}.
The time-dependent overlap is close to unity at all times if the 
normalized initial 
state is located in a classically regular region of phase space. In
contrast, if the initial state is in a chaotic region of the classical 
phase
space, the overlap falls off exponentially in time.

The foregoing lines of investigation concern quantum signatures of
classical dynamics. The inverse problem, namely, the
identification of
signatures of non-classical effects in the temporal behavior of
quantum mechanical expectation values (which, in turn,
could be regarded as effective classical dynamical variables 
in an appropriate phase space), has also
received attention. The dynamics of a quantum wave packet
governed by
a nonlinear
Hamiltonian provides adequate scope for such an investigation to be
carried out, as a wide variety of non-classical effects such as
revivals and fractional revivals (for reviews see, e.~g., 
\cite{aver, 
bluhm, robi}), as well as 
squeezing  
(for a review see, e.~g., \cite{dodo1}), are
displayed by the wave packet as it evolves in time.

While a generic
initial wave packet 
(corresponding to the state
$\ket{\psi(0)}$) governed by a nonlinear
Hamiltonian spreads rapidly during its evolution,
it could  return to its original
state (apart from an overall phase)  at 
integer multiples of a
revival time $T_{\rm rev}$, under 
certain conditions. This is signaled by
the return of $|\inner{\psi(0)}{\psi(t)}|^2$  to its initial
value of unity at $t = n T_{\rm rev}$.
The role played by the non-quadratic terms in the quantum Hamiltonian on 
the dynamics of a wave packet as it spreads and 
loses its original 
shape is best understood by examining the dynamics of appropriate 
expectation values and the consequences of Ehrenfest's theorem. 
Consider the 
simple one-dimensional example of 
a particle of 
mass $m$ subject to a 
potential $V(x)$. The Hamiltonian is
\begin{equation}
H = \frac{p^2}{2m} + V(x),
\label{ehrenfesthamiltonian}
\end{equation} 
where $x$ is the position operator and $p$ is the momentum operator.  
Ehrenfest's theorem states that  
\begin{equation}
\frac{d\aver{x}}{dt} = \frac{\aver{p}}{m},
\end{equation}
\begin{equation}
\frac{d\aver{p}}{dt}  = -\aver{\frac{\partial V(x)}{\partial x}}.
\end{equation}
For quadratic Hamiltonians, the latter equation gets simplified to
\begin{equation}
\frac{d\aver{p}}{dt}  = -\frac{\partial V(\aver{x})}{\partial x}.
\label{ehrenfestquad}
\end{equation}
Correspondingly, these expectation values 
follow the classical motion. For 
non-quadratic Hamiltonians, (approximate) 
classical motion arises only if the 
potential can be approximated to the quadratic form. In fact,  
certain techniques for the semi-classical treatment of the propagation of 
wave 
packets 
\cite{keller, deschamps} are based on the replacement
of the
exact Hamiltonian with a tractable, 
usually quadratic, approximation 
governing the evolution of the wave packet.  
Further, the quadratic 
potential turns out to be the most general 
potential under which a 
Gaussian wave packet continues to remain a Gaussian. 
In fact,   
there is an interesting parallel \cite{littlejohn} between the   
following two situations: 
The quadratic approximation given in Eq. 
(\ref{ehrenfestquad}), and the 
retention of the 
Gaussian shape under a quadratic potential, 
on the one hand; and  
the role of the second-order term in a series 
expansion for the classical 
Hamiltonian in the linearized behavior of classical orbits close to some 
reference orbit in phase space, on the other. 

The eventual spreading of a generic wave packet 
(which could initially follow 
classical motion for a brief period of time) 
is inevitable if a quadratic 
approximation does not hold good. It has been argued \cite{littlejohn} 
that such wave packet spreading in both position and momentum space should 
also be describable in a classical framework
by considering an ensemble of nearby classical orbits. Despite the 
spreading phenomenon, interesting quantum 
interference phenomena \cite{tara} can lead not only to wave 
packet revivals, but also to 
fractional revivals between two successive 
revivals at specific instants. The latter is
characterized by the splitting  
up of the initial wave packet into a number of
spatially distributed sub-packets,
each of which is similar to
the original wave packet. Classical analogs of 
revival phenomena can also be envisaged. For 
instance, one can model the individual energy 
eigenstates comprising the wave packet as 
an ensemble of racing cars on a 
circular track with different speeds \cite{nauenberg}. 
The quantum mechanical spreading of 
the wave packet  arises due to the difference 
in speeds. However, certain distinct 
time scales appear in the problem: The 
initial classical 
periodicity may be retained for a few laps, 
while for longer times no 
correlations between the various racers are observed --- 
but obvious patterns 
can recur, including the original form (revivals), as well as 
small similar groups of 
racers clumped together at certain instants 
(fractional revivals). Another 
interesting analogy exists between wave packet 
revivals (in {\it time}) and the 
recurrence of images (in {\it space}) 
in the classical Talbot effect. 
This is the repeated self-imaging 
of a diffraction grating at multiples of a 
certain fundamental distance 
from the source due to 
coherent interference of waves, akin to 
quantum revivals of a wave 
packet at multiples of the 
fundamental time $T_{\rm rev}$. Further, 
in between the appearance of two such 
self-images, Talbot images (which are 
superposed copies of the original) appear at definite spatial positions, 
akin to fractional revivals of the quantum wave packet 
at certain instants between two successive revivals (see, for instance, 
\cite{berry3, banaszek}). 

In this thesis, 
we focus on the quantum revival phenomena and other 
non-classical effects that arise close to instants of 
revivals and 
fractional revivals, and the 
information that can be obtained 
about these effects by examining the 
expectation values and higher moments 
(or cumulants) of operators corresponding to 
appropriately chosen observables. 
It is evident that, at full revivals 
of any initial state,
the expectation values of all 
observables return to their
initial values.
Thus, in a ``phase space'' where these expectation values
play the role of classical 
dynamical variables, the trajectories
are closed orbits, indicative of
the regular temporal evolution
of the corresponding dynamical system.
In principle, such a classical phase space 
would be infinite-dimensional, 
because the full information contained in a 
specific quantum state can 
only be obtained if the 
expectation values and all the 
higher moments of all 
operators pertaining to the system are included in the set of  
dynamical variables. A more practicable approach, 
however, is to use the time 
series of an available set of expectation 
values pertaining to the system,
in order to estimate the embedding 
dimension of the relevant phase 
space, as is done when carrying out a time-series analysis to determine 
the 
extent of randomness of a given set of signals.   

Interesting signatures of different
fractional revivals also manifest themselves in
the dynamics of the
expectation values of various operators. For 
one-dimensional systems with Hamiltonians that 
can be written in the form given in  
Eq.(\ref{ehrenfesthamiltonian}), 
the Ehrenfest equations can 
be generalized to include the temporal evolution 
of the second moments of 
$x$ and $p$. These are given by 
\begin{equation}
\frac{d\aver{x^2}}{dt}=\frac{1}{m}\aver{xp+px},
\end{equation}
\begin{equation}
\frac{d\aver{xp}}{dt}=\frac{d\aver{px}}{dt}=
\frac{\aver{p^2}}{m}+\aver{xF(x)},
\label{ehrenfestxp}
\end{equation}
and
\begin{equation}
\frac{d\aver{p^2}}{dt}=\aver{pF(x)+F(x)p},
\end{equation}
where $F(x)= -dV/dx$.
 These equations have been used to examine the role of 
quantal uncertainties 
in wave packet dynamics in the case of the free particle and 
the harmonic oscillator \cite{styer}.
We have examined the role of expectation values in a more general 
setting by 
considering a nonlinear Hamiltonian which 
is not of the form 
given in Eq. (\ref{ehrenfesthamiltonian}). 
We have established
that, by tracking the
time evolution of various
moments of certain operators, selective
identification of different fractional revivals can be achieved 
\cite{sudh1}.

While revival phenomena have been investigated extensively 
in 
the context of Rydberg atoms, both theoretically and experimentally 
(see, e.~g., \cite{aver1,park1,yeaz1,yeaz2}), 
for our purposes we have studied 
the dynamics of an initial 
wave packet (a single mode of an electromagnetic field) propagating in a 
Kerr-like medium.   
This is motivated by the fact that non-classical effects 
displayed by 
optical fields have also been investigated 
intensively for well over a decade now. The  
model that we consider (in the first part of this thesis) 
neatly captures the important role played by 
expectation values of higher moments of operators 
in wave packet dynamics: we examine 
the dynamics of the field mode 
under an effective Hamiltonian 
\begin{equation}
H = \hbar \chi a^{\dagger 2}\, a^2\,,
\label{singlemodehamiltonian}
\end{equation} 
where $a$ 
and $a^\dagger$ are photon destruction and creation operators.  
The starting point in obtaining this 
effective Hamiltonian \cite{tanas}, which is written in 
terms of the field operators alone (without explicitly
invoking the field-medium interaction terms), is the set of 
classical field equations pertaining to the 
optical Kerr effect, namely, 
the dependence of
the refractive index of the medium on the intensity of the light 
propagating through it. While this effect, 
displayed by certain nonlinear 
optical materials (Kerr media), can 
be explained within a classical framework,  
the field amplitudes are replaced in the standard manner by 
photon destruction and creation operators, in order to take into 
account the quantum nature of the radiation field. The 
effective quantum Hamiltonian is constructed subject to the
requirement that the Heisenberg equations for the field 
operators 
yield the correct dynamical equations in the classical limit. A 
`Kerr-like' 
medium is one for which the above effective Hamiltonian is a 
good approximate Hamiltonian. In 
this thesis, we shall use the words `Kerr medium' and `Kerr-like medium' 
interchangeably. 

We note, in passing, that fractional revivals of an initial 
Gaussian wave 
packet propagating under the Kerr Hamiltonian 
have been considered to be 
potentially useful in quantum computation. 
They play a significant role 
in the context of quantum cloning with continuous variables \cite{cerf}. 
Fractional revivals corresponding to the 
appearance of two and four 
superposed 
sub-packets are particularly useful in implementing 
one- and two-bit logic 
gates \cite{shapiro}. Further, by combining 
these superposed states with the vacuum using  
a $50\%$ beam splitter, the output state can be shown to 
be an entangled state (a Bell state) under certain conditions 
\cite{vanenk}.  

Another aspect which we have investigated 
in some detail in this thesis, in the context 
of wave packet evolution in the Kerr-like medium, 
is the precise role played by the 
initial state in determining the non-classical features 
exhibited by the time-evolved state. Taking the 
standard oscillator coherent state (abbreviated as CS 
throughout this thesis) as the reference initial state for 
the purposes of comparison, we have examined the extent to which  
the departure from coherence of different  
initial states affects the dynamics of the wave packet. 
In particular, we have identified, both qualitatively and 
(wherever possible) quantitatively, 
the manner in which effects such as 
revivals, fractional 
revivals and squeezing of the state at certain 
instants depend on the 
degree of coherence of the initial state. The  initial states 
considered for this purpose are the 
photon-added coherent states 
(abbreviated as PACS throughout this thesis). 
By repeated 
application of the photon creation operator on 
a CS, photons can be 
added to it systematically, and 
a quantifiable departure of the state from 
perfect coherence 
obtained as a consequence. The recent experimental 
production and complete characterization 
of the single-photon added coherent 
state by quantum state tomography 
\cite{zavatta} adds impetus to our investigation. 
The single-photon added 
coherent state is particularly interesting, 
as it exhibits marginal 
departure from coherence, and is 
intermediate between a one-photon Fock 
state (that is purely quantum mechanical) 
and a CS (a minimum uncertainty, ``classical'' state). 
Hence, in principle, this state offers a convenient 
tool to follow the smooth 
transition from the particle to the wave nature of light.We have shown  
that expectation values computed in the case of an initial PACS 
at various instants during its 
temporal evolution in a Kerr-like medium 
offer a means to assess the effect of 
imperfect coherence of the 
initial state upon its subsequent dynamics \cite{sudh2}. 

A detailed analysis of the squeezing and higher-order squeezing
properties of the wave packet in the neighborhood of
a two-sub-packet fractional revival provides
another method to 
ascertain whether the initial state is coherent or not 
\cite{sudh3}.  
A quantifier to assess 
the extent of non-classicality of a state can be identified  
\cite{ken} in terms of the Wigner function 
corresponding to that state. We have used this quantifier 
to also identify the manner in 
which non-classical effects in 
the dynamics of the wave packet are affected, 
when the initial state is itself a 
non-classical state such as a PACS.   
The Wigner function is a quasi-probability 
distribution function of (a pair of)  
phase space variables, 
and could become negative 
in some regions of the  
phase plane. It is well known that 
the Wigner function corresponding to a CS (which is 
considered to be a classical state) is positive 
in all regions of the phase plane. This remains true for 
the Wigner function of the squeezed vacuum state 
as well. It is therefore not unreasonable to 
regard the negativity of this  
function in some region of the phase plane 
as an indicator of the non-classical nature 
of the state concerned.  
Further, the Wigner function can be reconstructed 
from experimentally measured quantities: for instance,  
that of a
one-photon Fock state has been reconstructed using
a phase-randomized pulsed optical 
homodyne method \cite{lvovsky}.
The indicator \cite{ken} 
based on the Wigner function that we have used 
is merely one way of quantifying the extent of 
the non-classicality of a state, but it 
suffices for our 
purposes. The Sudarshan-Glauber $P$-function \cite{mandel}, for 
instance, would provide another 
measure of the non-classicality of a state. 

The dynamics of a single mode of the radiation field 
enables us to understand, 
to a considerable extent, 
the connections between the behavior of 
quantum expectation values and revival phenomena.
When two-mode interactions are considered, another 
significant aspect of quantum physics  appears --- 
namely, entanglement. We have examined the  
conditions under which  
analogs of revivals and fractional revivals 
of the initial state occur  
in this case. Revivals of two initially non-entangled modes governed by a 
nonlinear Hamiltonian have been examined in the literature \cite{sanz}. 
However, in order to study the full spectrum of randomness exhibited by 
the expectation values of relevant observables as they evolve in time, we 
need a system in which revival phenomena can either occur or be 
suppressed, depending on the values of the parameters in the Hamiltonian 
considered. A good candidate Hamiltonian for this purpose 
is obtained by a straightforward extrapolation of the single-mode 
example considered earlier, to explicitly take into account the atomic 
modes in the Kerr-like medium through which a single-mode electromagnetic 
field propagates, and the interaction between the 
atoms and the field. The Hamiltonian for the total system 
is given 
by \cite{agar1}
\begin{equation}
H = \hbar \omega \,a^\dagger a
+ \hbar \omega_0\,b^\dagger b + \hbar \gamma\,
b^{\dagger 2}\,b^2 +
\hbar g \,(a^\dagger b + b^\dagger a).
\label{2modehamiltonian}
\end{equation}
As before, the ladder 
operators $a$ and $a^\dagger$ pertain to the
field, while $b$  and $b^\dagger$
are the ladder operators for the
nonlinear oscillator modeling an atom of the medium.
$\gamma$ is the
anharmonicity parameter, 
and $g$ is the strength of 
the interaction
between the field and the medium.
It is easily verified that the operator
$(a^{\dagger} a + b^{\dagger} b)$ commutes with
$H$. We have analyzed this model 
for initial states that are
direct product states, in which the atom is in its
ground state, while the field is, respectively, in a
Fock state, a coherent state, and a photon-added coherent
state. Detailed results have been obtained on the precise
manner in which the moments of appropriate quadrature variables display
the counterparts of the revival and
fractional revival phenomena in this case. These phenomena arise for 
weak nonlinearity, i.e., for 
sufficiently small ($\ll 1$) values of the ratio
$\gamma/g$.
                                                           
We have shown that these revival phenomena also manifest 
themselves
in the entropy of entanglement of the system.
The latter may be taken to be either
the sub-system von Neumann entropy 
or the sub-system linear entropy, 
where by ``sub-system'' we mean either the 
field or the medium. These entropies are defined 
in terms of the corresponding 
reduced density operators. Striking differences arise
in the behaviour of the entropies for 
different initial states
of the field. These have been described
in detail in the thesis.

The situation that prevails for sufficiently 
large values of $\gamma/g$ is 
especially interesting. We have carried out a 
time-series and power 
spectrum
analysis \cite{abar, grass,fraser} of the
time series generated by the
values of the mean photon
number $\aver{a^{\dagger}\,a}$ 
computed over long intervals of time, 
and assessed the effect of the 
departure from coherence of the 
initial field mode on the ergodicity 
properties of the mean photon number \cite{sudh4}.  
The time series analysis shows that
a wide range of behavior can manifest
itself in the dynamics of the sub-system
represented by the field mode:
a progression occurs, from mere ergodicity
for a coherent initial state and weak
nonlinearity, all the way to
``chaotic'' behavior (as characterized 
by a positive Lyapunov
exponent) for a sufficiently
large departure from coherence of the
initial state (i.~e., a PACS with a relatively large
number of added photons) 
and a sufficiently large nonlinearity
(as given by the ratio $\gamma/g$). 
This is corroborated by the fact that 
similar behavior is
exhibited by $\aver{b^{\dagger} b}$, in such a manner that
$\aver{b^{\dagger} b} + 
\aver{a^{\dagger} a}$
remains constant, as required. 
The implications of ``chaotic'' 
dynamics, and what it really means in this 
context, are of special interest. 
A discussion of these aspects is given in the relevant 
chapter of the thesis.  

A summary of the contents of the contents of the 
rest of this thesis is as follows: 

{\bf Chapter 2} is devoted to a discussion of
the manner in which distinctive signatures of
revivals and fractional revivals  are displayed in the time evolution of
appropriate observables, which help in selectively identifying
different fractional
revivals \cite{sudh1}. The model Hamiltonian used is
that of a single mode
propagating in a Kerr-like medium, 
Eq. (\ref{singlemodehamiltonian}), 
while the wave packet considered
corresponds to a coherent initial state.

\noindent
{\bf Chapter 3} describes  how revivals and fractional revivals
of a wave packet propagating in the nonlinear medium
provide signatures of the degree of
coherence of the initial wave packet.
The investigation focuses on revival phenomena
in the case of an initial photon-added coherent state.
A comparison vis-\`a-vis
the corresponding
results obtained for an initial coherent state
enables the elucidation of the role played by
the coherence of the initial state in determining the
revival properties of the state \cite{sudh2}.

\noindent
{\bf Chapter 4} is concerned with
the squeezing and higher-order
squeezing
properties of a PACS as it propagates in the nonlinear medium. Once
again, the precise manner in which departure from coherence
of the initial state
affects its squeezing properties
at the instants of revivals and
fractional revivals is brought out \cite{sudh3}.

\noindent
{\bf Chapter 5} is based on the Hamiltonian 
in Eq. (\ref{2modehamiltonian}), that
describes the entanglement dynamics of a single-mode
electromagnetic field interacting with a
nonlinear medium, an atom
of the latter being modelled as an anharmonic oscillator.
Revival phenomena are now examined in terms
of the expectation values
of certain observables, as well as the sub-system
von Neumann entropy and the sub-system linear entropy.
The initial states considered are direct product states,
in which the atom in its ground
state, while the field is taken to be in either a number state,
or a CS, or a PACS. Once again, this leads to
an understanding of the role played by the initial state.

\noindent
{\bf Chapter 6} deals with the case in which
revivals of the
initial state are suppressed.
The dynamical behaviour of the mean photon number
is investigated \cite{sudh4} by means of a time-series
analysis, in order to
identify the effects of different initial states on the
entanglement and ergodicity properties of the system. 
This behavior is seen to
range from quasi-periodicity to exponential instability,
depending on the specific initial state
and the values of certain parameters
occurring in the Hamiltonian.

\noindent
{\bf Chapter 7} concludes the thesis with some brief remarks placing
this work in a broader perspective, and a list of open problems and 
avenues for further research.

%% file: manifest.tex
\chapter{Manifestations of wave packet revivals in the 
moments of observables - I}\label{manifest}

%----------------------------SECTION-------------------------------%
\section{Introduction}
%------------------------------------------------------------------%

Revivals and fractional revivals of a wave packet 
propagating in a nonlinear medium have been extensively investigated 
in the literature \cite{robi}. Since  revival phenomena  
arise from  quantum interference, it is of interest to examine  
precisely how  their  occurrence is captured in the expectation 
values of appropriate 
physical observables. Such investigations have led to a better 
understanding of different aspects of  revival phenomena in a wide 
variety of  
physical situations, ranging from      
wave packet propagation in the infinite square well \cite{robi1} to the 
long-time behavior of a  
quantum bouncer \cite{don}.  
In this Chapter, we discuss 
the manner in which distinctive signatures of 
revivals and fractional revivals  are displayed in the time evolution of 
certain observables, thereby facilitating the  selective identification of  
different fractional 
revivals \cite{sudh1}.  
It is necessary to bear in mind  the salient features of the  
specific quantum interference 
 between the basis states comprising the wave packet
when we analyze the dynamics of 
expectation values. For this purpose, we outline in the next Section 
the conditions under which revivals and fractional revivals of a wave packet 
occur.

\section{Revivals and fractional revivals in wave packet dynamics}
The broad features of wave packet dynamics are 
quite generic, 
regardless 
of the details of the physical system concerned, the initial 
wave packet 
considered, and the specific nonlinear Hamiltonian 
governing the time evolution. We recall from Chapter 1 that, given an 
initial state 
$\ket{\psi (0)}$, a revival is the return of the overlap function 
\begin{equation}
C(t)
=\left|\langle \psi(0)|\psi(t)\rangle\right|^2
\end{equation}
 to its initial
value of unity at specific instants of time.
Revivals and fractional revivals are essentially controlled by the
parameters occurring, respectively, in the first and second-order 
terms in the Taylor
expansion of the energy spectrum
$E_{n}$ about the energy $E_{n_{0}}$ corresponding to the peak of
the wave packet. Hence, it suffices to consider wave packet evolution 
governed by an effective 
Hamiltonian whose energy eigenvalues are at most quadratic functions 
of $n$. Further, one of our objectives is to examine the 
role played by the initial state 
in the  subsequent dynamics. Therefore, for purposes of comparison, an 
appropriate 
reference state must be selected.  
A good starting point  is the investigation of the revival phenomena
displayed by an initial coherent state (CS)
propagating in a Kerr-like medium, so as to facilitate subsequent 
comparison with the dynamics of initial states which depart in a  
quantifiable manner
from coherence.  We explicitly demonstrate below the 
role played by quantum interference in this context \cite{tara}.

The effective Hamiltonian for the propagation of a single-mode 
electromagnetic field in a 
Kerr medium is given by \cite{agar2, kita}
\begin{equation}
H = \chi \,a^{\dagger 2}\,a^2=\chi \,{\sf N}({\sf N}-1)
\label{kerrhamiltonian}
\end{equation}
in the usual
notation, where $a$ (respectively, $a^{\dagger}$ ) 
is the photon annihilation 
(respectively, creation) operator, ${\sf N} = a^{\dagger}a$ is 
the photon number operator,  
and  $\chi$ ($>0$) is essentially the 
third-order nonlinear susceptibility of the medium.
Here, and in the rest of this thesis, 
$\hbar$ has been set equal to unity. We shall 
restore this factor when it becomes essential to do so 
(e.~g, in Sec. 6.1 of Chapter 6).  
This Hamiltonian is also 
relevant in a very different physical context, namely, in describing the 
dynamics of a Bose-Einstein condensate in a potential well. In that case, 
$a$ and $a^{\dagger}$ are boson annihilation and creation operators, and 
$\chi$ characterizes the energy needed to overcome the inter-atomic 
repulsion in adding an atom to the population of the potential well 
\cite{grei}.  Many of the results that we derive may therefore be expected 
to be applicable in the context of atom optics and Bose-Einstein 
condensate, as well.

The initial state of the field  
$\ket{\psi(0)}$ is taken to be the   
CS $\ket{\alpha}$, where $\alpha$ is any complex number. It satisfies
\begin{equation}
a\ket{\alpha}=\alpha\ket{\alpha}.
\label{annihilationoperatoreigenstate}
\end{equation}
The Fock state representation of the CS, which we will 
use extensively, is  
given by
\begin{equation}
\ket{\alpha}=e^{-|\alpha|^2/2}\,\sum_{n=0}^{\infty}\frac{\alpha^n}{\sqrt{n!}}
\,\ket{n}.
\end{equation}
The unitary time evolution operator $U(t)$ corresponding to the 
Hamiltonian of Eq. (\ref{kerrhamiltonian}) is 
\begin{equation}
U(t) = \exp\,[-i \chi t{\sf N}({\sf N}-1)].
\label{unitaryoperator}
\end{equation}
Hence the state of the field at time $t$ is 
given by 
\begin{equation}
\ket{\psi(t)} = \exp\,[-i \chi t{\sf N}({\sf N} - 1)] \ket{\alpha}.
\label{wavepacketpsi(t)}
\end{equation}
The time evolution operator at instants $t = \pi/(k \chi)$ (where $k$ is an
 integer), namely,
\begin{equation}
U (\pi/k\chi) = \exp\,[- {\frac{i \pi}{k}} {\sf N}({\sf N}-1)],
\label{unitaryoperator}
\end{equation}
displays interesting periodicity properties. These follow from the fact 
that the 
eigenvalues of ${\sf N}$ are integers. We will use $N$ to denote 
these eigenvalues.  
For odd integer values of $k$,
\begin{equation}
\exp\,[-{\frac{i \pi}{k}} (N + k) (N + k - 1)] = 
\exp\,[{\frac{-i \pi}{k}}N(N-1)]. 
\label{periodicitypropertyoddm}
\end{equation}
Similarly, for even integer values of $k$,
\begin{equation}
\exp\,[-{\frac{i \pi}{k}} (N + k)^2] =  \exp\,[-{\frac{i \pi}{k}}N^2].
\label{periodicitypropertyevenm}
\end{equation}
As a consequence, $U(\pi/k \chi)$ can be expanded in a  Fourier 
series with $\exp\,(-2 \pi i j/k)$ as the basis functions,
in the form
\begin{eqnarray}
\exp\,[-\frac{i\pi}{k} N(N-1)]&=&\sum_{j=0}^{k-1}f_j \exp\,[-\frac{2\pi 
ij}{k} N]
\end{eqnarray}
for odd values of $k$, and
\begin{eqnarray}
\exp\,[-\frac{i\pi}{k} N^2]&=&\sum_{j=0}^{k-1}g_j \exp\,[-\frac{2\pi
ij}{k} N]
\label{fourierexpansion}
\end{eqnarray}
for even values of $k$, respectively,
 where the coefficients $f_j$ and $g_j$ are known.
Using the above equations and the property
\begin{equation}
e^{i\chi a^\dagger a}\ket{\alpha}=\ket{\alpha e^{i \chi}},
\end{equation}
we find
\begin{equation}
\ket{\psi(\pi/k \chi)}=\left\{\begin{array}{ll}
\sum_{j=0}^{k-1} f_{j}\ket{\alpha \,e^{-2\pi ij/k}},\quad&
k \quad{\rm odd;} \\[10pt]
\sum_{j=0}^{k-1} g_{j}\ket{\alpha \,e^{i\pi/k}\,e^{-2\pi ij/k}},\quad&
k \quad{\rm even.} \end{array} \right.
\label{superpositionofcs}
\end{equation}

It is evident that at time $T_{\rm rev} = \pi/\chi$, corresponding 
to 
$k = 1$, the initial state revives for the first time. Periodic 
revivals occur at integer multiples of $T_{\rm rev}$. 
Further,  between $t = 0$ and $t = T_{\rm rev}$ and
between two successive revivals, fractional revivals 
occur at instants $n\,T_{\rm rev} + T_{\rm rev}/k$.
At these instants, the initial wave packet evolves to a state that can 
be described 
as a finite superposition of `rotated' coherent states with definite 
amplitudes. 
For instance,
the state at time
$\pi/(2\chi)$ (corresponding to $k = 2$)
is a superposition of the two
coherent states $\ket{i\alpha}$ and $\ket{-i\alpha}$. In general,
$\ket{\psi(\pi/k \chi)}$ is a superposition of $k$ coherent 
states.
The
corresponding wave packet in position space is
a superposition of $k$ spatially distributed Gaussian wave packets.  
The periodicity property of $U$ further implies
that the wave function at times $t=\pi j/(k \chi),\, 1\leq j \leq k-1$
for a given $k$, is also a superposition of $k$ wave packets \cite{aver1}.

\section{Signatures of revivals and fractional revivals in quantum 
expectation values}

We now proceed to examine the manner in which fractional revivals are 
manifested in distinctive ways in the expectation values of the physical 
observables pertaining to the system at hand. As the system enjoys 
revivals with a period $T_{\rm rev}$, all such expectation values 
are 
periodic functions of $t$ with this fundamental period. 
A family of relevant observables is provided (in the context of
nonlinear media) by the field
quadrature $(a\,e^{i\varphi} + a^{\dagger}\,e^{-i\varphi})$ for
various values of the phase $\varphi.$ In the context of 
Bose-Einstein condensation, the
expectation value $\aver{a(t)}$ of the atom annihilation operator 
represents the condensate wave function. Its real and
imaginary parts (which correspond to the cases $\varphi = 0$ and
$\varphi = -\frac{1}{2}\pi$, respectively) are analogous to classical
phase space variables. Let us define the operators 
\begin{equation}
x = \frac{(a +
a^{\dagger})}{\sqrt{2}}\quad {\rm and} \quad p =\frac{(a -
a^{\dagger})}{i\sqrt{2}}.
\label{quadraturevariable}  
\end{equation}
Clearly, their expectation values alone do not suffice to reproduce the full
information contained in the wave function itself. In principle, 
an infinite set of
moments, comprising the expectation values of {\em all} powers of $x$
 and $p$ and
their combinations, is required for this purpose.  In this sense, the
quantum system is equivalent to an infinite-dimensional classical
system in which the role of the dynamical variables is played by the
set of expectation values. However, we emphasize that even the first few
moments can be shown to yield considerable
information on the behavior of the system. 

Recalling that $a\ket{\alpha} = \alpha \ket{\alpha}$, 
we define the $c$-number
function 
\begin{equation}
\alpha(t)=\langle{\psi(t)}|{a}{\ket{\psi(t)}}=
\langle{\alpha}|{e^{iHt/\hbar}\,a\,e^{-iHt/\hbar}}
{\ket{\alpha}},
\label{alphat1}
\end{equation}
so that $\alpha (0) \equiv \alpha$. 
For the case at hand, this simplifies after some algebra to 
\begin{equation}
\alpha(t)=\alpha \,e^{ -\modu{\alpha}^2(1-\cos 2\chi t)}\, 
\left[ \cos \,\big(\modu{\alpha}^2\,\sin \,(2\chi t)\big)
-i\sin\,\big(\modu{\alpha}^2\,\sin \,(2\chi t)\big)\right].
\label{alphat3}
\end{equation}
Thus $\alpha(t)$ is a periodic function of time with period 
$\pi/\chi$, as expected.
It is convenient to introduce the notation
\begin{equation} 
\alpha =\alpha_1+i\alpha_2 =\frac{(x_0 + i p_0)}{\sqrt{2}}
\label{x0p0}
\end{equation}
and 
\begin{equation}
\nu=\modu{\alpha}^2 = \frac{1}{2}(x_{0}^{2}+p_{0}^{2}).
\end{equation}
$x_0$ and $p_0$ represent the locations of the 
centers of the initial Gaussian wave packets
corresponding to the CS $\ket{\alpha}$ in position space and momentum
space, respectively.  
The expectation values of $x$ and $p$ at any time can then 
be obtained as explicit
functions of $t$ (as stated in Appendix B) in the form
\begin{equation}
\aver{x(t)}=
e^{-\nu\,(1-\cos 2\chi t)}\,
\big[ x_0 \,\cos \,(\nu \sin 2\chi t)+p_0\, 
\sin \,(\nu \sin 2\chi t)\big],
\label{xt}
\end{equation}
\begin{equation}
\aver{p(t)}=
e^{-\nu(1-\cos 2\chi t)}\,
\big[ -x_0 \,\sin\, (\nu \sin 2\chi t)+p_0\, 
\cos \,(\nu \sin 2\chi t)\big].
\label{pt}
\end{equation}

Expressions for the higher 
moments can be deduced readily from the general result
\begin{equation}
\aver{a^{\dagger r}\,a^{r+s}} = \alpha^{s}\,\nu^{r}
e^{-\nu\,(1-\cos \,2 s \chi t)}\,
\exp \left[
-i \chi \big(s(s-1) + 2rs\big)\,t - i\nu \,\sin\,2s \chi t\right],
\label{nthmoment}
\end{equation}
where $r$ and $s$ are non-negative integers.
This result is derived in Appendix A (see Eq. (\ref{nthmomentappendix})).
It is a special case of a more general relation derived there 
(Eq. (\ref{generalmappendix})).
Using this result, the second moments of $x$ and $p$ are 
found to be  given by
\begin{eqnarray}
2 \aver{x^2(t)}&=& 1 
+ x_{0}^{2} + p_{0}^{2} + e^{-\nu\,(1-\cos \,4\chi t)}
\,\big[ (x_{0}^{2} - p_{0}^{2})\,\cos\,  
(2\chi t + \nu \sin \,4\chi t)\nonumber\\ 
&+& 2 x_{0} p_{0} \,\sin\, (2\chi t + \nu\sin \,4\chi t)\big],
\label{xsqrdt}
\end{eqnarray}
\begin{eqnarray}
2 \aver{p^2(t)}&=& 1 
+ x_{0}^{2} + p_{0}^{2} - e^{-\nu\,(1-\cos \,4\chi t)}
\,\big[ (x_{0}^{2} - p_{0}^{2})\,\cos\,  
(2\chi t + \nu \sin \,4\chi t)\nonumber\\ 
&+& 2 x_{0} p_{0} \,\sin\, (2\chi t + \nu\sin \,4\chi t)\big].
\label{psqrdt}
\end{eqnarray}
For reasons already mentioned, the higher moments also carry much
information of direct interest. We give below the expressions for the 
third and fourth moments as well, since we need them to obtain the 
skewness and kurtosis in all the cases of interest.  
The third moments can 
be written compactly in the 
form \begin{eqnarray}
4\aver{x^3(t)}&=&  e^{-\nu 
(1-\cos 6\chi t)}\big[(x_0^3-3x_0 p_0^2)
\,\cos\, (6\chi t+\nu\sin 6\chi t)\nonumber\\
& +& (3x_0^2 p_0- p_0^3) 
\sin \,(6\chi t+\nu\sin 6\chi t)\big]\nonumber\\
&+&6\nu \,\big[\aver{x(t)}(1 + 
\cos\, 2\chi t) + \aver{p(t)} 
\,\sin\,2\chi t)\big],
\label{xcubedt}
\end{eqnarray}
\begin{eqnarray}
4\aver{p^3(t)}&=&  e^{-\nu 
(1-\cos 6\chi t)}\big[(x_0^3-3x_0 p_0^2)\,\sin\, (6\chi 
t+\nu\sin 6\chi t)\nonumber\\
 &-&(3x_0^2 p_0- p_0^3)
\cos \,(6\chi t+\nu\sin 6\chi t)\big]\nonumber\\
&+&6\nu\,\big[\aver{p(t)} (1+\cos 2\chi t)-
\aver{x(t)}\,\sin \,2\chi t\big].
\label{pcubedt}
\end{eqnarray}
The fourth moments are given by 
\begin{eqnarray}
8\aver{x^4(t)}&=&
e^{-\nu(1-\cos 8\chi t)}\big[(x_0^4+p_0^4-6x_0^2p_0^2)
\cos\,(12\chi t+\nu\sin 8\chi t)\nonumber\\
&+&4(x_0^3p_0-x_0p_0^3)\sin\,(12\chi t+\nu\sin\, 
8\chi t)\big]\nonumber\\
&+&8\nu e^{-\nu(1-\cos\, 4\chi t)}[(x_{0}^{2} - p_{0}^{2})\,
\cos\,(6\chi t + \nu \sin \,4\chi t)\nonumber\\
&+&2x_0p_0\sin\,(6\chi t + 
\nu \sin \,4\chi t)]\nonumber\\
&+&24\aver{x^2(t)}+3(x_0^2+p_0^2)^2-6,
\end{eqnarray}
\begin{eqnarray}
8\aver{p^4(t)}&=&
e^{-\nu(1-\cos 8\chi t)}\big[(x_0^4+p_0^4-6x_0^2p_0^2)
\cos\,(12\chi t+\nu\sin 8\chi t)\nonumber\\
&+&4(x_0^3p_0-x_0p_0^3)\sin\,(12\chi t+\nu\sin\,
8\chi t)\big]\nonumber\\
&-&8\nu e^{-\nu(1-\cos\, 4\chi t)}[(x_{0}^{2} - p_{0}^{2})\,
\cos\,(6\chi t + \nu \sin \,4\chi t)\nonumber\\
&+&2x_0p_0\sin\,(6\chi t +
\nu \sin \,4\chi t)]\nonumber\\
&-&24\aver{x^2(t)}+3(x_0^2+p_0^2)^2+24(x_0^2+p_0^2)+18.
\end{eqnarray}
With these expressions,  
the variances
of $x$ and $p$ as functions of $t$, as also  
the skewness and kurtosis in each case,  
can be obtained.
The uncertainty product (i.~e., the product of the respective standard 
deviations)
$\varDelta x \,\varDelta p$, which initially has the minimum value
$\frac{1}{2}$, is of special interest. We do not write down the
lengthy expressions for these quantities here, but we shall comment upon
their time variation in the sequel, when discussing the numerical results 
for specific values of the parameters.

\section{Moments of $x$ and 
$p$: A classical interpretation}
Before discussing our results in detail, we digress briefly to compare 
the explicit solutions found for the expectation
values of $x$ and $p$ 
in Eqs. (\ref{xt}) and (\ref{pt}) with the
solutions that would have been obtained for $x(t)$ and $p(t)$
{\it had} the system been a {\it classical} one, governed by the 
classical
counterpart of the normal-ordered Hamiltonian (restoring the factor $\hbar$) 
$H =\hbar\chi \,a^{\dagger 2}\,a^2$, namely,
\begin{equation} 
H_{\rm cl} = \frac{1}{4} (x^2 + p^2)^2.
\end{equation} 
Although the equations of motion corresponding to $H_{\rm cl}$ are
nonlinear, it is evident that $x^2 + p^2$ is a constant of the motion,
so that the phase trajectories are circles. However, the 
frequency of motion is dependent on the initial conditions (i.e., the
amplitude of the motion), being
equal to $\nu = \frac{1}{2}(x_{0}^2 + p_{0}^2)$. This is, of course,
a well-known feature of nonlinear oscillators. But we note that
the actual solutions
for $\aver{x(t)}$ and $\aver{p(t)}$ in Eqs. (\ref{xt}) and
(\ref{pt}) are more complicated than the classical ones for $x(t)$ and
$p(t)$ under $H_{\rm cl}$. This is a consequence of the 
quantum mechanical nature of the system, over and above the
nonlinearity of $H$. 
However, the expressions for $\aver{x(t)}$ and $\aver{p(t)}$
 can be given the following 
interesting interpretation in classical terms. 
Define the (non-canonical) pair of classical dynamical variables 
\begin{equation}
X = \aver{x}\,e^{\nu\,(1-\cos 2\chi t)}\,,\quad
P = \aver{p}\,e^{\nu\,(1-\cos 2\chi t)},
\label{XPdefn}
\end{equation}
and the {\em re-parametrized time} $\tau = \sin\, (2\chi t)$. 
The initial values of these variables are again
$x_{0}$ and $p_{0}$, respectively. Equations (\ref{xt}) and (\ref{pt})
can then be re-written in the suggestive form 
\begin{equation}
X = x_{0}\,\cos\,\nu\tau + p_{0}\,\sin\,\nu\tau\,,\quad
P = -x_{0}\,\sin\,\nu\tau + p_{0}\,\cos\,\nu\tau.
\label{XPtau}
\end{equation} 
But these are the solutions to the system of equations
\begin{equation}
dX/d\tau = \nu P\,,\,dP/d\tau = -\nu X,
\end{equation}
describing a nonlinear oscillator of frequency 
\begin{equation}
\nu = 
\frac{1}{2}(x_{0}^2 + p_{0}^2),
\end{equation}
 in terms of
the transformed variables $(X,P)$ and the re-parametrized time $\tau$. 
At the level of the {\em first} moments, therefore, the system is
effectively a nonlinear oscillator after a suitable transformation
of the relevant variables, together with  a re-parametrization of the time.

%-----------------------SECTION---------------------------------%
\section{Signatures of fractional revivals}
%---------------------------------------------------------------%

We now discuss our results, using representative numerical values for 
the various parameters in the problem. 
Turning to the time dependence of the various moments of
$x$ and $p$, there are two striking features 
that underlie the essential point we wish to make. 
First, the higher the order of the moment (or cumulant), the 
more rapid is its variation, since the leading (highest) frequency 
in the $k^{\rm th}$ moment is $2k\chi$.
Second, the time dependence is strongly controlled by the 
factor $\exp\,[-\nu\,(1-\cos \,2 k \chi t)], \, k = 1,2,\ldots,  $ that 
modulates the oscillatory terms. While this permits substantial time
variation for sufficiently small values of $\nu$, it acts as a strong
damping factor for large values of $\nu$, {\it except when
$\cos\,(2m\chi t)$ is near unity}. As one might expect, this happens precisely 
at revivals (when $t = n\pi/\chi$, an integer multiple of $T_{\rm
rev}$). But it also happens --- in the $k^{\rm th}$ moment and 
{\it not} in the lower moments --- at the fractional revival
times $t = (n + j/k)T_{\rm rev}$. Thus, by setting $\nu$ at a suitably
large value, we can ensure that the moments are essentially static,
bursting into rapid variation at specific instants of time before
reverting to quiescence. 

These points are illustrated in the figures.
Owing to an obvious symmetry of $H$, the
moments of $x$ and $p$ behave in an 
essentially similar manner,
especially if we start with the symmetric initial condition 
$x_0 = p_0$. Without significant loss of generality, we restrict ourselves 
to this case in what follows. We have set $\chi = 5$ in the numerical 
results to be presented, but this is irrelevant, as all the plots
correspond to $t$ measured in units of $T_{\rm rev}=\pi/\chi$. 
We find that for very small values ($ \ll 1$) of $x_{0}$ and $p_{0}$
(i.~e., of $\nu$), the
nonlinearity of $H$ does not play a significant role, and the behavior
of the system is much like that of a simple oscillator. Interesting
behavior occurs for larger values of $\nu$. We therefore present
results for three typical values of the parameters
representing the initial conditions, namely:
\begin{eqnarray*}
&({\rm a})& x_0 = p_0 = 1 \Rightarrow \nu = 1;\\ 
&({\rm b})& x_0 = p_0 = \sqrt{10} \Rightarrow \nu = 10; \\
&({\rm c})& x_0 = p_0 = 10 \Rightarrow \nu = 100. 
\end{eqnarray*}
These correspond, respectively, to small, intermediate, 
and large values of $\nu$.
In all the ``phase plots'', the point representing the state at $t =
0$ is labeled A. 

\begin{figure}
\includegraphics[width=5.7in]{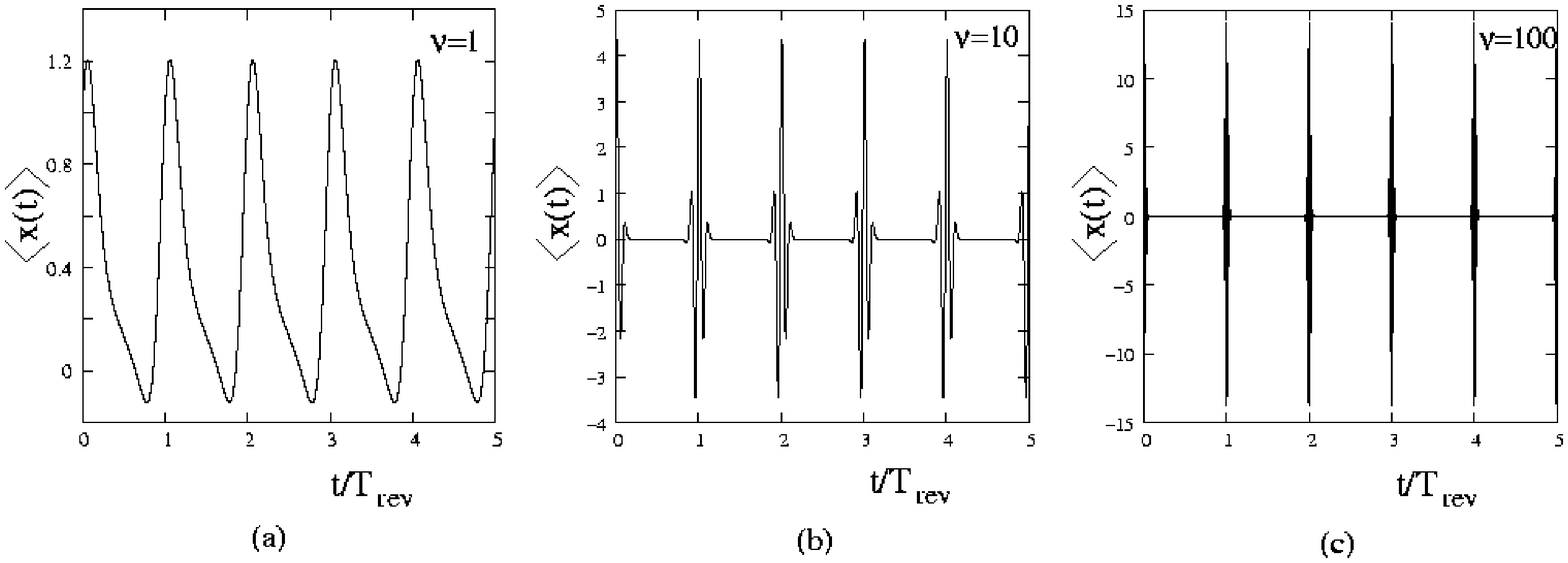}
\caption{$\aver{x}$ as a function of time}
\label{xversustime}
\end{figure}
Figures \ref{xversustime}(a)-(c) show the variation of $\aver{x(t)}$ 
as a function of $t$ for small, medium and large values of $\nu$.
(As already mentioned, $\aver{p(t)}$ displays similar behavior.)
For sufficiently large values of $\nu$, it is evident
that, except for times close to integer multiples of 
$T_{\rm rev}$, 
$\aver{x(t)}$ and $\aver{p(t)}$ essentially 
remain static at the
value zero. 
\begin{figure}
\includegraphics[width=6.2in]{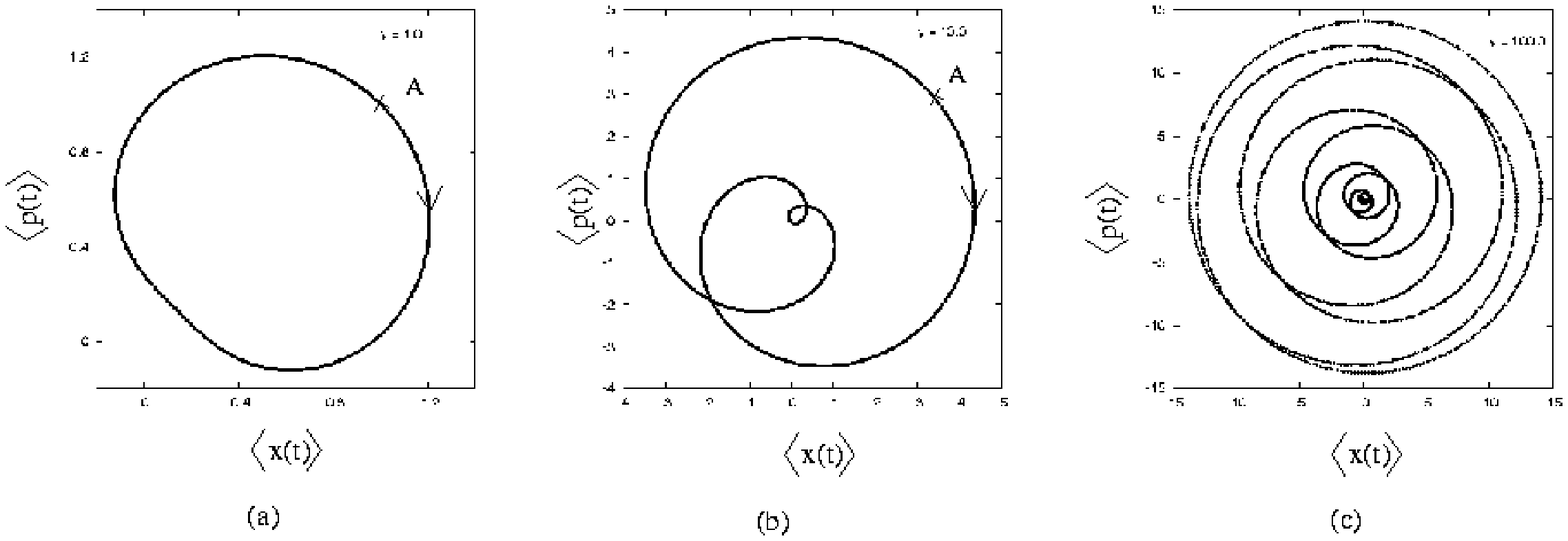}
\caption{``Phase plot'' of $\aver{p}$ versus $\aver{x}$}
\label{phaseplot}
\end{figure}
Figures \ref{phaseplot}(a)-(c) depict the corresponding 
``phase plots'' in the $(\aver{x}, \aver{p})$ plane. 
In Fig. \ref{phaseplot}(c), the representative point remains at the origin  
most of the time, except at times close to successive revivals,
when it rapidly traverses the rest of the curve before returning to
the origin. 

\begin{figure}
\includegraphics[width=5.7in]{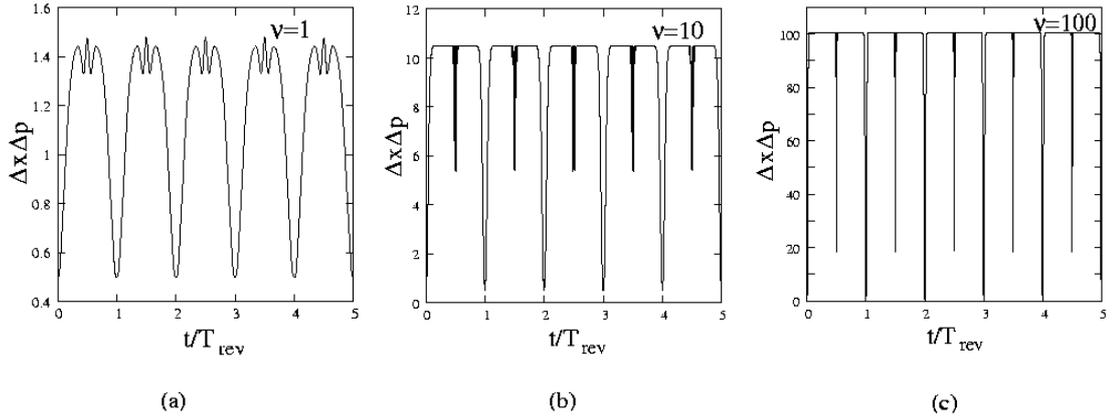}
\caption{Variation of the uncertainty product with time}
\label{deltaxdeltap}
\end{figure}

While sudden changes from nearly static values of $\aver{x(t)}$ 
and $\aver{p(t)}$ are thus signatures of revivals, the 
occurrence of fractional revivals is not captured in these mean 
values. The fractional revival 
occurring mid-way between successive revivals (e.~g., at $t=\pi/2\chi$ 
in the interval between $t = 0$ and $t = T_{\rm rev}$),  
when the initial wave packet reconstitutes itself into two separate 
wave packets of a similar kind, 
leaves its signature upon the second moments.
Figures \ref{deltaxdeltap}(a)-(c) show the variation with time of the 
uncertainty product 
$\varDelta x\, \varDelta p$. 
In each case, this product returns at every revival to its
initial, minimum, value ($=\frac{1}{2}$), rising to higher
values in between revivals. 
Once again, for sufficiently large values
of $\nu$, the product remains essentially static at the approximate
value $(\frac{1}{2} + \nu)$ for most of the time, but undergoes   
extremely rapid variation near revivals, and {\em also} near the fractional 
revivals occurring mid-way between revivals. 
During the latter,  
the uncertainty product drops to smaller values, but does not
reach the minimum value $\frac{1}{2}$. 

\begin{figure}
\includegraphics[width=6.2in]{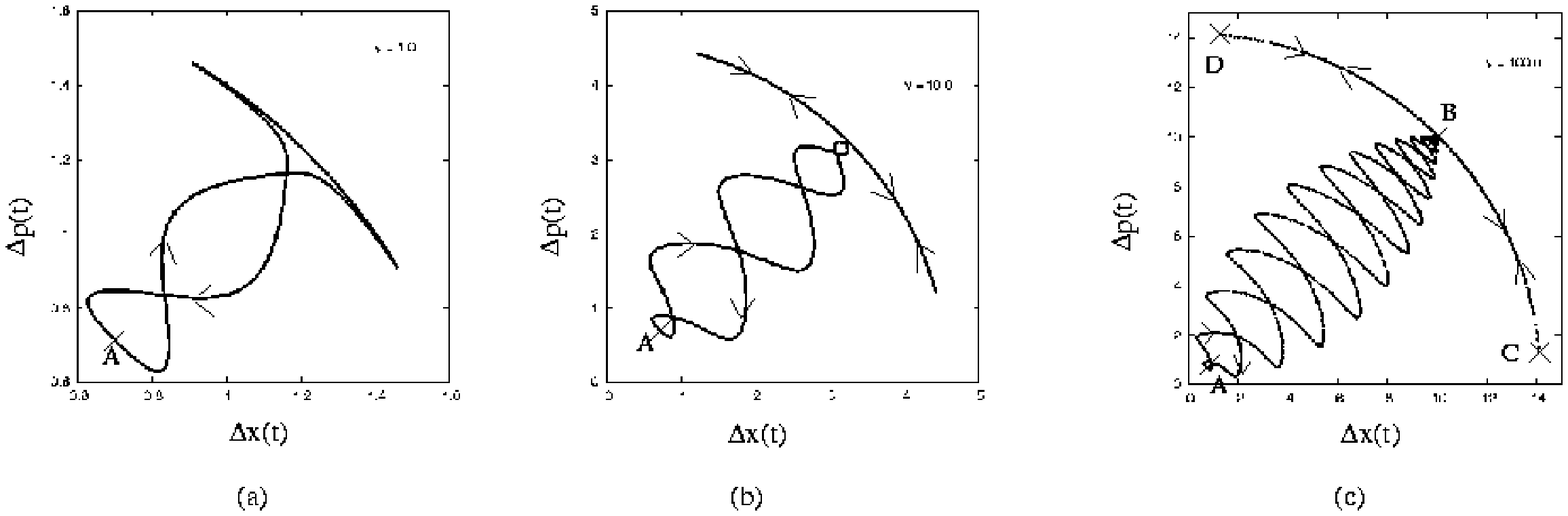}
\caption{``Phase plot'' of $\varDelta p$ versus $\varDelta x$
for $\nu=1, 10\, {\rm and}\, 100$, respectively.}
\label{deltapversusdeltax}
\end{figure}

There is a very striking 
difference in the behavior of the standard deviations 
near revivals as opposed
to their behavior near the foregoing fractional revivals. 
This is brought out in Figs. \ref{deltapversusdeltax}(a)-(c),
which is a ``phase plot'' of $\varDelta p$ versus $\varDelta x$.  
%\begin{figure}
%\includegraphics[width=6.2in]{deltapversusdeltax.eps}
%\caption{``Phase plot'' of $\Delta p$ versus $\Delta x$
%for $\nu=1, 10\, {\rm and}\, 100$, respectively.}
%\label{deltapversusdeltax}
%\end{figure}
For very small $\nu$, as in Fig. \ref{deltapversusdeltax}(a), 
$\varDelta x$ and 
$\varDelta p$ vary quite
gently around a simple closed curve. When $\nu$ is somewhat larger, as
in \ref{deltapversusdeltax}(b) which corresponds to $\nu = 10$, 
the plot begins to
show interesting structure. 
For much larger values of $\nu$ as in \ref{deltapversusdeltax}(c), 
the initial point A quickly moves out
on the zig-zag path about the radial $\varDelta p = \varDelta x$
 line
to the steady value represented by the point
B, and returns to A at every revival along the
complementary zig-zag path. 
Close to the fractional revival at $t = (n+\frac{1}{2})T_{\rm rev}$,
however, the representative point moves back and forth along the 
azimuthal path
BCDB rather than the zig-zag path: clearly, a kind of ``squeezing''
occurs, as one of the variances reaches a small value while the other
becomes large, and vice versa. (Of course the state of the system 
is far from a minimum uncertainty state throughout, except
at the instants $n T_{\rm rev}$.)  
\begin{figure}
\includegraphics[width=5.7in]{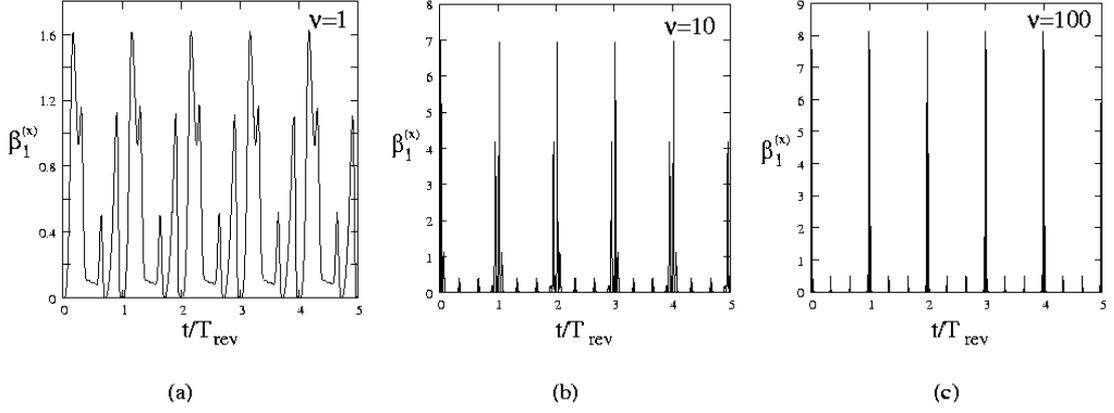}
\caption{Square of the skewness in $x$ as a function of time}
\label{skewness}
\end{figure}

The fractional revivals occurring at 
$t = (n + \frac{1}{3})T_{\rm rev}$ 
and $t = (n + \frac{2}{3})T_{\rm rev}$, at which  the initial wave packet 
is reconstituted into a superposition of {\it three} separate wave packets, 
are detectable in the third
moments of $x$ and $p$. To make this unambiguous, 
we may consider the
third moments about the mean values --- or, in standard statistical
notation, the square of the skewness, defined as 
\begin{equation} 
\beta_{1}^{(x)}
= \frac{\left\langle\big(x-\aver{x}\big)^3\right\rangle^2}
{\big(\aver{x^2}-\aver{x}^2\big)^3} 
\equiv \frac{\aver{(\delta x)^3}^2}
{(\varDelta x)^6},
\label{skew}
\end{equation}
and similarly for $\beta_{1}^{(p)}$. Figures \ref{skewness}(a)-(c) show the
variation of $\beta_{1}^{(x)}$ with $t$. 
%\begin{figure}
%\includegraphics[width=6.2in]{deltapversusdeltax.eps}
%\caption{``Phase plot'' of $\Delta p$ versus $\Delta x$
%for $\nu=1, 10\, {\rm and}\, 100$, respectively.}
%\label{deltapversusdeltax}
%\end{figure}
%\begin{figure}
%\includegraphics[width=5.7in]{skewness.eps}
%\caption{Square of the skewness in $x$ as a function of time}
%\label{skewness}
%\end{figure}
\begin{figure}
\includegraphics[width=5.7in]{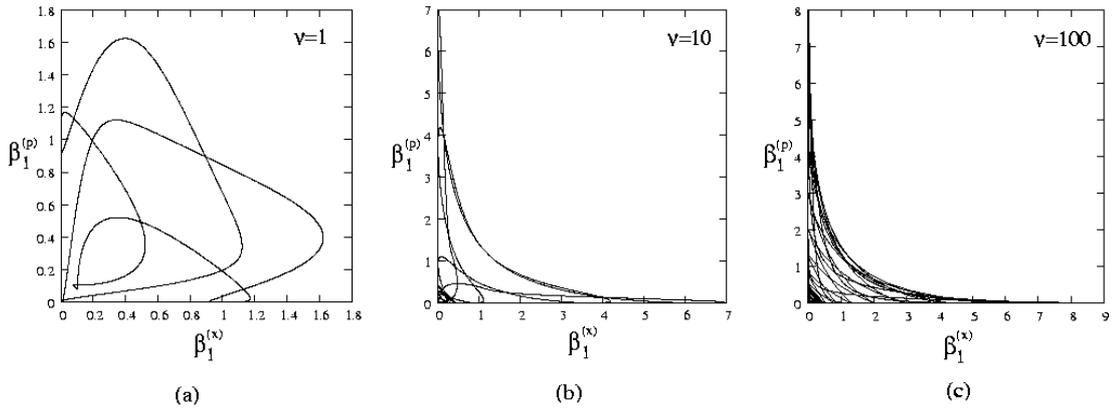}
\caption{The ``phase plot'' of $\beta_{1}^{(p)}$ versus 
 $\beta_{1}^{(x)}$}
\label{phaseplotbeta1}
\end{figure}
It is evident that, for
sufficiently large values of $\nu$, $\beta_{1}^{(x)}$ remains nearly
zero most of the time, except for bursts of rapid variation close to
revivals and fractional revivals. 
Both $\beta_{1}^{(x)}$ and $\beta_{1}^{(p)}$ actually vanish 
at $t = nT_{\rm rev}$ (since the Gaussian wave packet corresponding to 
the CS $\ket{\alpha}$ has no skewness in either $x$ or $p$),
 but they remain non-zero at $t = (n + \frac{1}{3}j)\,T_{\rm rev}\,, 
j = 1,2$.
More detailed information is obtained
from a ``phase plot'' of $\beta_{1}^{(p)}$ versus 
$\beta_{1}^{(x)}$, depicted in Figs. \ref{phaseplotbeta1} (a)-(c).

\begin{figure}[htpb]
\includegraphics[width=5.7in]{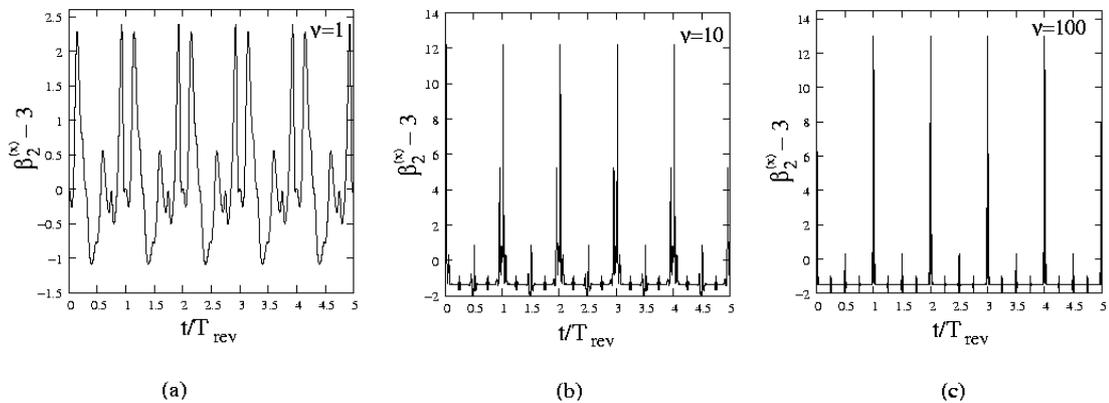}
\caption{The excess of kurtosis of $x$ as a function of time}
\label{kurtosis}
\end{figure}

Finally, we consider fractional revivals corresponding to 
$m = 4$, when {\it four} superposed 
wave packets appear.  
These are
detectable in the behavior of the fourth moments of $x$ and 
$p$. 
Equivalently, we may use the excess of kurtosis $(\beta_{2} - 3)$,  
where the kurtosis of $x$ is defined as 
\begin{equation}
\beta_{2}^{(x)} 
\,=\,\frac{\aver{\big(x-\aver{x}\big)^4}}{
\big(\aver{x^2}-\aver{x}^2\big)^2}\,\equiv\,
\frac{\aver{(\delta x)^4}}{(\varDelta x)^4},
\label{kurt}
\end{equation}
with a similar definition for $\beta_{2}^{(p)}$. The excess of
kurtosis is the measure of the departure of a distribution from a 
Gaussian.  
Figures \ref{kurtosis}(a)-(c) depict how $\quad(\beta_{2}^{(x)} -3)$ 
varies with time.  
\begin{figure}[htpb]
\includegraphics[width=5.7in]{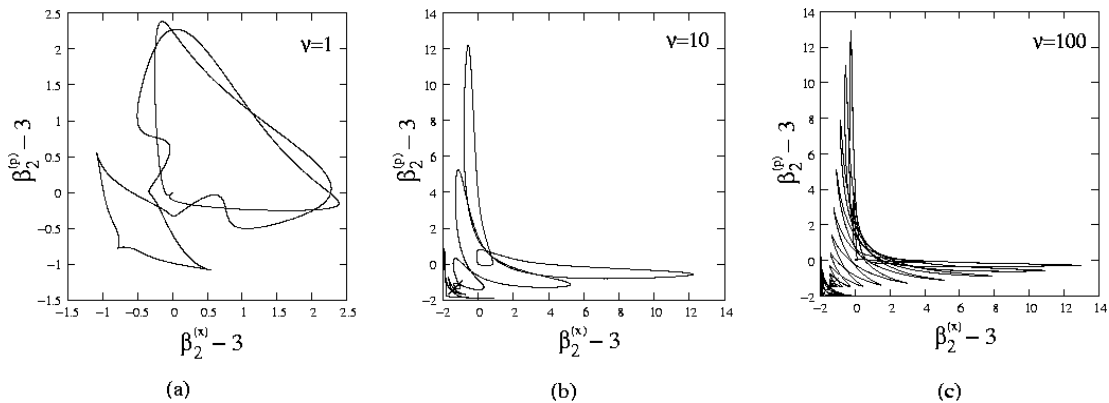}
\caption{``Phase plot'' of    
$(\beta_{2}^{(p)} -3)$ versus $(\beta_{2}^{(x)}-3)$}
\label{phaseplotbeta2}
\end{figure}
For sufficiently large $\nu$, both 
$(\beta_{2}^{(x)} -3)$ and $(\beta_{2}^{(p)}-3)$
remain essentially static near the value 
$-\frac{3}{2}$ most of the time. They 
vary rapidly near revivals, vanishing 
at $t = n\,T_{\rm rev}$ because the
wave packet is a Gaussian both in position space and in momentum space 
at these instants of time. As is clear from Fig. \ref{kurtosis}(c), they 
also vary rapidly near the fractional revivals
at $t = (n + j/4)T_{\rm rev}\, ({\rm where} \, j = 1, 2, 3)$, 
oscillating about the
``steady value'' $-\frac{3}{2}$.
 Once again, a ``phase plot'' of    
$(\beta_{2}^{(p)} -3)$ versus $(\beta_{2}^{(x)}-3)$ 
helps identify
features that distinguish between the three fractional revivals
concerned. These are shown in Figs. \ref{phaseplotbeta2} (a)-(c).

\section{Concluding remarks}
We have shown that distinctive signatures of 
the different fractional revivals of a suitably prepared 
initial wave packet  
are displayed in the mean values and higher moments 
of appropriate observables. The complicated 
quantum interference effects that lead
to fractional revivals can thus be captured in the dynamics of these
expectation values, which may be regarded as effective dynamical variables  
in a classical phase space. While direct experimental measurement of the 
expectation 
values considered, particularly the higher moments of the quadrature 
variables, is certainly not easy, it is not, {\it \`a  priori}, impossible.
Moreover, the essential point of our analysis is to 
bring out explicitly an interesting aspect of quantum interference: 
the link between the occurrence of a $k$-sub-packet 
fractional revival and the behavior of the $k$th moment of 
appropriate 
observables, emphasizing (as in the rest of this thesis) the role of 
quantum expectation values in the investigation of non-classical effects. 
Again, it is evident that the specific observables signaling the 
occurrence of revivals and fractional revivals will vary 
from system to system, and it may not be easy, in practice, to identify the 
correct observables in all cases. However, the special nature of the 
 interference between basis 
states at certain definite instants leads to  revival 
phenomena, and this feature is closely linked to sudden 
changes 
in 
these
expectation values at those instants. This feature is not 
necessarily traceable in all cases to the occurrence of an overall 
{\it exponential} 
factor in the expressions for the 
expectation values, although our analysis seems to indicate that
a factor involving  an exponential of 
periodic functions certainly plays a crucial role.
Wave packet propagation in a Kerr-like medium 
provides us with a 
clean framework for illustrating these aspects, as it involves just 
a single
pair 
of 
ladder operators, and the spectrum is not degenerate, in general.  
In the next Chapter we therefore extend our investigations to initial 
photon-added 
coherent states 
propagating in a Kerr-like medium, and draw attention to the differences 
that arise due to the departure from perfect coherence of the initial state.

%% file: pacs.tex
\chapter{Manifestations of wave packet revivals in the moments of 
observables - II}
\label{pacs}
%----------------------------SECTION-------------------------------%
\section{Introduction}
%------------------------------------------------------------------%

\begin{figure}[htpb]
\begin{center}
\includegraphics[]
{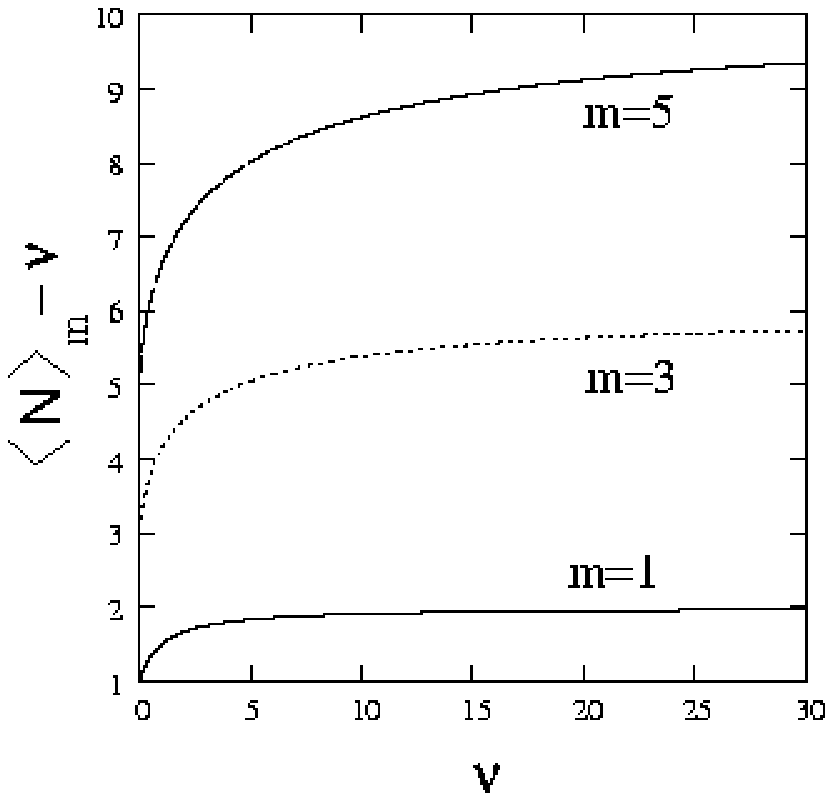}
\caption{$(\aver{{\sf N}}_{m} -\nu)$ versus $\nu$ for $m=1, 3$ and $5$.}
\label{nmminusnu}
\end{center}
\end{figure}

In this chapter, we study the  manner in which   
signatures of  the revival phenomena which are 
manifested in 
the expectation values of  observables   
depend on the extent of coherence enjoyed by the initial wave 
packet.   
For this purpose, we need to compare the details of the revival 
phenomena exhibited by an appropriate non-Gaussian initial state with 
that of an initial CS propagating in 
the nonlinear medium. As mentioned in Chapter 1, a suitable candidate for 
the initial state is the photon-added 
coherent state (PACS):   
 it possesses  
the useful property of a precisely quantifiable, as well as  tunable, 
degree of departure
from perfect coherence.  
This property follows from the fact that a PACS can  
be obtained in principle by repeated addition of photons to a CS. 
The normalized $m$-photon-added coherent 
state $\added{\alpha}{m}$ 
is defined as \cite{agar2} 
\begin{equation}
\added{\alpha}{m}
=\frac{(a^\dagger)^m\ket{\alpha}}{\sqrt{\expect{\alpha}{a^m 
a^{\dagger m}}{\alpha}}}
=\frac{(a^\dagger)^m\ket{\alpha}}{\sqrt{
m!\,L_{m}(-\nu)}}.
\label{photonadded} 
\end{equation}
where $m$ is a positive integer, $\nu = |\alpha|^2$ as before,  
and $L_{m}(-\nu)$ is 
the Laguerre polynomial
 of order $m$. Setting $m = 0$, we retrieve the CS $\ket{\alpha}$.
The mean photon number in the 
PACS $\ket{\alpha,m}$ is 
given by \cite{agar2}
\begin{equation}
\aver{{\sf N}}_{m} = \frac{(m+1)\,L_{m+1}(-\nu)}{L_{m}(-\nu)} - 1, 
\label{expectN}
\end{equation}
When $\nu = 0, \,\aver{{\sf N}}_{m}$  reduces to $m,$ as required. 
When $\nu \gg m$, we can show that  
$\aver{{\sf N}}_{m} = \nu + 2m + \mathcal{O}(\nu^{-1}).$ 
The plot of $(\aver{{\sf N}}_{m} - \nu)$ versus $\nu$   
in Fig. \ref{nmminusnu} shows how the 
ordinate saturates to the value $2m$ 
as $\nu$ increases. 
The extent of departure 
of a PACS from perfect coherence   
clearly becomes more 
pronounced  with increasing $m$.
A PACS also possesses other interesting properties 
which may be kept 
in mind when we investigate its dynamics. For instance, in contrast to a 
CS,   
it exhibits  phase-squeezing and sub-Poissonian statistics \cite{tara} .
The latter implies that the 
standard deviation of the photon
number operator ${\sf N}$ behaves 
like $\aver{\sf N}^{\frac{1}{2}-\beta}$ rather
than $\aver{\sf N}^{\frac{1}{2}}$, the exponent $\beta$ being a calculable
decreasing function of $m$. 

The PACS is also a {\it nonlinear} 
coherent state in the sense that it is an eigenstate of a 
nonlinear annihilation operator \cite{siva}, namely,
\begin{equation}
\left(1- \frac{m}{1+a^\dagger a}\right)a\ket{\alpha,m}= \alpha 
\ket{\alpha,m}.
\label{nonlinearcs}
\end{equation}
The state $\ket{\alpha,m}$ can also be viewed in yet another way. Instead of
the Fock basis $\{\ket{n}\}$, we may consider the unitarily 
transformed basis $\{\ket{n,\alpha}\}$
formed by the {\it generalized coherent states}
\begin{equation} 
\ket{n,\alpha}=e^{\alpha \,a^\dagger-\alpha ^{*}\, a}\,\ket{n}.
\end{equation}
Equivalently, for a given $n$, the generalized coherent state 
$\ket{n,\alpha}$ is simply the state    
$(a^\dagger-\alpha^{*})^{n} \ket{\alpha}$, 
normalized to unity. 
It can now be shown that, for a given $m$,  
the state $\ket{\alpha,m}$ is a {\it finite} superposition of the 
form $\sum_{n=0}^{m}
c_{n} \ket{n,\alpha}$.

\begin{figure}
\begin{center}
\includegraphics[width=2.5in]{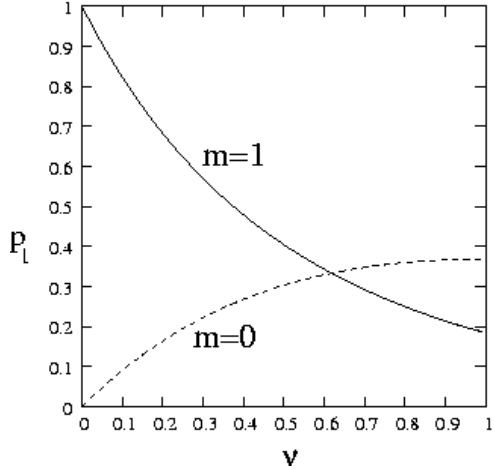}
\caption{Probability of single-photon content 
versus $\nu$ in a CS ($m=0$)
and in a 1-photon-added CS ($m=1$).}
\label{probability}
\end{center}
\end{figure}

We further note that the single-photon-added coherent state
recently generated in experiments \cite{zavatta} has a higher probability 
${\rm p}_1$
of single-photon content than an ideal CS
for sufficiently small values of $\nu$, 
as may be seen from Fig. \ref{probability}. (It is straightforward
 to show that this happens for $\nu<(\sqrt{5}-1)/2\simeq0.618$.)
 In this sense, a PACS
would seem to be preferable to a CS for single-photon experiments.

With this brief introduction to the properties of a PACS, we proceed 
to examine 
the details of its dynamics as it propagates in a nonlinear medium. 
 The latter is once again modeled by the Hamiltonian of 
Eq. (\ref{kerrhamiltonian}).
   
%-------------------------------SECTION----------------------------------%

\section{Dynamics of photon-added coherent states propagating in a 
nonlinear medium}

%------------------------------------------------------------------------%

As in the case of an initial CS, we examine the moments 
and variances of the hermitian operators 
$x$ and $p$,
defined in Eq. (\ref{quadraturevariable}), as the
single-mode field evolves from the initial PACS of 
Eq. (\ref{photonadded}) under
the Kerr-like Hamiltonian. 
We shall use the convenient notation
\begin{equation}
\aver{x(t)}_{m} = 
\expect{\alpha\,,\,m}{e^{iHt/\hbar}\,x\,e^{-iHt/\hbar}}
{\alpha\,,\,m},
\label{expectm}
\end{equation}
with an analogous definition for 
$\aver{p(t)}_{m}\,.$ 
As $H$ is diagonal in the
number operator ${\sf N}$, it follows 
that the mean photon number 
$\aver{{\sf N}}_{m}$, as well as its higher moments, 
and hence the sub-Poissonian statistics of the 
photon number, remain unaltered in time. 

We recall from Chapter 2 that, at the level of the first moments, the 
dynamical equations 
in the case of an initial CS are those of a classical nonlinear 
oscillator, with a certain 
re-parametrization of time. 
In contrast, the time dependences of 
$\aver{x(t)}_{m}$ 
and $\aver{p(t)}_{m}$ (i.~e., averages obtained in the time-evolved 
$m$-photon-added coherent state 
with $m \neq 0$) differ in striking ways from the
foregoing, even for small values of $m$.   
The revival time $T_{\rm rev}$ 
remains equal to $\pi/\chi,$ of course, but   
in the intervals between revivals the
time evolution is considerably more involved than the
expressions in Eqs. (\ref{xt}) and (\ref{pt}) for the case $ m = 0$. 
This implies that
even a small departure from coherence in the
initial state and from Poissonian number statistics 
leads to a very different time 
evolution of the system and the phase squeezing it exhibits. 
We have calculated analytically 
the exact expressions for $\aver{x(t)}_{m}$ 
and $\aver{p(t)}_{m}\,$ in Appendix B,
 and these are given  in Eqs. (\ref{xtmappendix})
 and (\ref{ptmappendix}). The initial values of these expectation
 values in the PACS $\ket{\alpha,m}$ are
\begin{equation}
\aver{x(0)}_{m} =
\frac{L_{m}^{1}(-\nu)}{L_{m}(-\nu)}\,x_{0}\,,\,\,
\aver{p(0)}_{m} =
\frac{L_{m}^{1}(-\nu)}{L_{m}(-\nu)}\,p_{0}\,,
\label{xpinitial}
\end{equation}
where $L_{m}^{1}(-\nu) = d\,L_{m+1}(-\nu)/d\nu$ is an 
associated Laguerre polynomial, $x_0=\sqrt{2}\, {\rm Re}\, \alpha$, 
and $p_0=\sqrt{2} \,{\rm Im}\, \alpha$. 
Analogous to the variables $X$ and $P$ in the case of the CS given by Eq. 
(\ref{XPdefn})  of Chapter 2, 
let us define 
\begin{equation}
\left.\begin{array}{ll}
X_{m}(t)=\aver{x(t)} _{m}\,
\exp\,[\nu\,(1-\cos 2\chi t)],\\  
P_{m}(t)=\aver{p(t)} _{m}\,\exp\,[\nu\,(1-\cos 2\chi t)].
\end{array}\right\}
\label{XPm}
\end{equation} 
(Thus $X_0$ and $P_0$ are just the quantities $X$ and $P$ of the preceding 
chapter.)
The solutions for these quantities may then be written in the following
compact and suggestive form (derived in Appendix B, 
Eqs. (\ref{mxtptappendix}) and Eqs. (\ref{zmappendix})):   
\begin{eqnarray}
\left.\begin{array}{ll}
X_{m}(t)= x_{0}\,{\rm Re}\,z_{m}(t)+p_{0}\,{\rm Im}\,z_{m}(t),\\
P_{m}(t)= p_{0}\,{\rm Re}\,z_{m}(t)-x_{0}\,{\rm Im}\,z_{m}(t),
\end{array}\right\}
\label{mxtpt}
\end{eqnarray}
where 
\begin{equation}
z_{m}(t)=
\frac{L_{m}^{1}(-\nu\,e^{2i\chi t})}{L_{m}(-\nu)}
\,\exp\,[i\,(2 m \chi t+\nu\,\sin\,2\chi t)]\,.
\label{zm}
\end{equation} 
A number of differences between these results and those for the case
$m = 0$ are noteworthy. First, $|z_{m}|$ varies with $t$, in contrast
to $|z_{0}(t)|
\equiv 1.$ (Note also that $z_{m}(0) = 
L_{m}^{1}(-\nu)/L_{m}(-\nu) \neq
1.$) The time dependence of $X_{m}$ and $P_{m}$ involves the sines and
cosines of the set of arguments $(2\chi \,l t + \nu\,\sin\,2\chi t)$, 
where $l = m,\ldots\,,\,2m.$ 
Thus, not only are
higher harmonics present, but the arguments also involve {\em secular}
(linear) terms in $t$ added to the original ($\nu\,\sin\,2\chi t.$).  This
important difference precludes the possibility of 
subsuming the time dependence into that of an effective nonlinear
oscillator by means of a re-parametrization of the time, in contrast to  
the case $m = 0.$  

We present the rest of our results with the help of figures based on
numerical computation. As in Chapter 2, we have set $\chi = 5$ for
definiteness. (Our results on wave packet 
dynamics will hold good for any other value of $\chi$ as well. 
The numerical value of $\chi$ merely sets a time scale.) 
Again, as before, we restrict ourselves 
to the case $x_0= p_0$ (there is no significant loss of generality as
a result of this symmetric choice of parameters). As before, 
the presence of the
overall factor $\exp\,[-\nu\,(1-\cos 2\chi t)]$ in the expressions for 
$\aver{x(t)}_{m}$ 
and $\aver{p(t)}_{m}$ implies that, for sufficiently large values of
the parameter $\nu$, the expectation values remain essentially static
around the value zero, and burst into rapid variation
only in the neighborhood of revivals. Smaller values of $\nu$ enable us to
resolve the details of the time variation more clearly.  

\vspace{.5cm}
\begin{figure}[htpb] 
\begin{center}
\includegraphics[width=6.4in,height=2.6in]{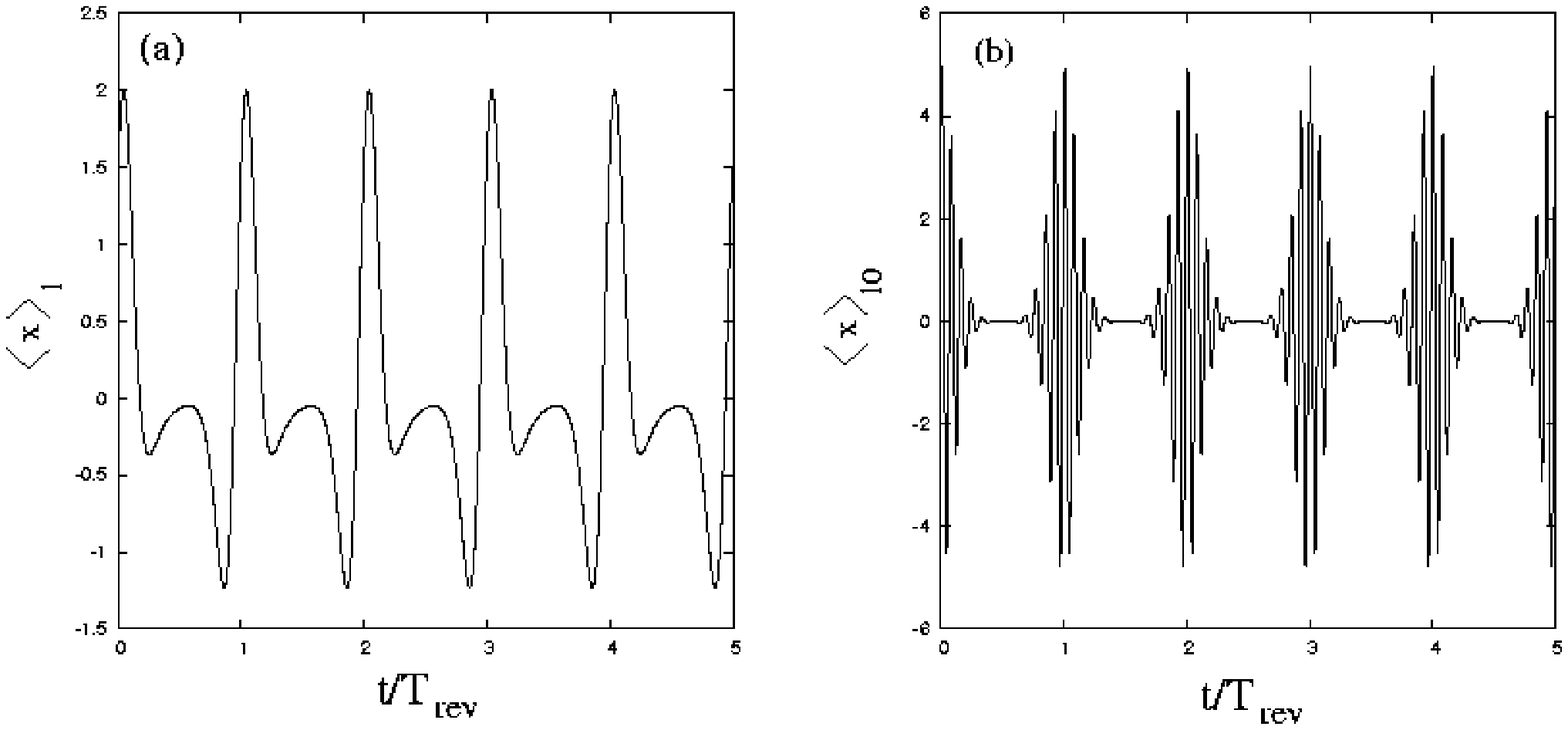} 
\caption{$\aver{x(t)}$ as a function of $t/T_{\rm rev}$ for a PACS (a)  
$\ket{\alpha,1}$ 
and (b) $\ket{\alpha,10}$, with $\nu=1$.  
}
\label{xversustimem1and10}
\end{center}
\end{figure}
Figures \ref{xversustimem1and10} (a) and (b) are, respectively,  plots of 
the expectation values 
$\aver{x(t)}_{1}$ and $\aver{x(t)}_{10}$ versus 
$t/T_{\rm rev}$ for parameter values  
$x_0=p_0=1$ (i.e., for $\nu=1$). The revivals at integer values of 
$t/T_{\rm rev}$ are manifest. With increasing $m$ (or a decreasing
degree of coherence in the initial state), the relatively smooth 
behavior of $\aver{x(t)}_{0}$ 
gives way to increasingly
rapid oscillatory behavior in the vicinity of revivals. 
This is to be compared with 
Fig.
\ref{xversustime} (a), the corresponding plot for an initial CS.
The range over which the expectation value varies also
increases for larger values of $m$.

\begin{figure}[htpb] 
\begin{center}
\includegraphics[width=5.6in,height=2.8in]{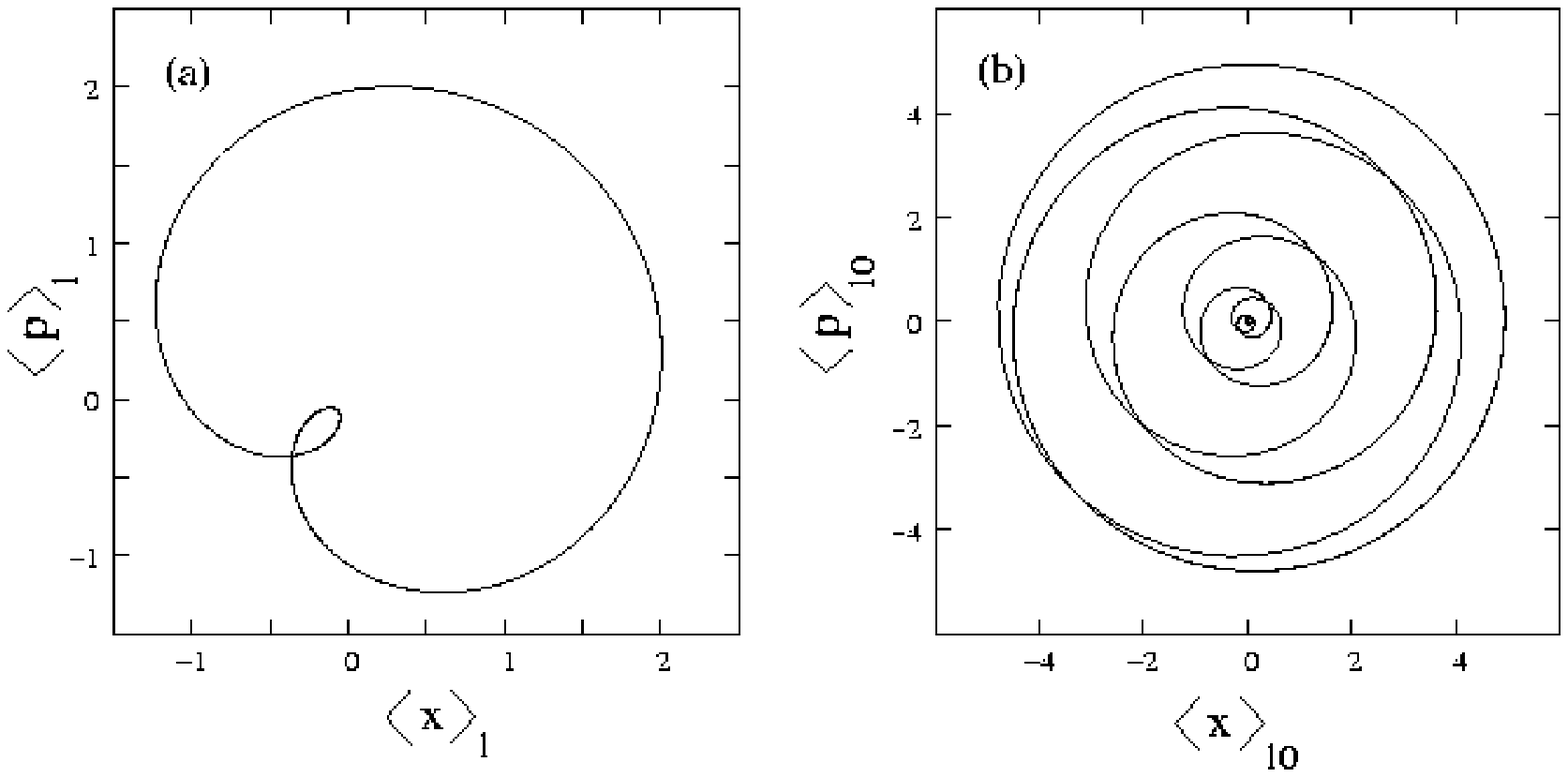} 
\caption{``Phase plot" of $\aver{p}$ vs $\aver{x}$ for an initial PACS (a) 
$\ket{\alpha,1}$ and (b) $\ket{\alpha,10}$, with $\nu=1$.  
}
\label{pversusxm1and10}
\end{center}
\end{figure}

Essentially the same sort of behavior is shown by  
$\aver{p(t)}_{m}\,.$ However, a ``phase plot'' of
$\aver{p(t)}_{m}$ versus $\aver{x(t)}_{m}\,$
in Figs. \ref{pversusxm1and10} (a) and (b) reveals complementary aspects 
of
such oscillatory behavior with increasing $m$, showing how the
oscillations in the two quantities go in and out of phase with each
other. The entire closed curve in each case is traversed in a time
period $T_{\rm rev}$.

As in the case of an initial CS, here too the initial state 
undergoes fractional revivals in the interval
between any two successive revivals
at the instants 
$T_{\rm rev}/k$ 
where $k = 2,3,\ldots\,,$ and also, for any
given $k$, at the instants 
$j\,T_{\rm rev}/k$ 
where $j =
1,2,\ldots\,,(k-1).$ 
 These fractional revivals at 
the instants $j\,T_{\rm rev}/k$ also show up 
in the rapid pulsed variation of the 
$k^{\rm th}$ moments of $x$ and $p,$ and not in the lower
moments.  However, there is a crucial difference if we use  
$\added{\alpha}{m}$ as the initial state: then, even for relatively
small values of $m$, the signatures of fractional revivals appear  
for values of $\nu$
that are not large, in contrast to what happens when the initial state
is the coherent state $\ket{\alpha}.$ 
(Recall that $\aver{{\sf N}}_{m}$ 
is determined by $\nu$ according
to Eq. (\ref{expectN}).) 

For illustrative purposes we investigate the specific case of the 
fractional revival at 
$t=\frac{1}{2}T_{\rm rev}\,.$ 
This corresponds to the appearance of 
two spatially separated similar wave packets, i. e., a single 
qubit in the language 
of logic gate operations. 
\begin{figure}
\begin{center}
\includegraphics[width=3.5in]{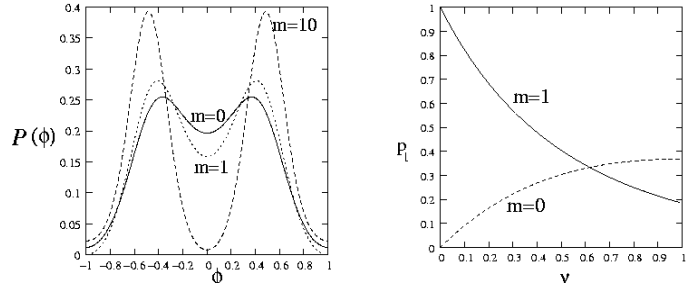}
\caption{Phase distribution $P(\phi)$ versus $\phi$
in units of $\pi$, for $\nu = 0.3$ and $m=0, 1$ and $10$.}
\label{Phase}
\end{center}
\end{figure}

At this instant the overlap between the two sub-packets in the wave 
function corresponding to the state 
$\ket{\psi(T_{\rm rev}/2)}$ decreases significantly with an increase 
in $m$. 
This feature emerges clearly in a plot of the phase distribution
$P(\phi)$ versus the angular variable $\phi$ (shown
in Fig. \ref{Phase} for $\nu = 0.3$
and  $m = 0, 1$ and $10$). This phase distribution is defined as
 \cite{schleich}
\begin{equation}
P(\phi)=\frac{1}{2\pi}\big\vert\inner{\phi}{\psi(T_{\rm
rev}/2)}\big\vert^2,\quad \ket{\phi}=
\sum_{n=0}^{\infty}e^{i n\phi}\ket{n}.
\end{equation}
In this sense, a PACS is better suited than a CS for performing single
qubit
operations treating
the two sub-packets as qubits $\ket{0}$ and $\ket{1}$, respectively.
Similarly, at $t = \frac{1}{4}T_{\rm rev}$, the overlap between the four 
sub-packets is significantly smaller in the case of an initial PACS than
in the case of an initial CS, and this feature becomes more pronounced with 
increasing $m$. The sub-packets therefore behave as genuine orthogonal 
basis states in the former case even for relatively small values of 
$\nu$. Hence 
the quantum logic approach to wave packet control for the purpose of 
implementing two-qubit operations, outlined in Ref. \cite{shapiro},  
is more feasible with an initial state that is a PACS rather than a CS.

\begin{figure} 
\begin{center}
\includegraphics[width=3.0in,height=3.0in]{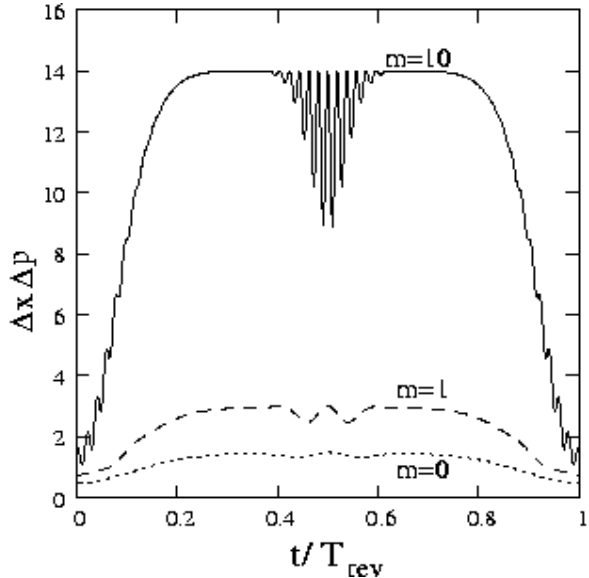} 
\caption{Variation of the uncertainty product with $t/T_{\rm rev}$ for an
 initial CS ($m=0$) and initial PACS $\ket{\alpha,1}$ and 
$\ket{\alpha,10}$ with $\nu=1$.}
\label{deltaxdeltapm01and10}
\end{center}
\end{figure}

Plots of the 
product $\varDelta x\, \varDelta p$ of the standard deviations of $x$ and $p$ 
versus $t/T_{\rm rev}$ over a full cycle  
are shown in 
Fig. \ref{deltaxdeltapm01and10} for initial states given, respectively, by  
the coherent state $\ket{\alpha}$ (dotted curve), the photon-added
state $\added{\alpha}{1}$ (dashed curve),
and  the 10-photon-added state $\added{\alpha}{10}$\, (bold curve), 
for 
$\chi=5,\,x_0=p_0=1,$ so that $\nu = 1$.
It is seen that hardly any trace of the fractional revival is 
evident 
in the case of an initially coherent state, in marked contrast to the case of
the photon-added states, in which 
the fractional revival is signaled by oscillations 
whose frequency and amplitude increase quite rapidly with increasing
$m$. This effect gets masked for 
larger values of the parameter $\nu$,
when these oscillations are relatively insensitive to the value of
$m$.  
\begin{figure}[htpb]
\begin{center}
\includegraphics[width=5.0in]{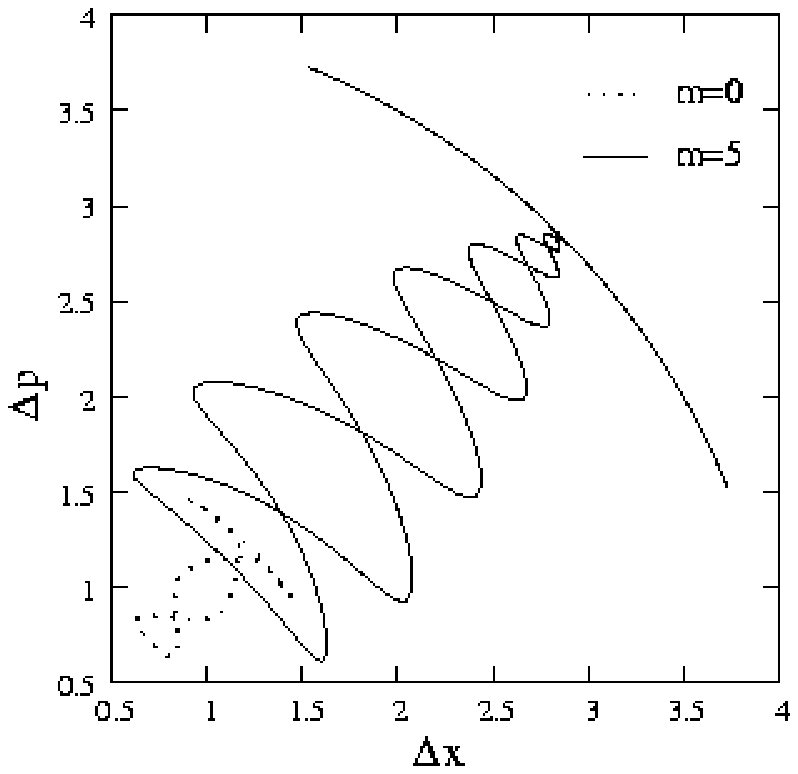}
\caption{``Phase plots'' of  $\varDelta p$ and $\varDelta x$ for $\nu=1$,
for  an initial
CS $(m=0)$ and an  initial PACS $\ket{\alpha,5}$.}
\label{phaseplotdeltapdeltax}
\end{center}
\end{figure}
Another striking feature that provides a clear distinction 
between the revivals at
$t = n\,T_{\rm rev}$ and the fractional revivals at $t =
(n+\frac{1}{2})\,T_{\rm rev}$ (where $n$ is a positive integer) 
is illustrated in 
Fig. \ref{phaseplotdeltapdeltax}, which is a
plot of $\varDelta p$ versus $\varDelta x\,.$ The dotted and full lines
in the same plot correspond to $m = 0$ and $m = 5,$ respectively, making 
ready comparison 
possible.
We recall from Chapter 2, Figs. \ref{deltapversusdeltax} (a)-(c)
 that for an initial CS these standard deviations start from an initial 
value $\frac{1}{2}$.
As $t$ increases, they rapidly build
up, oscillating about the radial $\varDelta p = \varDelta x$ line with an
initially increasing, and then decreasing, amplitude. A maximum value
of $\varDelta x$ and $\varDelta p$ is attained, at which these quantities 
then remain nearly static, till the onset of 
the fractional revival at $T_{\rm rev}\,.$ They then begin to 
oscillate rapidly once again, but this time
in a {\em tangential} direction, swinging back and forth along an arc 
with an amplitude that initially increases and then decreases to zero:
in other words, the individual standard deviations fluctuate 
rapidly in the
vicinity of the fractional revival (while $[(\varDelta x)^2 + (\varDelta
p)^2]^{1/2}$ remains essentially unchanged in magnitude), 
in marked contrast to what happens at a
revival. It is evident from Fig. \ref{phaseplotdeltapdeltax} that all these features are very 
significantly
enhanced and magnified for non-zero values of $m$, 
relative to what happens for the
case $m = 0.$   
\begin{figure}[htpb]
\begin{center}
\includegraphics[width=5.8in]{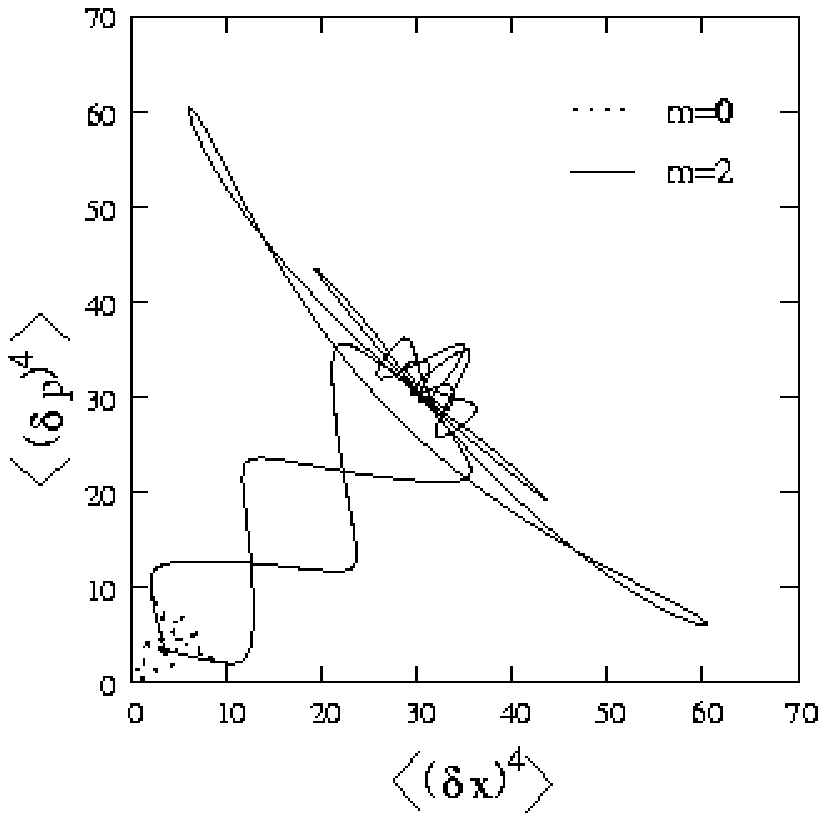}
\caption{``Phase plots'' of $\aver{(\delta p)^4}$ and
 $\aver{(\delta x)^4}$ for $\nu=1$,
for an initial CS and
an initial PACS $\ket{\alpha,2}$.}
\label{phaseplotdeltap4deltax4}
\end{center}
\end{figure}

Similar signatures of fractional revivals for higher values of $k$ 
can be picked out  by using initial states $\added{\alpha}{m}$, even with
relatively small values of $\nu$. The  
oscillations in the moments of observables become more pronounced
with increasing $m$. For instance, signatures of the 
fractional revivals at $t = 
\frac{1}{4}j\,T_{\rm rev}\,,\,\,j = 1,\,2,\,3$ are clearly discernible
in the behavior of the fourth moments of $x.$
Figure \ref{phaseplotdeltap4deltax4} is a plot of 
$\aver{(\delta p)^4}$
versus $\aver{(\delta x)^4}$ where $\delta x = x - \aver{x}\,,\,\,
\delta p = p - \aver{p},$ for $\nu = 1.$ The dotted and full lines
correspond to the initial states $\ket{\alpha}$ and $\added{\alpha}{2}$,
respectively.

%----------------------------------SECTION--------------------------%
\section{Concluding remarks}
%-------------------------------------------------------------------%

We have shown that distinctive signatures of wave packet revivals and 
fractional revivals are displayed in the temporal evolution of 
appropriate variables, enabling identification of departure from 
coherence of the initial state. These distinctions between an initial CS 
and an initial PACS are, however, clearly seen only for small values of 
$\nu$. This is related to the fact that, for small values of $\nu$, the 
probability of the $m$-photon 
content of an $m$-photon-added coherent state is substantially larger than 
that of the CS. 
Another difference between these states emerges during their temporal 
evolution: the addition of even a single 
photon washes out the squeezing 
property displayed by 
an initial CS at specific instants. In the next Chapter, we examine the 
squeezing and 
higher-order squeezing properties of these states as they propagate in a 
Kerr-like medium.

%% file: squeeze.tex
\chapter{Squeezing and higher-order squeezing in wave 
packet dynamics}
\label{squeeze}

%------------------------SECTION----------------------------------
\section{Introduction}
%-----------------------------------------------------------------

In this chapter, we investigate  
the squeezing and higher-order 
squeezing  
properties of an initial PACS as it propagates in a Kerr-like medium. In 
particular, we examine  
the precise manner in which departure from coherence 
of the initial state affects its squeezing properties
in the neighborhood of revivals and 
fractional revivals.   
 
For ready reference, we first summarize the conditions under which a 
generic state 
is said to display squeezing or higher-order squeezing \cite{wall1}. 
Consider two arbitrary Hermitian operators 
$A$ and $B$ with a commutator
\begin{equation}
[A,B]=iC.
\label{commutation}
\end{equation}
The generalized uncertainty relation satisfied by these operators
in any state is
\begin{equation}
\varDelta A\, \varDelta B \geq \frac{1}{2}|\aver{C}|.
\label{uncertaintyAB}
\end{equation}
The state is said to be squeezed in the 
observable $A$ if 
\begin{equation}
(\varDelta A)^2<\frac{1}{2}|\aver{C}|.
\label{squeezeA}
\end{equation}
A similar definition holds for $B$.

In particular, for the quadrature variables $x$ and $p$ defined in Eq. 
(\ref{quadraturevariable}), this implies that 
$x$ (or $p$) is squeezed if $\varDelta x$ (or $\varDelta p$) $<1/\sqrt{2}$
in the state concerned. 
Several states satisfying this squeezing property  have been 
investigated in 
the literature \cite{dodo1}.
Some examples are the squeezed vacuum state, squeezed number states, 
and  photon-added coherent states.
Proposals for the generation, and  detection
of squeezed states, as well as their potential applications,
 have  been outlined in the literature
\cite{wall2,loudon}.

Of direct 
relevance to us is the squeezing property displayed by the $m$- 
photon-added coherent state $\ket{\alpha, m}$. 
A general result for this state is as follows. 
Consider the hermitian operators
\begin{equation}
A_{\varphi} = \frac{a e^{i \varphi} + a^{\dagger} 
e^{-i \varphi}}{\sqrt{2}}\quad{\rm and}\quad
B_{\varphi} = \frac{a e^{i \varphi} - a^{\dagger} 
e^{-i \varphi}}{i\sqrt{2}},
\end{equation}
satisfying the commutation relation $[A_{\varphi},B_{\varphi}]=i$ for
any value of $\varphi$.
For $\varphi=0$, $A_{\varphi}$ and $B_{\varphi}$ reduce to 
quadrature variables 
$x$ and $p$.
It can be shown that the variance of $A_{\varphi}$ in the state 
$\ket{\alpha,m}$ is given by \cite{agar2}
\begin{eqnarray}
(\varDelta A_{\varphi})^2&=&\bigg\{2\nu\Big( 
L_m^{2}(-\nu)L_m(-\nu)-\big[L_m^{1}(-\nu)\big]^2\Big)
 \cos\,\big(2(\theta+\varphi)\big)
-2\nu\big[L_m^{1}(-\nu)\big]^2\nonumber\\
&-&\big[L_m(-\nu)\big]^2+2(m+1)L_{m+1}(-\nu)L_m(-\nu)\bigg\}\bigg/
\bigg\{2\big[L_m(-\nu)\big]^2\bigg\},
\end{eqnarray}
where $\alpha = \nu^{1/2} e^{i \theta}$ as before.
Setting $m = 0$ in the above equation and simplifying, 
we retrieve the corresponding expression for a CS $\ket{\alpha}$,
namely, $(\varDelta A_{\varphi})^2 =\frac{1}{2}$. 
As is well known, the
CS is not a squeezed state. It is in fact a minimum uncertainty 
state in  the observables $x$ and $p$ with 
$\varDelta x = \varDelta p=1/\sqrt{2}$.
Choosing the phases such that $\theta + \varphi = \pi$, a plot of
$(\varDelta A_{\varphi})^2$ 
versus $|\alpha|$ shows that almost 
$50\%$ squeezing can be obtained over a 
wide range of values of $|\alpha|$, 
for non-zero values of $m$.

The concept of higher-order squeezing was first introduced by Hong and 
Mandel \cite{hong1}.
A state is said to be squeezed to order $2q$ $(q = 1,2,3,\ldots)$ in the 
operator $A$  
if $\aver{(\delta A)^{2q}}$  in that state is less than the 
corresponding 
value obtained for a CS, where $\delta A = A-\aver{A}$.
For the 
quadrature variable $x$, this reduces to the 
requirement  
\begin{equation}
\aver{(\delta x)^{2q}} <\frac{1}{2^q}(2q-1)!!
\end{equation}
Another type of higher-order squeezing, called  
amplitude-squared squeezing, was 
defined first by Hillery \cite{hill}, and generalized 
 subsequently by Zhang {\it et al}. 
to $q^{\rm th}$-power 
amplitude-squeezing \cite{zhang}. (Amplitude-squared squeezing 
corresponds to the case $q=2$.)
One considers the two quadrature variables 
\begin{equation}
Z_1= \frac{(a^q+a^{\dagger q})}{\sqrt{2}}\quad{\rm and}\quad
Z_2= \frac{(a^q-a^{\dagger q})}{i\sqrt{2}}\quad (q=1,\,2,\,3,\ldots).
\label{Z1Z2}
\end{equation}
The generalized uncertainty principle now reads  
\begin{equation}
(\varDelta Z_1)^2 \,(\varDelta Z_2)^2 \geq
\textstyle{\frac{1}{4}}\,\big|\aver{\,[Z_1,Z_2]\,}\big|^2.
\label{uncertprinciple}
\end{equation}
The state is said to be $q^{\rm th}$-power amplitude-squeezed  
in the variable $Z_1$ if
\begin{equation}
(\varDelta Z_1)^2 < \textstyle{\frac{1}{2}}\,
\big|\aver{\,[Z_1\,,\,Z_2]\,}\big|.
\label{Z1squeezing}
\end{equation}
Amplitude squeezing in $Z_2$ is similarly defined. 
Let us write 
$[a^q\,,\, a^{\dagger q}]=F_q({\sf N})$. 
This is a 
polynomial of order 
$(q-1)$ in the number operator ${\sf N}$ (see Appendix C, 
Eq. (\ref{fqnappendix})),
given by
\begin{eqnarray}
F_q({\sf N})=q!\,
\bigg[1+\sum_{n=1}^{q-1}\binom{q}{n}\frac{1}{n!}\,
\Big\{{\sf N}({\sf N}-1)\cdots
\big({\sf N}-(n-1)\big)\Big\}\bigg].
\label{fqnchapter}
\end{eqnarray}
Let
\begin{equation}
D_q(t) =\frac{(\varDelta Z_1)^2-
\frac{1}{2}\aver{F_q({\sf N})}}{\frac{1}{2}\aver{F_q({\sf N})}}.
\label{Dqone}
\end{equation}
(The time-dependence has been indicated explicitly 
in $D_q(t)$ to remind us that the
expectation values involved are those in the instantaneous state of
the system.) It is easily seen that the state is $q^{\rm th}$-power
amplitude-squeezed in $Z_1$ if $-1 \leq D_q < 0$. 
We can rewrite Eq. (\ref{Dqone}) in terms of $a^q$ and 
$a^{\dagger q}$ as 
\begin{equation}
D_q(t) 
=\frac{2\,\left[{\rm Re}\,\aver{a^{2q}} 
-2 \,\big({\rm Re} \,\aver{a^q} \big)^2+ \aver{a^{\dagger q} \,a^{q}}\right]}
{\aver{F_q({\sf N})}}.
\label{Dqtwo}
\end{equation} 

We now examine the manner in which the extent of 
coherence of the initial state affects its subsequent squeezing 
properties. There is a considerable literature
on this subject.
For instance,  
dynamical squeezing of a PACS arising as a result of the 
time-dependence of the 
frequency of the electromagnetic field oscillator in a cavity has been 
investigated \cite{dodo2}.   
The non-classical effects exhibited by an initial CS propagating in a
Kerr-like medium \cite{yurke}, its Hong-Mandel squeezing 
properties   
\cite{gerry}, and its $q^{th}$-power amplitude-squeezing properties 
\cite{du}
have  been discussed in the literature.  
We are interested specifically in the squeezing properties of wave 
packets propagating in a Kerr-like medium. In the next 
Section, we therefore  
summarize the results  of \cite{du} in this context, obtained for an 
initial CS. This will facilitate 
subsequent 
comparison between our work on the squeezing properties of an initial PACS 
 \cite{sudh3} with those 
of an initial CS.  

%------------------------SECTION------------------------------
\section{The case of an initial coherent state}
%------------------------------------------------------------

When the initial CS evolves under the Hamiltonian of Eq. 
(\ref{kerrhamiltonian}),  the (time-dependent) 
expectation values on the RHS 
of Eq. (\ref{Dqtwo}) can be evaluated \cite{du}  
using Eq. (\ref{nthmoment}). The result is the expression
for $D_q(t)$ given in Eq. (\ref{Dqtcsappendix}), Appendix C.

We are interested, in particular, in examining 
whether fractional revivals are accompanied by 
any significant degree of squeezing and higher-order squeezing. For
this purpose, we focus on $D_q(t)$ at 
a 2-sub-packet fractional revival. 
Setting $t = \pi/(2\chi)$ in 
Eq. (\ref{Dqtcsappendix}) (Appendix C) and simplifying, we get
\begin{equation}
D_q(T_{\rm rev}/2)=\frac{2\nu^q}{\aver{F_q({\sf N})}}\Big\{1+
(-1)^{q-1}\cos\,2q\theta
-2e^{-4\nu\sin^{2}q\pi/2}\cos^{2}q\theta\Big\}.
\end{equation}
As before, $\theta$ is the argument of $\alpha \,(= \nu^{1/2}\,e^{i\theta}).$
It is evident that $D_q$ vanishes at this instant 
for all even values of $q$, implying that 
no even-order squeezing of the
state accompanies this fractional revival. For odd values of $q$,
however, we find  
\begin{equation}
D_q(T_{\rm rev}/2) =\frac{4\nu^q}{\aver{F_q({\sf N})}}
\big(\sin^2 q\theta - e^{-4\nu} \cos^2 q\theta\big).
\label{Dqthree}
\end{equation}
\begin{figure}[htpb] 
\includegraphics[width=5.6in,height=2.8in]{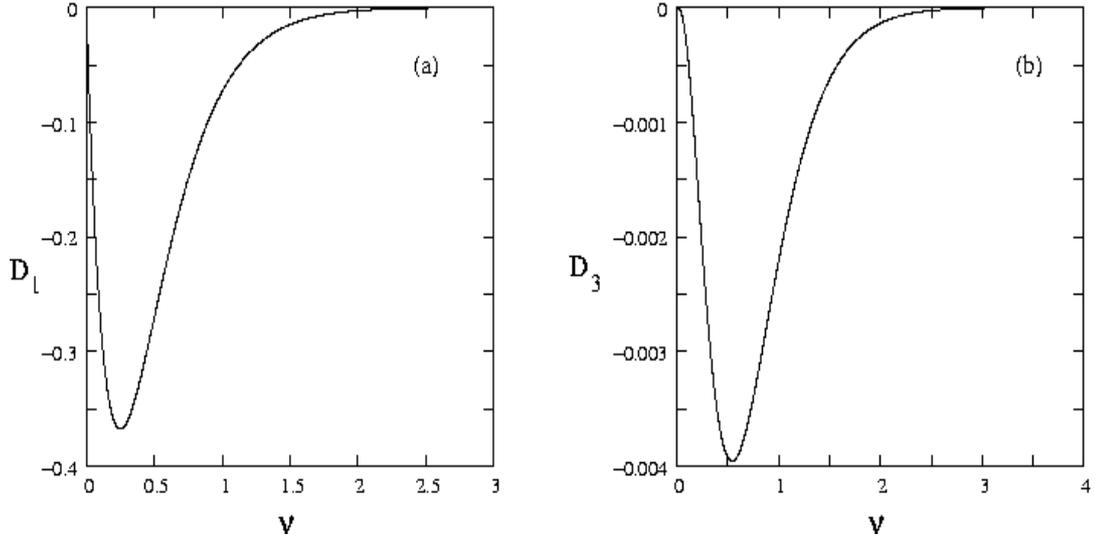}
\caption{$D_q(\frac{1}{2}T_{\rm rev})$ 
versus $\nu$ for an initial CS (with $\theta=0$), 
for (a) $q=1$ and (b) $q=3$. (Note the different ordinate scales in the
two cases.)}
\label{Dqnu}
\end{figure}
Thus squeezing (or higher-order squeezing) 
occurs at this instant, provided $D_q < 0$, i.e.,  
$|\tan\,q\theta| < e^{-2 \nu}$. We illustrate this in Figs. 
\ref{Dqnu}(a) and (b), where $D_1$ and $D_3$  
are plotted as functions of $\nu$ for an initial CS with $\theta = 
0$. (Once again, we have set $\chi = 5$ in  
all the numerical results presented in this Chapter.) 
\begin{figure}[htpb]
\includegraphics[width=5.6in,height=2.8in]{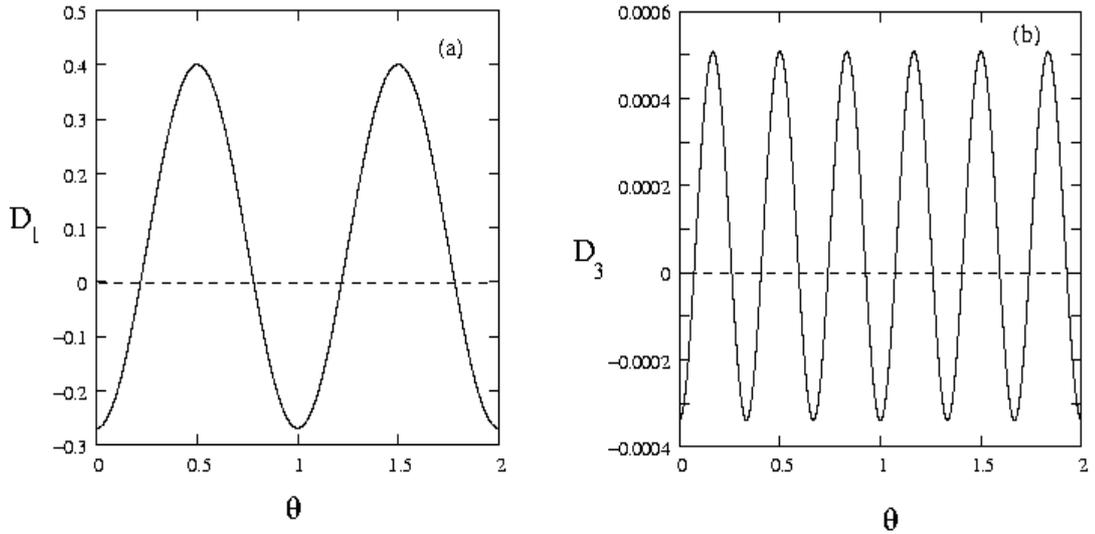}
\caption{$D_q(\frac{1}{2}T_{\rm rev})$ 
versus $\theta$ for an initial CS (with $\nu=0.1$), 
for (a) $q=1$ and (b) $q=3$.}
\label{Dqtheta}
\end{figure}
We  have also plotted $D_1$ and $D_3$ as 
functions of $\theta$ for a fixed value of $\nu\,(=0.1)$ 
in Figs. \ref{Dqtheta}(a) and (b), showing how squeezing occurs for
certain ranges of the argument of $\alpha$, when $D_q$ becomes
negative.  

\begin{figure}[htpb]
\begin{center}
\includegraphics[width=3.5in]{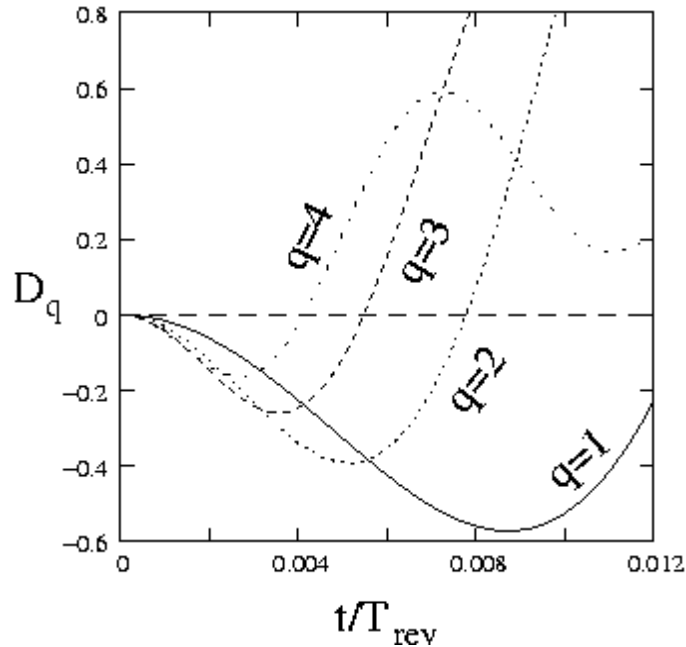}
\caption{$D_q$
versus $t/T_{\rm rev}$ for an initial CS with $\nu=10$,
for $q=1, 2, 3$ and $4$.}
\label{Dqalpha}
\end{center}
\end{figure}

An interesting feature is that the CS spreads and loses its coherence 
almost immediately after it starts propagating in the medium. This is 
borne out by the fact that it becomes squeezed even within a very short 
time of propagation \cite{du} as seen in Fig. \ref{Dqalpha}, 
which 
is 
a plot of 
$D_q$ versus $t/T_{rev}$ for an initial CS with $\nu = 10$. 
Further, we see that close to $t = 0$ the state not only displays  
squeezing, but also  higher-order squeezing, for both odd and even integer 
values of $q$.  

In the light of these results, we are  in a position to investigate the 
dynamics of an initial PACS and compare it with that of an initial CS.
%------------------------SECTION----------------------------------
\section{The case of an initial PACS}
%-----------------------------------------------------------------

\begin{figure}[htpb]
\includegraphics[width=5.6in,height=2.8in]{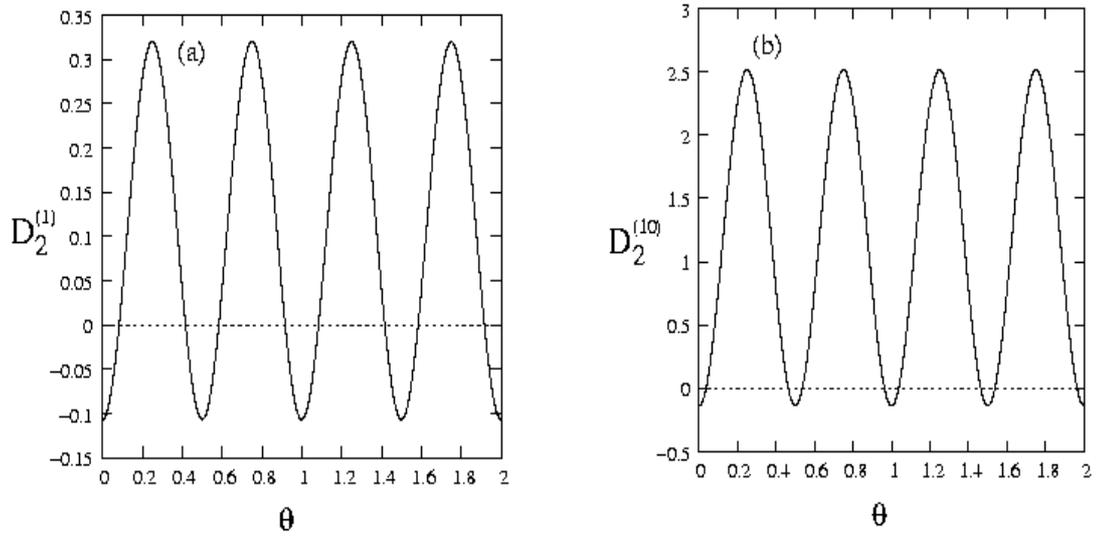}
\caption{$D_2^{(m)}(\frac{1}{2}T_{\rm rev})$ 
versus $\theta$ (with $\nu = 2$), 
for an initial PACS with (a) $m=1$ and (b) $m=10$.}
\label{D2mtheta}
\end{figure}

We show that if the initial state departs even marginally from 
coherence, as in a PACS with a small value of $m$, 
the results of the preceding Section change significantly. Writing $D_q$ as  
$D_q^{(m)}$ (for ready identification) when the expectation 
values in Eq. (\ref{Dqtwo}) are evaluated
for an initial state 
$\ket{\alpha , m}$, we find the following general result:
\begin{eqnarray}  
\frac{1}{2}L_m(-\nu)\aver{F_q({\sf N})}_m D_q^{(m)}(t) &=&
e^{-\nu(1-\cos4\chi qt)}\sum_{n=0}^{m}\binom{m+2q}{n+2q}
\,\frac{\nu^{n+q}}{n!} \nonumber\\
&\times& \cos\,\Big[2(2m+2n + 2q -1)\chi qt+\nu\sin4\chi qt
-2q\theta\Big]\nonumber\\
&-& \frac{2e^{-2\nu(1-\cos2\chi qt)}}{L_m(-\nu)}
\biggl\{\sum_{n=0}^{m}\binom{m+q}{n+q}\,\frac{\nu^{n+q/2}}{n!}\nonumber\\
&\times&\cos\,\Big[(q-1+2m+2n)\chi qt+\nu\sin\,2\chi qt-q\theta\Big]
\biggr\}^2\nonumber\\
&+&\sum_{n=n_{\rm min}}^{q}\binom{q}{n}\frac{m!}{(m-q+n)!}\,\nu^nL_m^n(-\nu),
\label{Dqmone}
\end{eqnarray}
where $n_{\rm min} = \max\,(0\,,\,q-m)$. The derivation of the 
expression above is given in Appendix C. It is the generalization of the 
expression for $D_q\equiv D_q^{(0)}$ obtained for a coherent state.
As before, we examine $D_q^{(m)}$ at $t = \frac{1}{2}T_{\rm rev}$ for
possible  squeezing and higher-order squeezing of the 
state at that instant. The foregoing
expression reduces at this instant of time to 
\begin{eqnarray}
\frac{1}{2}L_m(-\nu)\aver{F_q({\sf N})}_m D_q^{(m)}(T_{\rm rev}/2)& =& 
(-\nu)^{q}\,L_m^{2q}(-\nu)\,\cos\,2q\theta\nonumber\\
&-&\frac{2e^{-2\nu(1-\cos q\pi)}}{L_m(-\nu)}\nu^{q}
\Big\{L_m^q\big((-1)^q\nu\big)\Big\}^2
\cos^2 q\theta \nonumber\\
&+&\sum_{n=n_{\rm min}}^{q}\binom{q}{n}\frac{m!}{(m-q+n)!}
\,\nu^nL_m^n(-\nu).\nonumber\\
&&
\label{Dqmtwo}
\end{eqnarray}
We use this to analyze various cases numerically.
In contrast to what happens for an initial CS, it turns out that there
is no {\it odd}-power amplitude-squeezing for an initial
PACS. Even-power amplitude-squeezing does occur, though, for
sufficiently large values of $\nu$. 
\begin{figure}[htpb]
\includegraphics[width=5.6in,height=2.8in]{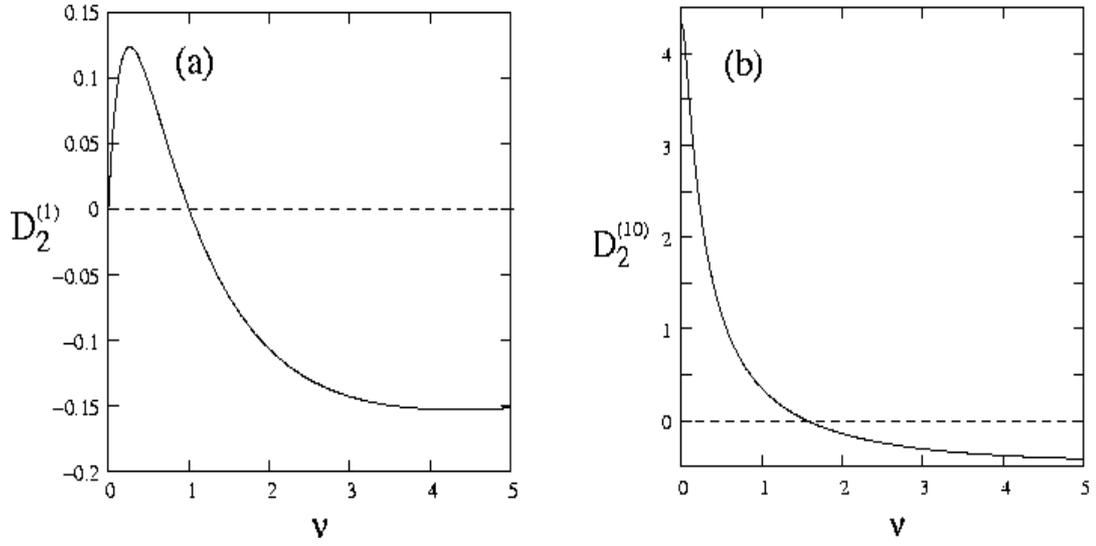}
\caption{$D_2^{(m)}(\frac{1}{2}T_{\rm rev})$
versus $\nu$ (with $\theta=0$), 
for an initial PACS with (a) $m=1$ and (b) $m=10$.}
\label{D2mnu}
\end{figure}
This is illustrated in
Figs. \ref{D2mnu}(a) and (b), which show the ranges of $\nu$ for which
$D_2^{(m)}$ falls below zero. Figures \ref{D2mtheta}(a) and (b) depict the
variation of $D_2^{(m)}$ with the phase angle $\theta$ for a fixed value of
$\nu$.

\begin{figure}
\begin{center}
\includegraphics[width=3.0in,height=3.0in]{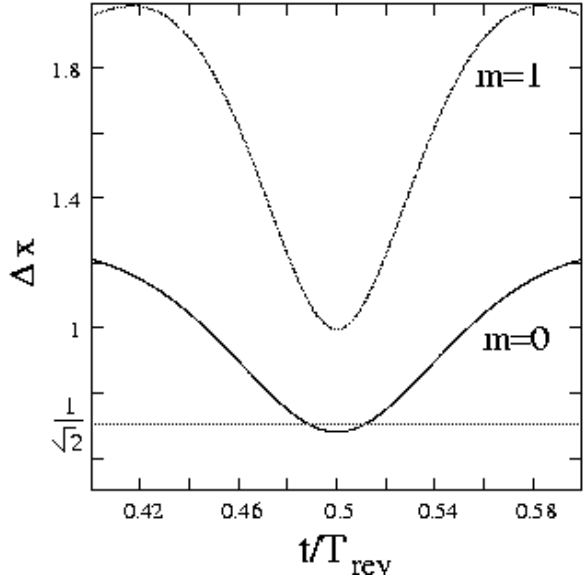}
\caption{$\varDelta x$ versus time in units of $T_{\rm rev}$, 
in the case $x_0 =
  \sqrt{2}\,,\,p_0 = 0$.}
\label{deltaxt}
\end{center}
\end{figure}

Turning to the extent of squeezing as a function of time,
in Fig. \ref{deltaxt} we compare the temporal variation of the 
standard deviation 
$\varDelta x$ for an initial CS ($m = 0$) and an initial PACS ($m =
1$), with $\alpha = 1$ (hence $\theta = 0$).  
The horizontal dashed line demarcates the level below which 
the state is squeezed. It is evident that squeezing in the rigorous 
sense accompanies the
fractional revival at $t = 
\frac{1}{2}T_{\rm rev}$ 
when the initial
state is a CS. This feature is suppressed when it is a PACS, although  
$\varDelta x$ does dip down considerably around this fractional
revival. 

\begin{figure}
\begin{center}
\includegraphics[width=5.6in,height=2.8in]{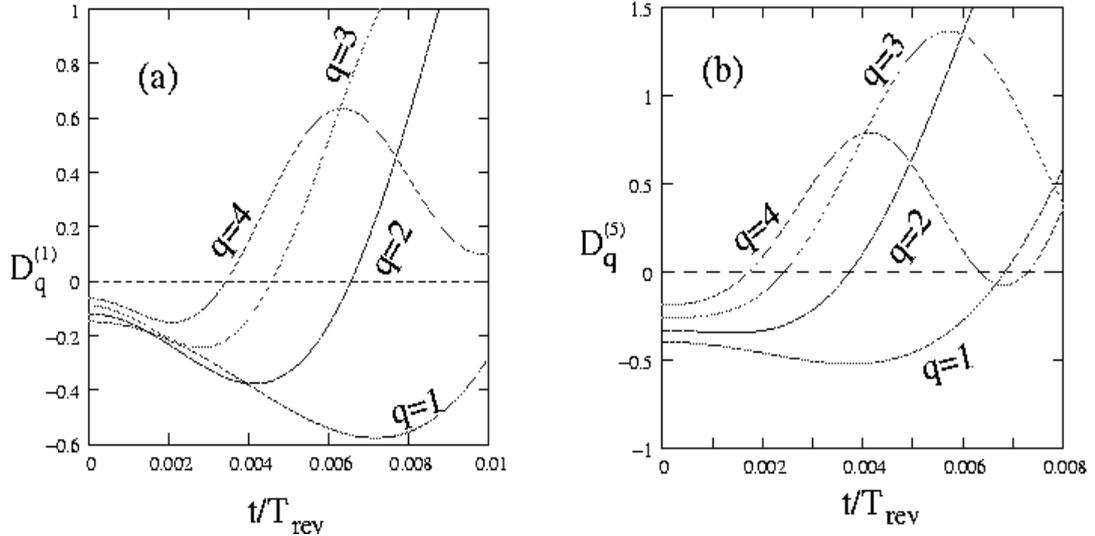}
\caption{$D_q^{(m)}$
versus $t/T_{\rm rev}$ for an initial PACS (a) $\ket{\alpha, 1}$ and (b) 
$\ket{\alpha, 5}$,  with $\nu=10$,
for $q=1, 2, 3$ and $4$.}
\label{Dqpacs1and5}
\end{center}
\end{figure}

To complete the picture, we have also plotted, in Figs. \ref {Dqpacs1and5} 
(a) and (b),  $D_q^{(m)}$ versus 
$t/T_{rev}$ 
for the initial states $\ket{\alpha, 1}$ and $\ket{\alpha, 5}$ 
respectively, for $\nu = 
10$ and $q = 1,2,3$ and $4$.

We have focused on $q^{\rm th}$-power amplitude-squeezing at
fractional revivals, as this turns out to  provide rather more
discriminatory signatures of higher-order squeezing effects 
than the other alternative, namely, 
Hong-Mandel squeezing. However, a few remarks on the
latter are in order here. 
The relevant variables in the case of Hong-Mandel squeezing
are
\begin{equation} 
\Big[\frac{a+a^\dagger}{\sqrt{2}}\Big]^{q} =x^q\quad {\rm and} 
 \quad \Big[\frac{a-a^\dagger}{i\sqrt{2}}\Big]^{q}=p^q.
\end{equation} 
For $q=1$, of course, Hong-Mandel 
squeezing is the same as amplitude-squeezing, but the two kinds of
squeezing differ for $q \geq 2$. 
\begin{figure}
\begin{center}
\includegraphics[]{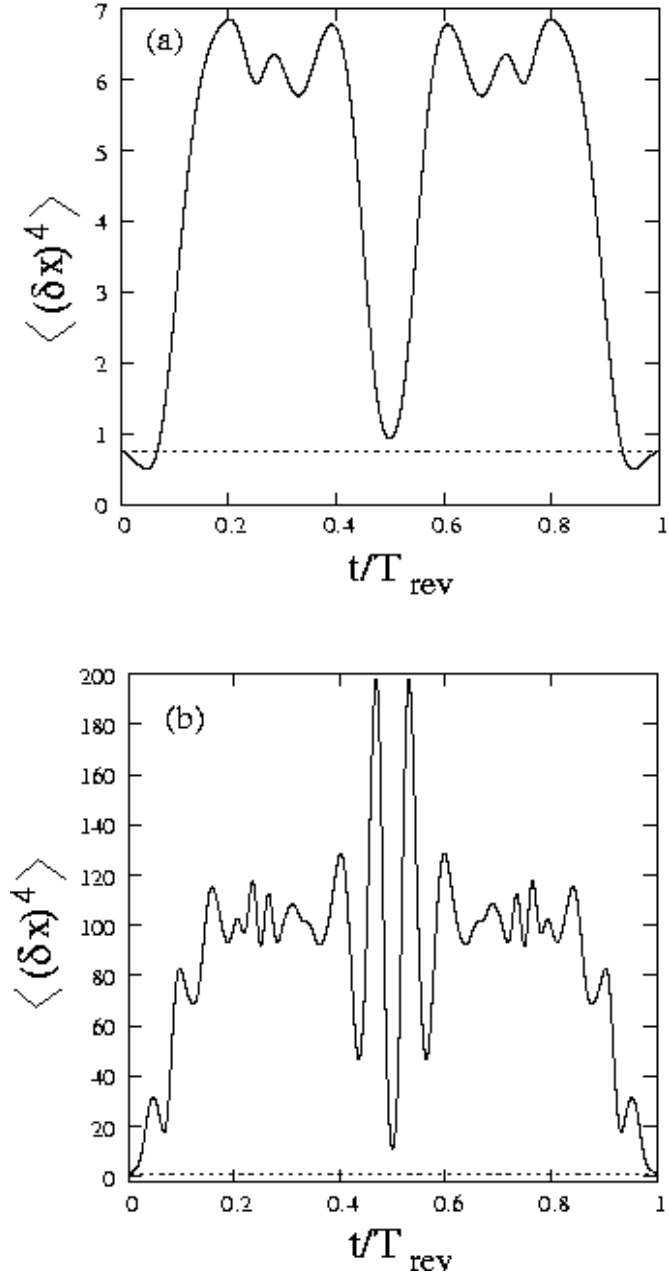}
\caption{$\aver{(\delta x)^4}$ 
versus time in units of $T_{\rm rev}$ for initial states
 (a) $\ket{\alpha}$  and (b) $\ket{\alpha,5}$, 
with $\nu = 1$. Note the very different
ordinate scales in the two cases. Hong-Mandel squeezing in $x$
occurs below the dotted horizontal line.}
\label{4thmoment}
\end{center}
\end{figure}
Figures \ref{4thmoment}(a) and (b) show how 
$\aver{(x-\aver{x})^4}\equiv\aver{(\delta x)^4}$,   
the fourth moment of $x$ about its mean value, varies  
over a revival period for an initial CS and PACS, respectively. 
The horizontal dotted lines indicate the bound on 
$\aver{(\delta x)^4}$,   
below which fourth-order
Hong-Mandel squeezing occurs in this quadrature. 
An initial CS exhibits such squeezing near revivals, and 
comes close to doing so near the fractional revival at  
$\frac{1}{2}T_{\rm rev}$, but does not actually do so. 
An initial PACS does not display such higher-order squeezing at any
time, although 
$\aver{(\delta x)^4}$ attains its lowest value at
revival times. However, fractional revivals are marked by 
rapid oscillations of $\aver{(\delta x)^4}$,  
these being most pronounced around the $2$-sub-packet fractional
revival. These features are enhanced further in the case of 
initial states with larger values of
$m$. 

\section{The Wigner function and the non-classicality indicator}

Finally, we examine the 
Wigner functions corresponding to the wave packets 
at instants of fractional revivals, to quantify  
non-classical behavior during their time evolution. It has been
suggested \cite{ken} that the ``extent'' to which the 
Wigner function becomes
negative (as a function of its complex argument) may serve as 
a measure of
the non-classicality of the state concerned. 
The normalized Wigner function $W(\beta\,;\,t)$ 
(where $\beta \in \mathbb{C}$)  
is defined in the case at hand as follows \cite{perina}. One
first defines the symmetric characteristic function
\begin{equation}
K(z,t)={\rm Tr}\,\big[\rho(t)\,e^{za^{\dagger}-z^*a}\big],
\end{equation}
where $z \in \mathbb{C}$, and $\rho(t)$ is the density matrix. Then
$W(\beta\,;\,t)$ is the two-dimensional Fourier transform given by
\begin{equation}
W(\beta\,;\,t)=\frac{1}{\pi^2}\int d^2 z\,K(z,t)\,e^{za^{\dagger}-z^*a}.
\end{equation}
Using the representation for the density matrix in the Fock basis,
i.~e.,
\begin{equation}
\rho (t) =\sum_{l,n=0}^{\infty}\rho_{ln}(t) \ket{l}\bra{n},
\end{equation}
one arrives at the following
representation for the Wigner function in terms of the density 
matrix elements $\rho_{ln}(t)$ \cite{brune}:
\begin{eqnarray}
W(\beta\,;\,t)
&=&\frac{2}{\pi}e^{-2|\beta|^2}\,{\rm Re}\,\biggl\{\,\sum_{l=0}^{\infty}
\sum_{n=l}^{\infty}
(-1)^l(2-\delta_{l\,n})\,(l!/n!)^{1/2}\nonumber\\
&&\times\,(2\beta)^{n-l}\,
\rho_{l n}(t)\,
L_l^{n-l}\big(4|\beta|^2\big)\biggr\}.
\label{wignerfn}
\end{eqnarray}

For an initial 
coherent state  $\ket{\alpha}$, we have the standard result
\begin{equation}
\rho_{ln}(0) =
\frac{\alpha^{*n}\,\alpha^l}
{\sqrt{l!\, n!}}\,e^{-|\alpha|^2}.
\label{rhocs}
\end{equation}
The corresponding Wigner distribution is given by the well-known
expression \cite{gerry2}
\begin{equation}
W(\beta\,;\,0)=\frac{2}{\pi}e^{-2|\alpha-\beta|^2}.
\label{wignerfncs}
\end{equation}
This is positive definite everywhere in the complex $\beta$-plane,
justifying the appellation ``classical'' for an oscillator coherent
state $\ket{\alpha}$.
The time-dependent density matrix element in this case is given by
\begin{equation}
\rho_{ln}(t) =
\frac{\alpha^{*n}\,\alpha^l}
{\sqrt{l!\, n!}}\,e^{-|\alpha|^2}\,e^{-i\chi[l(l-1)-n(n-1)]t}.
\label{rhotcs}
\end{equation}
Using this in Eq. (\ref{wignerfn}), we compute $W(\beta,t)$ numerically.
\begin{figure}
\begin{center}
\includegraphics[width=5in, height=7in]{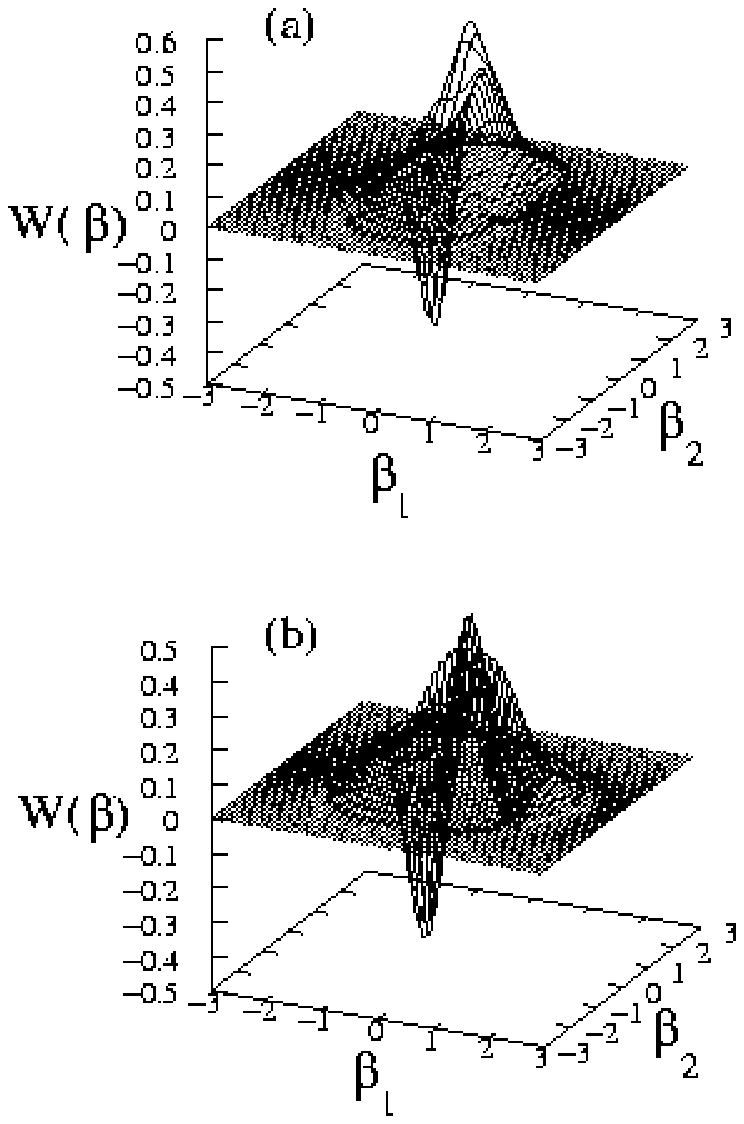}
\caption{Wigner function corresponding to an initial state 
$\ket{\alpha}$ with $\alpha = 1$, at (a) $t=\frac{1}{2}T_{\rm rev}$ and (b) 
$t=\frac{1}{3}T_{\rm rev}$. Here, and in the succeeding figures, 
$\beta_1 = {\rm Re}\,\beta\,,\,
\beta_2 = {\rm Im}\,\beta$.} 
\label{wignercs}
\end{center}
\end{figure}
\begin{figure}
\begin{center}
\includegraphics[width=4.5in, height=9in]{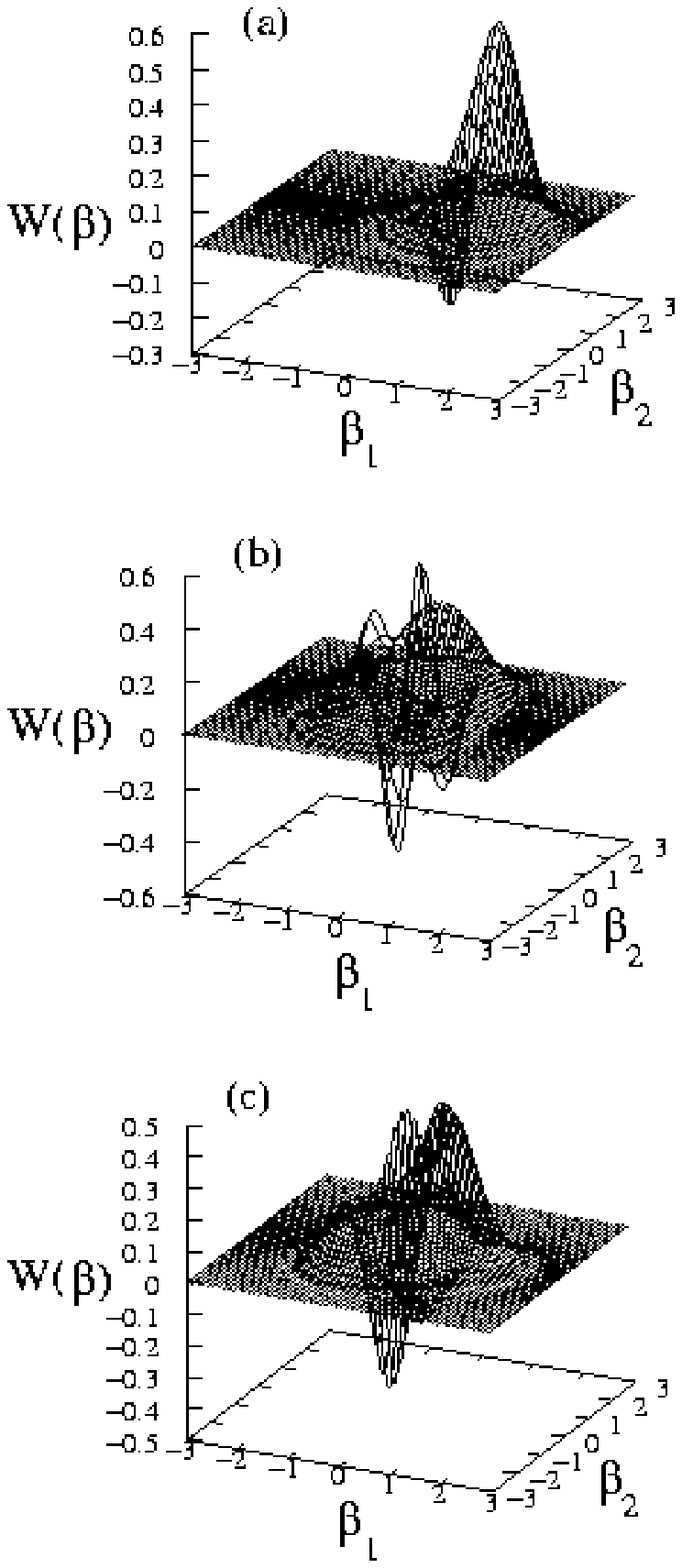}
\caption{Wigner function corresponding to an initial state
$\ket{\alpha,1}$ with $\alpha = 1$, 
at (a) $t=0$,\, (b) $t=\frac{1}{2}T_{\rm rev}$ 
and \,(c) $t=\frac{1}{3}T_{\rm rev}$. }
\label{wigner1}
\end{center}
\end{figure}
Figures \ref{wignercs}(a) and (b) show how the 
Wigner function for an initial CS 
behaves at the 
instants of the 2-sub-packet 
and 3-sub-packet fractional revivals,
respectively. 
\begin{figure}
\begin{center}
\includegraphics[width=5.5in,height=9in]{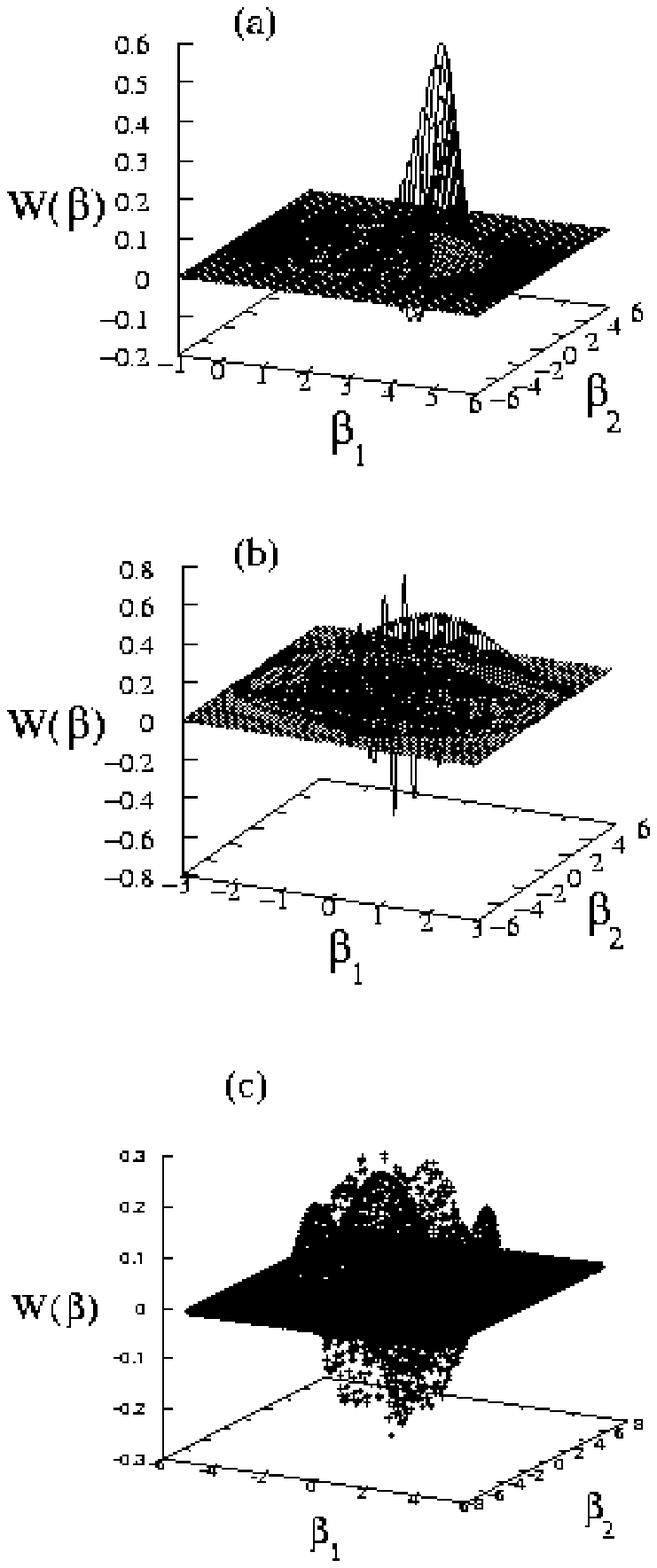}
\caption{Wigner function corresponding to an initial state
$\ket{\alpha,10}$ with $\alpha = 1$,
at (a) $t=0$,\, (b) $t=\frac{1}{2}T_{\rm rev}$ 
and (c) $t=\frac{1}{3}T_{\rm rev}$.}
\label{wigner10}
\end{center}
\end{figure}

For the initial photon-added coherent state
$\ket{\alpha,m}$, one finds 
\begin{equation}
\rho_{l n}(0) =\frac{e^{- \nu}}{m!\,L_m(- \nu)}
\frac{\alpha^{l-m}\,{\alpha^*}^{n-m}\sqrt{l!\, n!}}{(l-m)!\,(n-m)!}\,.
\label{rhopacs}
\end{equation}
Correspondingly, the Wigner function at $t = 0$  
can be expressed in the closed form \cite{agar2,dodo2} 
\begin{equation}
W(\beta\,;\,0)=\frac{2\,(-1)^m}{\pi L_m(- \nu)}\,
L_m\big(|2\beta-\alpha|^2\big)\,
e^{-2|\alpha-\beta|^2}.
\label{wignerfnpacs}
\end{equation}
It is to be noted that this is no longer positive definite for  all
complex $\beta$, reflecting the fact that this initial state 
is no longer
``totally'' classical, as it departs
from perfect coherence due to the photons that have been ``added'' to 
$\ket{\alpha}$ to produce the PACS. 
The time-dependent density matrix element in this case is given by
\begin{equation}
\rho_{l n}(t) =\frac{e^{- \nu}}{m!\,L_m(- \nu)}
\frac{\alpha^{l-m}\,{\alpha^*}^{n-m}\sqrt{l!\, n!}}{(l-m)!\,(n-m)!}\,
e^{-i\chi\big[l(l-1)-n(n-1)\big]t},
\label{rhotpacs}
\end{equation}
where $l,n \geq m$. (The matrix element $\rho_{ln}(t)$ vanishes
identically for $l,n<m$.) Once again, using this in Eq. (\ref{wignerfn}), 
we compute the Wigner function numerically.
Figures \ref{wigner1}(a), (b) 
and (c) are plots of $W(\beta\,;\,t)$ for the case 
$m=1$ at $t=0,\,\frac{1}{2}T_{\rm
  rev}$ and $\frac{1}{3}T_{\rm rev}\,$, respectively.  
The  corresponding plots for the case $m=10$ 
are shown in Figs. \ref{wigner10}(a), (b) and (c). With increasing
$m$, the oscillations of
the Wigner function in the $\beta$-plane between positive and negative
values become more pronounced, and the region of 
non-classicality becomes more extensive.

\begin{figure}
\begin{center}
\includegraphics{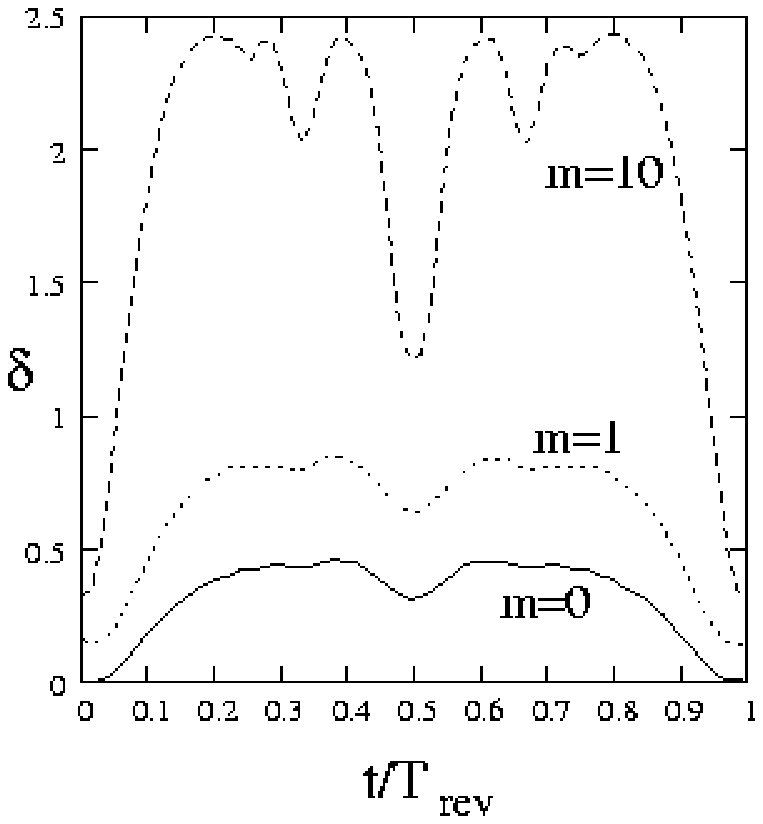}
\caption{Non-classicality indicator 
$\delta$ versus time in units of $T_{\rm rev}$,
for initial states $\ket{\alpha,m}$ with $\alpha=1$.}
\label{wignerdelta}
\end{center}
\end{figure}
To get an idea of the degree of non-classicality as a continuously
varying function of time
for each of the different initial states we have considered, 
it is instructive to 
consider the non-negative quantity $\delta$ defined as \cite{ken}  
\begin{eqnarray}
\delta (t) &=&\int \!d^{\,2}\beta\,
\Big(\big|W(\beta\,;\,t)\big|
-W(\beta\,;\,t)\Big)\nonumber\\
&=&\int \!d^{\,2}\beta\,\big|W(\beta_1\,;\,t)\big|-1.
\end{eqnarray}
The larger the value of $\delta$, the greater is the extent of
non-classicality of the state concerned, although of course 
$\delta$ alone does not give a complete picture of the oscillations of
the Wigner function. In Fig. \ref{wignerdelta} 
we have plotted 
$\delta$ versus $t$ for initial states $\ket{\alpha},\,\ket{\alpha,1}$, 
and $\ket{\alpha,10}$ with $\alpha=1$. 
It is clear that in the interval 
between $t=0$ and $t = T_{\rm rev}\,$,  $\delta$ 
has the most pronounced local minimum at 
the $2$-sub-packet 
fractional revival, followed by the local minima at the $3$-sub-packet 
and $4$-sub-packet fractional revivals. This 
feature becomes increasingly 
prominent for larger values of $m$, showing that 
the extent of 
non-classicality also increases with $m$, as expected.

%---------------------------SECTION----------------------------
\section{Concluding remarks}
%--------------------------------------------------------------
We have shown that the crucial difference between an initial CS and an 
initial PACS propagating through a Kerr-like medium arises at the 
instant $\frac{1}{2}T_{\rm rev}\,.$, when the former displays squeezing 
and the latter does not. A possible experimental 
set-up to examine the squeezing 
properties of the state at $\frac{1}{2}T_{\rm rev}\,$ involves  
passing it through a Kerr 
medium with a known value of $\chi$ (the third-order nonlinear 
susceptibility), and therefore $T_{\rm rev}$, and using the time evolved 
state at  $\frac{1}{2}T_{\rm rev}\,$ as the input field. 
Superimposing this on the field from a local oscillator,  
and performing  balanced homodyne detection to 
measure the variance in $x$, would determine the extent of squeezing.
Thus, the successful experimental generation of non-classical states of light  
together with the  observation of higher-order moments \cite {breit} of 
appropriate quadrature variables would enable the probing of the rich  
structure of wave packet dynamics in a nonlinear medium.

%% file: twomode.tex
\chapter{Ergodicity properties of entangled two-mode states - I} 
\label{twomode}
\section{Introduction}
In the preceding chapters, we have examined the manner in which signatures 
of revivals and different     
fractional revivals of a wave packet manifest themselves in 
the dynamics of the 
expectation values of appropriate operators. We have shown that, by 
tracking the 
time evolution of various 
moments of  certain operators, selective 
identification of 
different fractional revivals can be achieved \cite{sudh1}.
Again, since the precise nature of the initial state concerned plays a 
crucial role 
in determining its 
subsequent dynamics and the non-classical features it exhibits, these 
signatures help 
assess the extent of coherence of the initial state 
\cite{sudh2}. Further, it has been shown that 
the squeezing and higher-order squeezing 
properties of certain quadrature variables in the neighborhood of 
a two sub-packet fractional revival of the wave packet provide 
quantifiable
measures of the degree of departure from coherence of the initial 
state \cite{sudh3}. 

In this chapter, we extend these investigations to the dynamics of
entangled states.   
To be  specific, we consider a 
physically relevant generalization of the 
example considered earlier, and now study  
the dynamics of a single-mode electromagnetic field 
interacting with the atoms of a 
nonlinear medium \cite{agar1},  taking into account the 
interaction 
between the field mode and the medium.  
Although we consider initial states that are 
direct product states of the field and atom modes,  entanglement of 
these 
two modes sets in during temporal evolution. However, we have shown 
that for 
certain ranges of values of the parameters and coupling constants in the 
governing Hamiltonian, the modes disentangle 
at specific instants  during the 
evolution, and the state of the system  
returns to its initial form, apart from an overall phase. This 
feature is clearly the 
analog of wave packet revivals in the dynamics of a 
single mode. The entropy of 
entanglement of the system provides a measure of the extent of revival of the 
original state. At the instants of revivals (these being integer
multiples of a revival time which we refer to as $T_{\rm rev}$ in 
this two-mode
example as well), 
this entropy vanishes.

It is evident that, as in the single-mode case, the dynamics of the quantum
state between $t = 0$ and $t = T_{\rm rev}$ would lead  to closed orbits 
in a classical phase space in which quantum mechanical 
expectation values are regarded as 
the dynamical variables. The entropy of entanglement 
is also seen to  reduce 
significantly in certain cases at fractions 
$\frac{1}{2},\, \frac{1}{3}, \,\frac{2}{3}$ and $\frac{1}{4}$ 
of $T_{\rm rev}$, signaling the appearance of 
the counterparts  of fractional revivals  
at these instants. We identify suitable 
operators whose 
expectation values carry clear signatures of this phenomenon.    

We have also investigated the role of the (non-entangled) 
initial state,  and 
the extent of departure of the initial field mode from coherence, on  
the subsequent revival 
properties of the state. In particular, the link 
between entanglement of states and their  squeezing 
properties \cite{sudh5} has been examined. The 
entropy of entanglement at any instant increases
with increasing   departure from coherence of the initial field mode, 
and revivals do not
occur. 
This aspect is studied by means of a time-series analysis 
of the expectation 
value of a certain observable. The details of this analysis 
and the  results obtained are presented in the next chapter.

In the next section, we describe the model 
\cite{agar1} we use 
to study the dynamics of two-mode entanglement. 
Earlier results on the 
collapse and revival phenomena in this case, and their dependence 
on the initial state and parameter values, are summarized.
In Section 5.3, we show how the
extent to which a state revives at any instant    
is related to the instantaneous sub-system von Neumann entropy 
(SVNE) and the sub-system linear entropy (SLE).     
Certain operators whose expectation 
values carry signatures of the different fractional revivals 
are also identified. We have 
examined the dynamics of
several initial states, taking the oscillator representing 
an atom of the medium  to be in the ground 
state while the field is, respectively, in a Fock state, a CS, and a PACS. 
This enables us to analyze systematically  
the effects of different initial field 
modes on the dynamics. The relationship between the squeezing 
exhibited by  
the state of 
the system and the initial field mode is also brought 
out. 
We conclude the chapter with some brief comments.    

\section{Single-mode field in a nonlinear medium:\\ 
The model} 
\label{interaction}

Sanz {\it et al.} \cite{sanz} 
have examined the  
non-classical effects that arise in the dynamics of two entangled modes 
governed by a nonlinear 
Hamiltonian, in the framework of an exactly 
solvable case: 
two modes of an electromagnetic field interacting in a 
Kerr-like medium. 
Taking the initial state 
to be a direct product, either of two Fock states or of two coherent states,  
the periodic exact revival of these states has been established,  
and the manner 
in which 
these properties show up in the collapse and revival 
phenomena displayed by the expectation values of  observables has 
been investigated. The collapses are marked by expectation values 
remaining constant over a certain interval, while the revivals are signaled 
by rapid pulsed variations of these quantities. This is clearly 
analogous to the signatures of revivals  displayed in the single-mode case 
that we have discussed in the preceding chapters. 

As stated in Chapter 1,  our objective is to investigate the  
full range of regular and irregular dynamical behavior 
exhibited by quantum mechanical expectation 
values. For this purpose, we require a system where 
revivals, near revivals, as well as no revivals of the initial state 
can all occur, for 
different parameter values.
As already mentioned, a good candidate Hamiltonian for our 
purposes is the one that  
describes the interaction of   
a single-mode field of frequency 
$\omega$  with the atoms of the nonlinear medium  through which 
it propagates. The medium is modeled \cite{agar1} by an anharmonic oscillator 
with frequency $\omega_0$ and anharmonicity parameter $\gamma$. The  
Hamiltonian of the total system is given 
by
\begin{equation}
H =  \omega \,a^\dagger a + \omega_0 \,b^\dagger b +  \gamma\, 
b^{\dagger 2} b^2 +  g \,(a^\dagger b + b^\dagger a).
\label{coupledhamil1}
\end{equation}
$a$ and $a^\dagger$ are the annihilation and creation operators 
pertaining to the field, while  $b$ and $b^\dagger$ are the
corresponding  atomic oscillator operators. 
The coupling constant $g$ is a measure 
of the strength of the coupling between the 
field modes and the atom modes. 
The Fock basis is given by 
$\{\ket{n}_a 
\otimes \ket{n'}_b\}$, where $n$ and $n'$ are the eigenvalues of  
$a^\dagger a$ and 
$b^\dagger b$, respectively. 
The photon number operator 
${\sf N}
= a^{\dagger}a$ no longer commutes with $H$. However, 
the {\it total} number operator
${\sf N}_{\rm tot}= a^\dagger a + 
b^\dagger b$ {\it does} commute with $H$. 
Hence we may write the basis states as 
$\ket{N-n}_a \otimes \ket{n}_b$, 
using $N$ to label the eigenvalue 
of ${\sf N}_{\rm tot}\,$. (We remark that, 
in the discussion of the 
one-mode case in the preceding chapters, we have used 
$N$ in places to denote 
the eigenvalue of the photon number operator 
${\sf N}$. No confusion 
should arise as a result, 
as we specify the notation explicitly  
wherever required.) 
For notational simplicity, 
let us write   
\begin{equation} 
\ket{N-n}_a \otimes \ket{n}_b \equiv  \ket{N-n\,;\, n}.
\label{prodstate1}
\end{equation} 
It is evident that
$\bra{N-n\,;\, n} H \ket{N'-n'\,;\,n'} = 0$, if $N \neq N'$.  
Hence, for each given value of $N$, 
the Hamiltonian $H$ can be diagonalized 
in the space of the states $\ket{N-n\,;\, n}$, where $n 
= 0,\,1,\,\ldots ,N$. Let the eigenvalues and eigenstates of $H$ be 
$\lambda_{Ns}$ and $\ket{\psi_{Ns}}$, respectively, 
where $s = 0,\,1,\,\ldots ,N$  for a given $N$,
and $N = 0,\,1,\,\ldots \,{\it ad \ inf.} $ 
It is convenient to expand $\ket{\psi_{Ns}}$ 
in the basis $\{\ket{N-n\,;\, n}\}$ as
\begin{equation}
\ket{\psi_{Ns}} = \sum_{n = 0}^{N}\, d_n^{Ns}\,\ket{N-n\,;\, n},
\label{}
\end{equation}
so that $d_n^{Ns} = \langle{N-n\,;\, n}\,|\,\psi_{Ns}\rangle$.

The time-dependent density operator $\rho(t)$ corresponding to 
different choices of the initial state $\ket{\psi(0)}$ can now be 
obtained. For instance, in the simple case in which 
$\ket{\psi(0)} 
= \ket{N}_a \otimes \ket{0}_b \equiv \ket{N\,;\, 0}$,   
we have 
\begin{eqnarray}
\rho(t) &=& \sum_{s = 0}^{N}\,\sum_{s' = 
0}^{N}\,\exp{[-i(\lambda_{Ns} - \lambda_{Ns'})t]} 
\,\langle{\psi_{Ns}}|N\,;\, 0 \rangle \nonumber\\
&\times& \langle{N\,;\, 0}|{\psi_{Ns'}}\rangle\,
\ket{\psi_{Ns}}\bra{\psi_{Ns'}}.
\label{}
\end{eqnarray}
Such expressions for the time-dependent density matrix  and 
the
corresponding reduced density matrices for the 
two sub-systems 
(i.~e., the field and the medium, respectively) 
are required in order to evaluate
the expectation values and entropies 
to be considered in the next section. Further details are 
given in Appendix D.  

We now summarize some of the key results of Ref. \cite{agar1}  
pertaining to 
the collapse and revival phenomena exhibited by the field,
as these are of relevance to our study. 
When $g=0$, 
the Hamiltonian in Eq. (\ref{coupledhamil1}) is in diagonal form 
as it stands. If $\gamma=0$, $H$ is exactly diagonalizable, 
and there is periodic exchange of energy 
between the field and the atomic 
oscillator.
For non-zero values of the ratio $\gamma/g$ of 
the respective strengths of the nonlinearity and the 
field-atom interaction, collapses and revivals could occur 
over certain intervals of time, in between these 
periodic exchanges of energy.  
This phenomenon translates into the behavior of expectation 
values of certain observables as well. For instance, during  a 
collapse of the energy exchange over an interval of time, the mean 
photon number $\aver{a^\dagger a}$ essentially remains   
constant. A revival of the energy exchange  
is signaled by rapid oscillations of the mean photon number 
about this steady value, over the relevant time interval. 

Further, when the atomic oscillator is initially in its 
ground state, while the 
field starts  either in a Fock state or in a coherent state, one 
finds the the following results:\, 
(a)\, For weak nonlinearity ($\gamma/g \ll 1$),  collapses 
and revivals of the mean 
photon number occur almost periodically in time, for both  
kinds of initial field states. The revival time 
can be shown to be  {\it approximately}
equal to 
$2\pi/\gamma$ in the former case, and $4\pi/\gamma$ in the latter. 
\,(b) \,For $\gamma/g \sim 1$, such collapses and revivals 
occur more irregularly if the field is initially in a coherent state,  
compared to the case when it is initially in a Fock state. \,
(c) \,For large nonlinearity ($\gamma/g \gg 1$), 
these collapses and revivals do 
not occur.
Bearing these results in mind, in the next section we examine the 
manner in which the foregoing  collapse 
and revival phenomena are mirrored in the 
entropy of entanglement of the system. We  
identify  observables which carry signatures of these collapses 
and revivals, and discuss 
the influence of the departure from coherence of 
the initial state of the field 
on the extent to which it revives subsequently. 

\section{Entanglement properties}
\label{entanglement}

Let us now examine the detailed dynamics of three different 
classes of initial states 
which are direct products of the field and atom states, evolving under the 
Hamiltonian in Eq. (\ref{coupledhamil1}). 
As stated earlier, the initial state of 
the atom  is taken to be the oscillator ground state $\ket{0}_b$, while that 
of the field is, respectively, (a) 
a Fock state $\ket{n}_a\,,\,n = 0,\,1,\,\ldots \,$;
\, (b) a CS $\ket{\alpha}_a\,$; 
and \,(c) an $m$-photon-added CS 
$\ket{\alpha,m}_a\,,\,\,m = 1,\,2,\,\ldots $\,.  
(Recall that the suffixes $a$ and 
$b$ correspond to the electromagnetic field and  
the atoms of the medium, respectively.)
Extending the notation in Eq. (\ref{prodstate1}), we 
shall write 
\begin{equation}
\ket{\alpha,m}_a \otimes \ket{n}_b \equiv \ket{(\alpha,m)\,;\,n}
\label{prodstate2}
\end{equation}
for the direct product state concerned.

As we are dealing with a pure bipartite system,  
it is natural to consider the 
time-dependences of $S_k$\,, the sub-system 
von Neumann entropy (SVNE), 
and $\delta_k$\,, the sub-system linear entropy 
(SLE), where the suffix $k$ stands for either $a$ or $b$, 
depending on the sub-system considered. 
These quantities are defined as
\begin{equation}
S_k(t) = -{\rm Tr}_k\,\big(\rho_k(t) \,\ln \,\rho_k(t)\big)
\label{svne1}
\end{equation}
and
\begin{equation}
\delta_k(t) = 1 - {\rm Tr}_{k}\,\big(\rho_k^2(t)\big),
\label{lne1}
\end{equation}
where $\rho_k(t)$ is the time-dependent reduced 
density operator for 
the sub-system concerned. In terms of the set of 
eigenvalues of $\rho_k(t)$, we have
\begin{equation}
S_k(t) = - \sum_{i} \lambda_k^{(i)}(t) \,\ln \,\lambda_k^{(i)}(t),\,\,
\delta_k(t) = 1 - \sum_{i} \,\big[\lambda_k^{(i)}(t)\big]^2,
\label{svnelne}
\end{equation}
where the summation runs over all the eigenvalues $\lambda_k^{(i)}(t)$.  
Once again, as we have a pure bipartite system, the SVNEs for the 
two sub-systems are in fact equal to each other at all times, as are the
SLEs. (This used as one of the checks on the numerical routines employed.) In 
Appendix D, we have outlined the steps leading to the derivation 
of the matrix elements of the reduced density matrices 
$\rho_a(t)$ and $\rho_b(t)$, required for the numerical computation 
of $S_k(t)$ and $\delta_k(t)$.

We are interested, in this chapter, in examining revival phenomena. 
We therefore restrict ourselves to   
the case of  weak nonlinearity, i.~e., 
$\gamma/g <<1$, as this is the situation in which 
these phenomena occur. For 
illustrative purposes, we set the values of the
parameters at $\omega = \omega_0 = 1$, and 
$\gamma = 1,\,\,g = 100$, so that   
$\gamma/g = 10^{-2}$, in all the numerical results reported 
in the rest this section. Note that the revival time then corresponds to 
$gt\approx200\pi$ for an initial Fock state of the field, and 
$gt\approx400\pi$ for an initial coherent state of the field.

\begin{figure}[htpb]
\begin{center}
\includegraphics[]{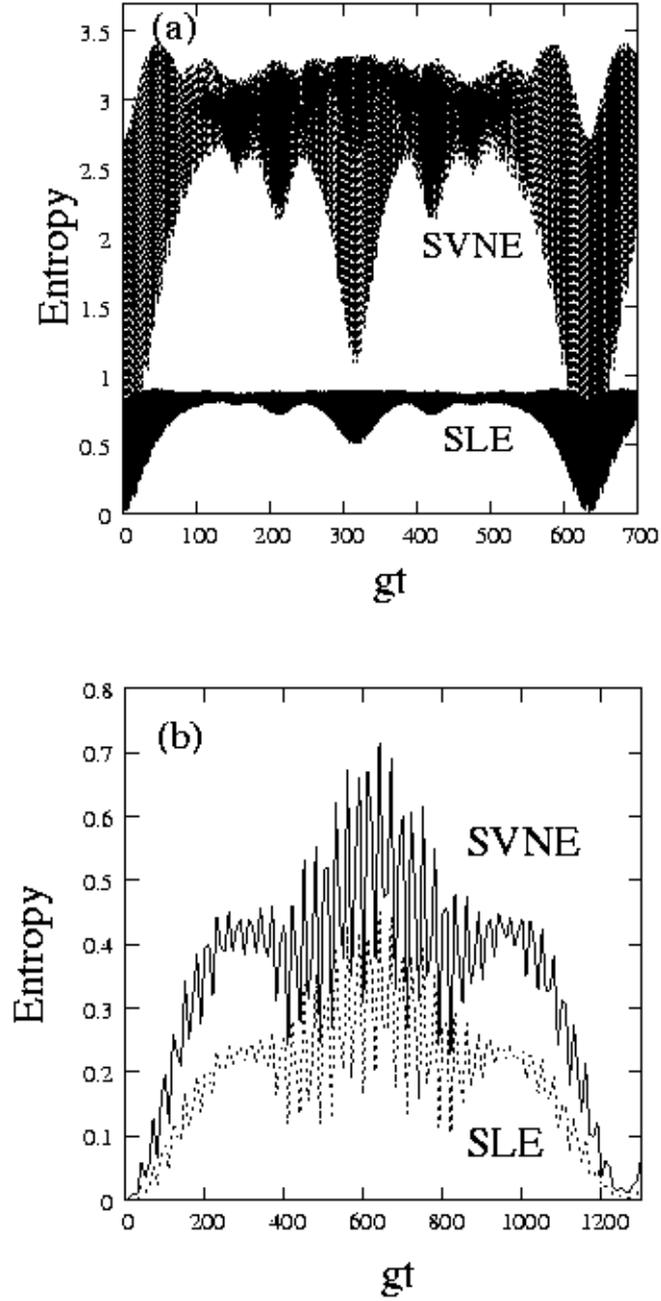}
\caption{SVNE and SLE versus $gt$   
for  
(a) an initial Fock state $\ket{10\,;\, 0}$, and 
(b)~~an initial coherent state $\ket{\alpha\,;\, 0}$ with $\nu = 1$.
 The ratio  $\gamma/g = 10^{-2}$ in 
Figs. \ref{entropy10cross0andalphacross0qbyg.01nu1} through 
\ref{deltaxam0nu5}.}
\label{entropy10cross0andalphacross0qbyg.01nu1}
\end{center}
\end{figure}

\begin{figure}[htpb]
\begin{center}
\includegraphics[width=4.0in]
{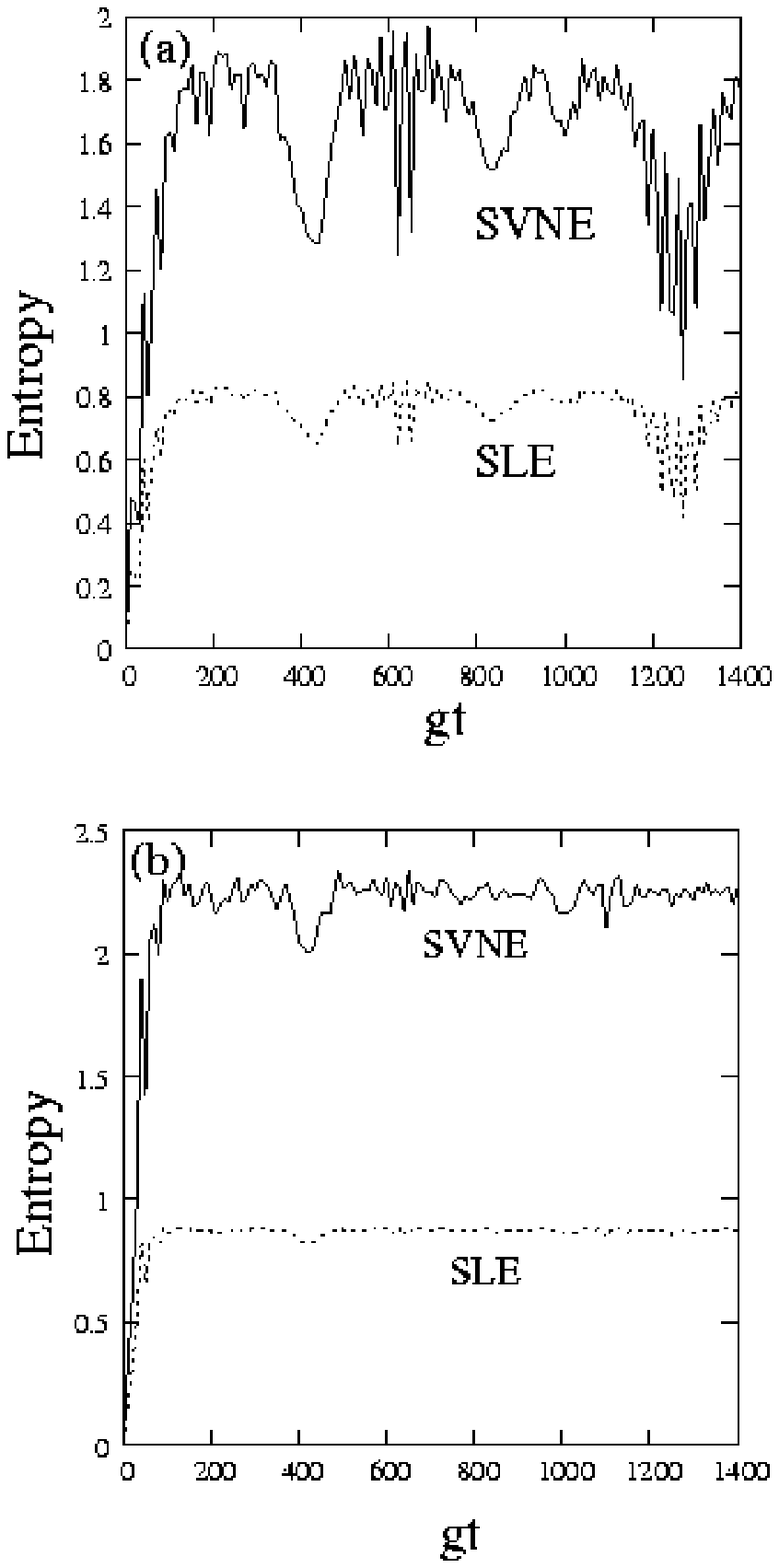}
\caption{SVNE and SLE versus $gt$ for an initial 
state $\ket{(\alpha,5)\,;\, 0}$ for (a) $\nu = 1$ and (b) $\nu = 5$.}
\label{entropyalpha5qbyg.01nu1andnu5}
\end{center}
\end{figure}

Figures \ref{entropy10cross0andalphacross0qbyg.01nu1} (a) and (b) 
depict plots of the SVNE and the SLE versus $gt$ for  
respective initial states  $\ket{10\,;\,0}$  
and  $\ket{\alpha}_a \otimes \ket{0}_b \equiv 
\ket{\alpha\,;\, 
0}$ with the parameter value $\nu = 1$.
The band-like appearance of the plots in 
Fig. \ref{entropy10cross0andalphacross0qbyg.01nu1} (a)  
arises from the extremely rapid oscillations of the ordinates. 
The corresponding plots for an initial state  
$\ket{\alpha,5}_a \otimes  \ket{0}_b \equiv \ket{(\alpha,5)\,;\, 0}$
in the cases $\nu = 1$ and $\nu = 5$ are shown   
in Figs. \ref{entropyalpha5qbyg.01nu1andnu5} (a) and (b) respectively.  
In all the cases above, the SVNE (the upper plot in each figure) 
is larger than the SLE 
(the lower plot in each figure) at any instant of time. 
It is evident that both SVNE and SLE display roughly similar 
oscillatory behavior in time. 
However, certain striking differences arise in 
the time evolution of the SVNE and 
SLE, depending on the actual initial state considered.
If the field is initially in a Fock state or a CS, 
the entropies return to zero at regular intervals of time  
(see Fig. \ref{entropy10cross0andalphacross0qbyg.01nu1}),
signaling a revival of the initial state. (Note that 
the revival times are indeed approximately equal to 
$2\pi$ and $4\pi$, respectively, recalling that we have set  
$\gamma = 1$ and $g = 100$.)   
In contrast, if the initial state of the field is a PACS, 
 the entropies do not exhibit such exact revivals, although
they do show substantial oscillatory behavior near $T_{\rm rev}$.  
 Further, with an increase in the value of 
$\nu$, the oscillations in the SVNE and SLE die down. This effect is 
enhanced for larger values of $m$, as  seen in  
the rapid increase and saturation of both the SVNE and SLE for an initial 
state $\ket{(\alpha,5)\,;\,0}$ for $\nu = 5$, in contrast 
to the corresponding plots for $\nu = 1$: 
see Figs. \ref{entropyalpha5qbyg.01nu1andnu5} (a) and (b). 

It is also clear that the SVNE and SLE display 
marked oscillatory behavior near 
$\frac{1}{2}T_{\rm rev}, \frac{1}{3}T_{\rm rev}$ and 
$\frac{1}{4}T_{\rm rev}$. 
This behavior may be regarded as the counterpart, in the  coupled system,
of the fractional revivals seen in the single-mode case 
(as described
in Chapters 2 and 3). Again, these oscillations 
die down in amplitude with increasing 
$m$ when the initial state of the field is a PACS, and are most 
pronounced  if the field is initially in a Fock state. 

We now examine whether  
signatures of the features seen above appear 
in the time evolution  
of  expectation values of observables. For this 
purpose, we define the  quadratures  
\begin{equation}
\xi = (x_a + x_b)/2, \,\, \eta = (p_a + p_b)/2, 
\label{xiandeta}
\end{equation}
where 
\begin{equation}
x_a = (a + a^\dagger)/\sqrt{2}, \,\,x_b = (b + b^\dagger)/\sqrt{2}
\label{xaxb}
\end{equation}
and 
\begin{equation}
p_a = (a - a^\dagger)/(i\sqrt{2}), 
\,\,p_b = (b - b^\dagger)/(i \sqrt{2}). 
\label{papb}
\end{equation} 
If the field is initially in a Fock state, both $\aver{\xi}$ and 
$\aver{\eta}$ vanish identically at all times.  
\begin{figure}[htpb]
\begin{center}
\includegraphics[]
{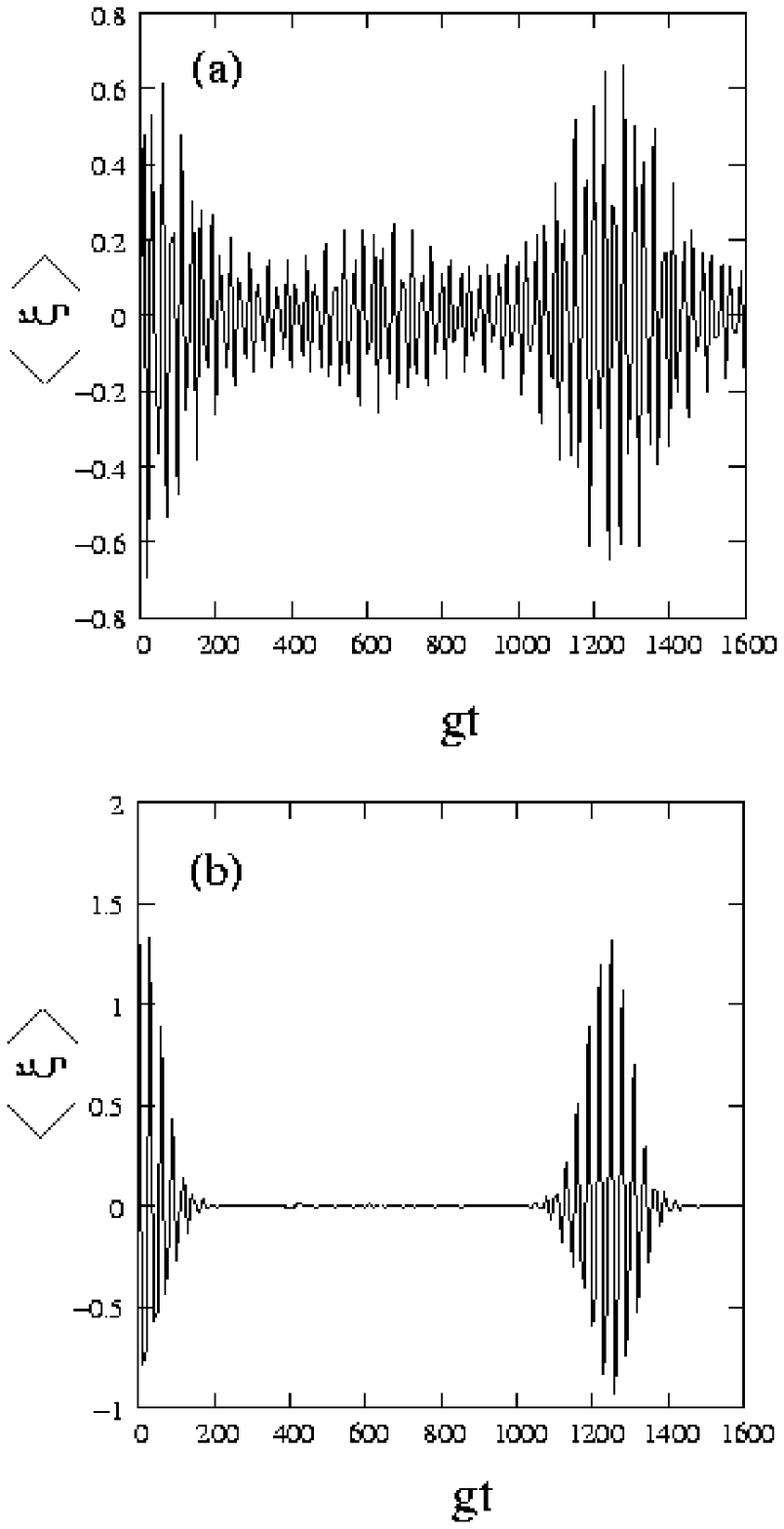}
\caption{$\aver{\xi}$ versus $gt$ for an  
initial state $\ket{\alpha\,;\, 0}$ 
with (a) $\nu = 1$ and (b) $\nu = 5$, respectively. }
\label{X1alphacross0qbyg.01nu1andnu5}
\end{center}
\end{figure}
Setting $\omega =\omega_0 = 1,\, \gamma =1$ and $g = 100$ 
as before (for ready comparison with the behavior of the SVNE 
and SLE  discussed above,),  
we have plotted $\aver{\xi}$  versus $gt$ for an initial state 
$\ket{\alpha\,;\,0}$ with $\nu = 1$ and $5$, respectively, in Figs.
\ref{X1alphacross0qbyg.01nu1andnu5} (a) and (b).
Figure \ref{X1alphacross0qbyg.01nu1andnu5} (a)  
shows that $\aver{\xi}$  
displays rapid pulsed oscillations near $t = 4\pi$, similar to
its behavior near $t = 0$. This mark of a revival 
is consistent with the behavior of 
the SVNE and SLE in this case, {\it cf.} Fig. 
\ref{entropy10cross0andalphacross0qbyg.01nu1} 
(b). The collapses 
are not sharp, in the 
sense that $\aver{\xi}$ is not constant over the time interval between 
successive revivals --- oscillatory bursts 
occur in between, with a slight enhancement of these
oscillations around the fractional revival at 
$\frac{1}{2}T_{\rm rev}$.  
In contrast, the collapses in between revivals 
are much more
complete for larger values of $\nu$, as seen in Fig. 
\ref{X1alphacross0qbyg.01nu1andnu5} 
(b), consistent with 
the corresponding behavior of the SVNE and SLE 
in this case.   
\begin{figure}[htpb]
\begin{center}
\includegraphics[]
{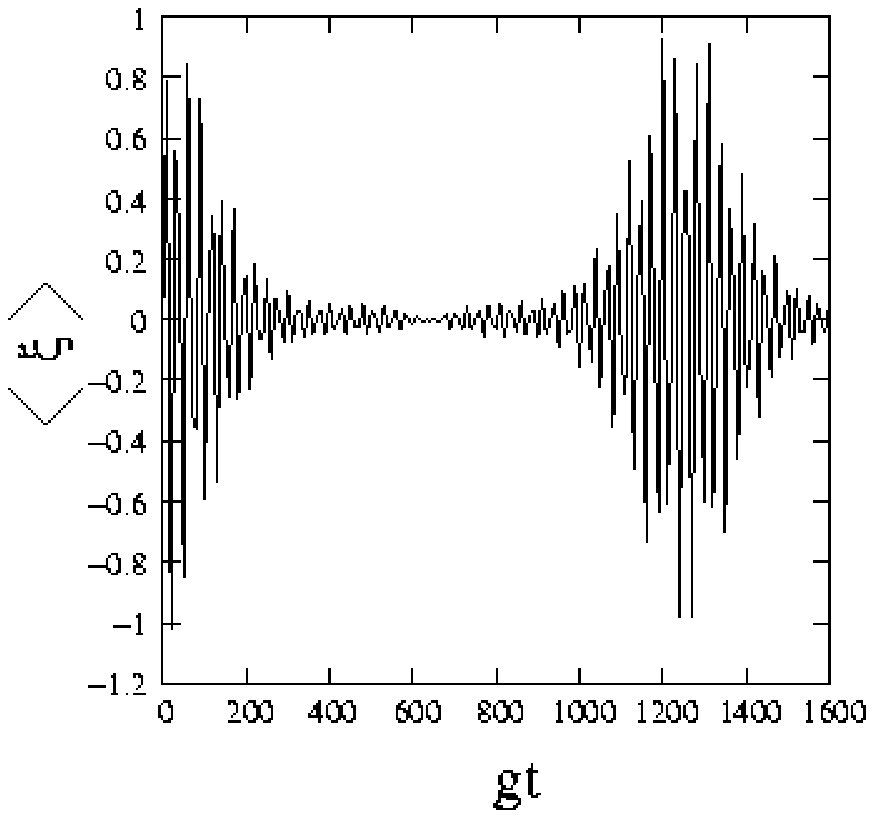}
\caption{$\aver{\xi}$ versus $gt$ for 
an initial state  
$\ket{(\alpha,1)\,;\,0}$ with 
$\nu = 1$.}
\label{X1alpha1cross0qbyg.01nu1}
\end{center}
\end{figure}

An interesting feature is that 
these collapses become much sharper
for even a marginal departure from 
coherence of 
the initial state of the field. $\aver{\xi}$ remains  
virtually constant 
over the duration of the collapse, 
and then bursts into rapid oscillations 
close to revivals, as seen in 
Fig. \ref{X1alpha1cross0qbyg.01nu1} 
which corresponds to 
$m = 1$ and $\nu = 1$. As in the case of 
single-mode dynamics, the  amplitude 
of the oscillations in the neighborhood of $T_{\rm rev}$  
decreases 
significantly with an increase in $m$.  
Thus, for small values of 
$\nu$, it is easy to distinguish between 
an initial CS and an initial PACS. The 
expectation value of  $\eta$ also displays these 
signatures. While the 
sub-system variables $x_a$, $x_b$, $p_a$ and 
$p_b$ do exhibit revivals in their 
expectation values, their higher moments 
do not capture the occurrence of 
fractional revivals. However, 
the higher moments of the combinations 
$\xi$ and $\eta$ {\it do} 
carry distinguishing signatures to 
selectively pin-point the analogs of the 
different fractional revivals that 
occur in the single-mode case. Hence $\xi$ and $\eta$ are 
the appropriate dynamical variables for the interacting  
system under consideration. 
\begin{figure}[htpb]
\begin{center}
\includegraphics[]
{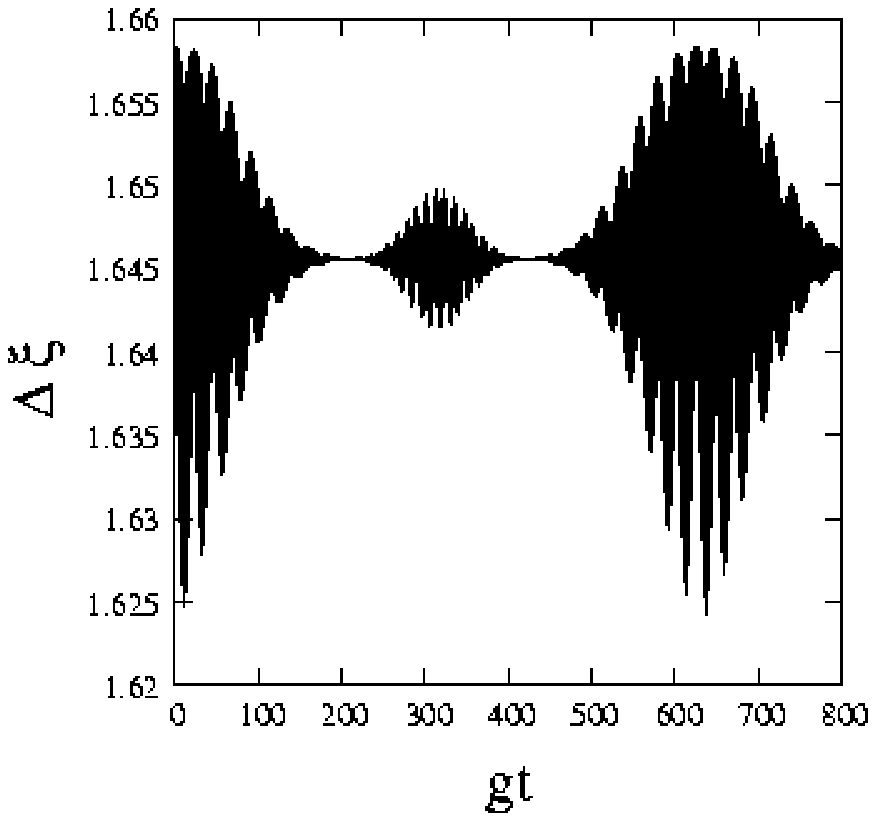}
\caption{$\varDelta \xi$ versus $gt$ for an initial state 
$\ket{10\,;\,0}$.}
\label{deltaX110cross0qbyg.01}
\end{center}
\end{figure}

\begin{figure}[htpb]
\begin{center}
\includegraphics[]
{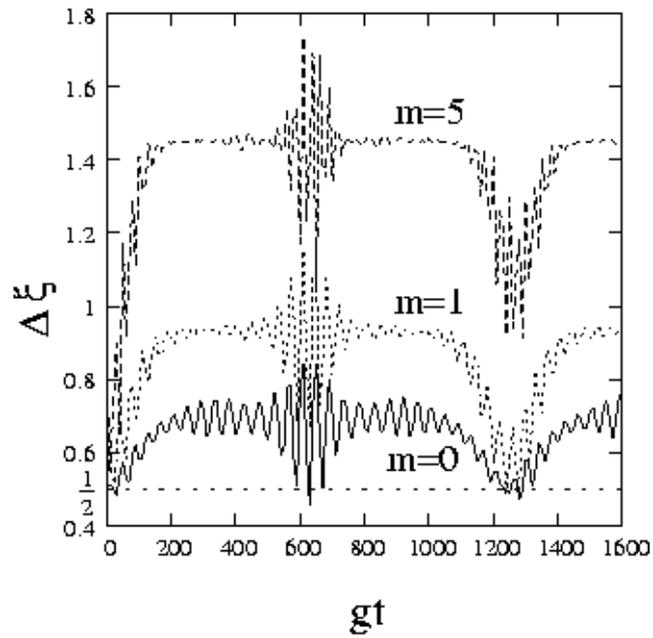}
\caption{$\varDelta \xi$ versus $gt$ for an initial 
state $\ket{(\alpha,m)\,;\, 0}$ with
$\nu = 1$ and $m = 0, 1$ and $5$, respectively.}
\label{deltaX1alphacomma01and5cross0qbyg.01nu1}
\end{center}
\end{figure}

The standard deviation $\varDelta{\xi}$ of $\xi$ 
reflects the occurrence of the dips 
in the plots of the 
SVNE and SLE at $\frac{1}{2}T_{\rm rev}$.
Consider, first, the case when the initial state is of the form
$\ket{N;\,0}$. Then a plot 
of $\varDelta{\xi}$ versus $gt$ shows a  
 burst of rapid oscillations at 
$t \approx \pi\, (\approx \frac{1}{2}T_{\rm rev})$. 
Fig. \ref{deltaX110cross0qbyg.01} illustrates this for $N=10$. (Note
 that the case $N=1$ is exceptional in this regard, since $\varDelta \xi$ 
reduces to a constant in this instance.) 
This feature holds for an initial CS or  PACS as well,  
as is evident (see Fig. 
\ref{deltaX1alphacomma01and5cross0qbyg.01nu1})  
from the  sudden burst of 
oscillations in $\varDelta \xi$ 
around $t \approx 2\pi\approx\frac{1}{2}T_{\rm rev}$ 
for initial states  
$\ket{\alpha\,;\, 0}$,
$\ket{(\alpha,1)\,;\, 0}$ and $\ket{(\alpha,5)\,;\, 0}$
(recall that $T_{\rm rev} \approx 4\pi$ in this case). 
%\begin{figure}[htpb]
%\begin{center}
%\includegraphics[]
%{chapter5/deltaX110cross0qbyg0.01.eps}
%\caption{$\Delta \xi$ versus $gt$ for an initial state 
%$\ket{10\,;\,0}$\,  
%($\gamma/g = 10^{-2}$).}
%\label{deltaX110cross0qbyg.01}
%\end{center}
%\end{figure}

\begin{figure}[htpb]
\begin{center}
\includegraphics[]
{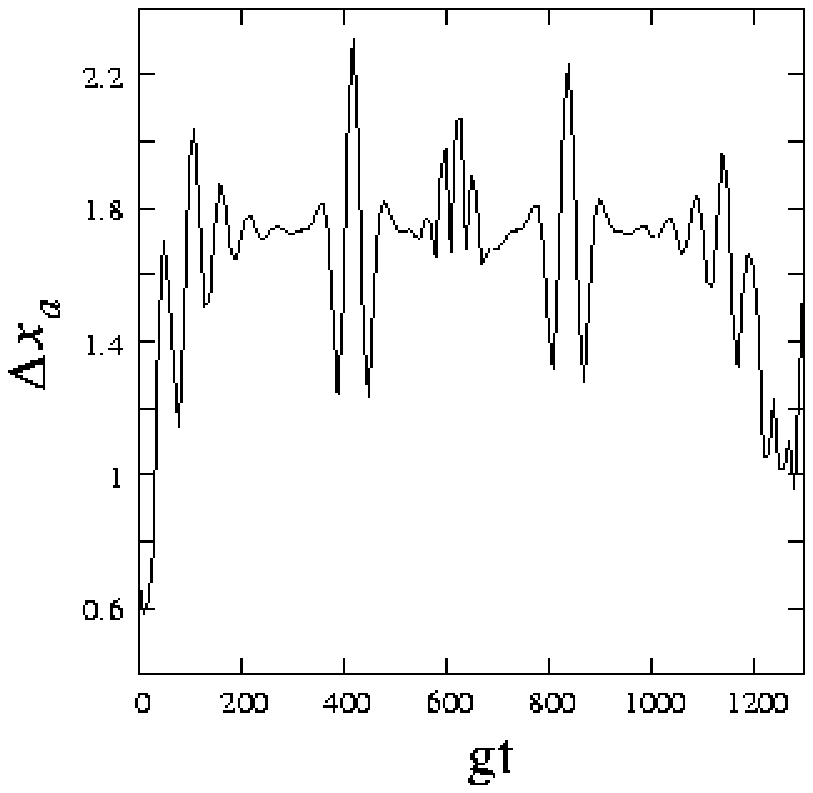}
\caption{$\varDelta x_a$ versus $gt$ for an initial
state $\ket{\alpha\,;\, 0}$ with
$\nu = 5$.}
\label{deltaxam0nu5}
\end{center}
\end{figure}

We note that {\it squeezing} occurs in the neighborhood of  
$\frac{1}{2}T_{\rm rev}$
when the initial state 
of the field is a coherent 
state: $\varDelta \xi$ drops below the 
value $\frac{1}{2}$ (the horizontal
dotted line in 
Fig. \ref{deltaX1alphacomma01and5cross0qbyg.01nu1}) 
in the case when the initial state is 
$\ket{\alpha \,;\, 0}$,  
in contrast to what happens for an initial 
PACS $\ket{(\alpha,m)\,;\,0}$. 
While this is similar to  
the squeezing property seen in the  
single-mode case, such parallels 
do not hold in the case of higher-order 
squeezing. The relevant quadrature variables 
in this case (analogous to those in 
Eq.(\ref{Z1Z2})) are
\begin{equation}
Z_1=\frac{(a^q+a^{\dagger q}+b^q+b^{\dagger q})}{2\sqrt{2}},\quad
Z_2=\frac{(a^q-a^{\dagger q}+b^q-b^{\dagger q})}{2i\sqrt{2}}.
\label{Z1Z2coupled}
\end{equation}
For $q = 1$, $Z_1$ and $Z_2$ reduce  to $\xi$ and $\eta$, respectively.  
The indicator of higher-order squeezing is defined, analogous to  
$D_q(t)$ in Eq. (\ref{Dqone}) of Chapter 4, as
\begin{equation}
D_q(t) =\frac{(\varDelta Z_1)^2-
\frac{1}{2}|\aver{[Z_1,\,Z_2]}|}{\frac{1}{2}|\aver{[Z_1,\,Z_2]}|},
\label{Dq2mode}
\end{equation}
 where $\varDelta Z_1$  denotes the standard deviation in the variable 
$Z_1$ defined in Eq. (\ref{Z1Z2coupled}). We note that 
$[Z_1\,,\,Z_2]$ is no longer a polynomial 
function of the (total) number 
operator ${\sf N}_{\rm tot}$ alone. 
Extending the notation used 
in Chapter 4,  we shall write 
$D_{q}^{(m)}$ for $D_q$ in the case of an initial state 
$\ket{(\alpha,m)\,;\,0}$, and $D_{q}^{(N)}$ to denote $D_q$ for an 
initial state $\ket{N\,;\,0}$. 
In contrast 
to the single-mode case, even for weak nonlinearity ($\gamma/g = 
10^{-2}$) and $\nu = 1$, amplitude-squared 
squeezing ($q = 2$) is absent at $t 
= \frac{1}{2}T_{\rm rev}$ whether the 
field is in a Fock state, or a CS, or a PACS.  
For $N = 10$, and an initial state $\ket{10\,;\, 0}$, we have plotted 
$D_2^{(N)}$ versus time in Fig. \ref{d2n10}. The state is never squeezed, 
as $D_2^{(N)}$ is always positive.  
Similar plots for an initial state $\ket{\alpha\,;\, 0}$ and 
$\ket{(\alpha, 1)\,;\, 0}$ are given in Fig. {\ref{d2m0and1}. While  
$D_2^{(0)}$ is almost zero at certain instants for a coherent
 initial field 
state $\ket{\alpha}_a\,$, 
it is clear that it does not drop below 
the horizontal 
line indicating the value below which amplitude-squared squeezing 
occurs. 
\begin{figure}[htpb]
\begin{center}
\includegraphics[]
{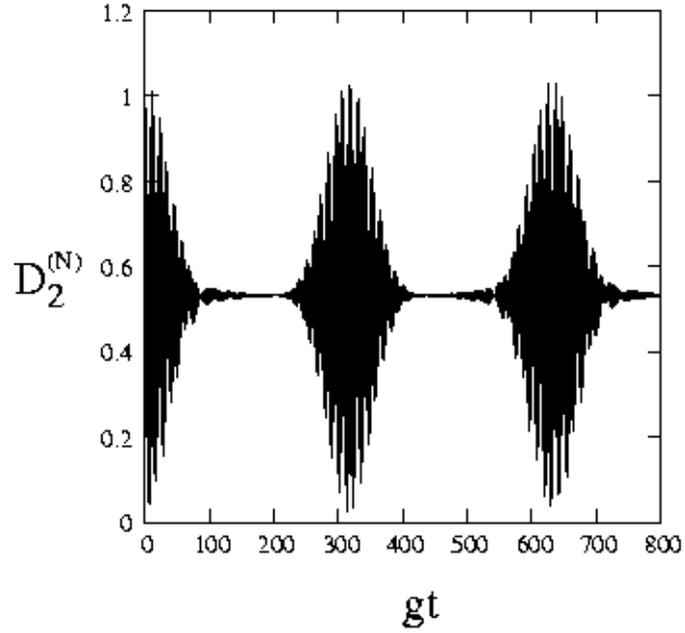}
\caption{$D^{(N)}_2 $ versus $gt$ for an initial
state $\ket{N\,;\, 0}$ with $N = 10$ .}
\label{d2n10}
\end{center}
\end{figure}
\begin{figure}[htpb]
\begin{center}
\includegraphics[]
{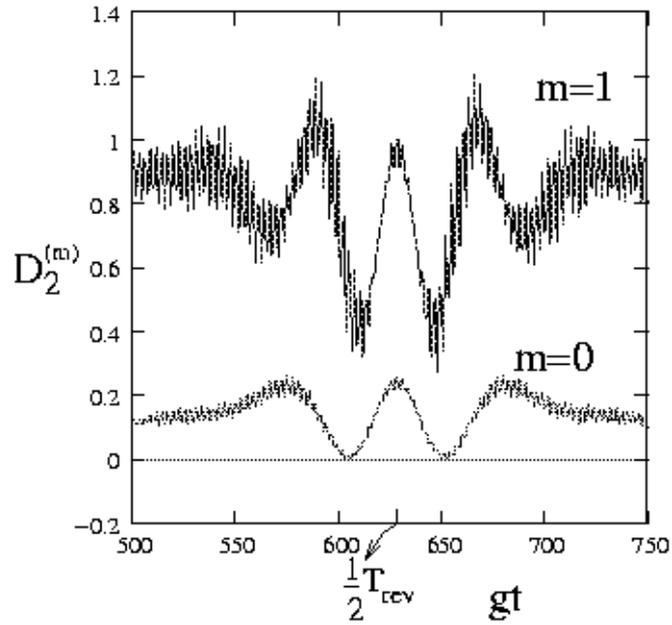}
\caption{$D^{(m)}_2 $ versus $gt$ for an initial 
state $\ket{\alpha\,;\, 0}$ and $\ket{(\alpha,1)\,;\, 0}$   with
$\nu = 1$.}
\label{d2m0and1}
\end{center}
\end{figure}
\begin{figure}[htpb]
\begin{center}
\includegraphics[width=5.5in]
{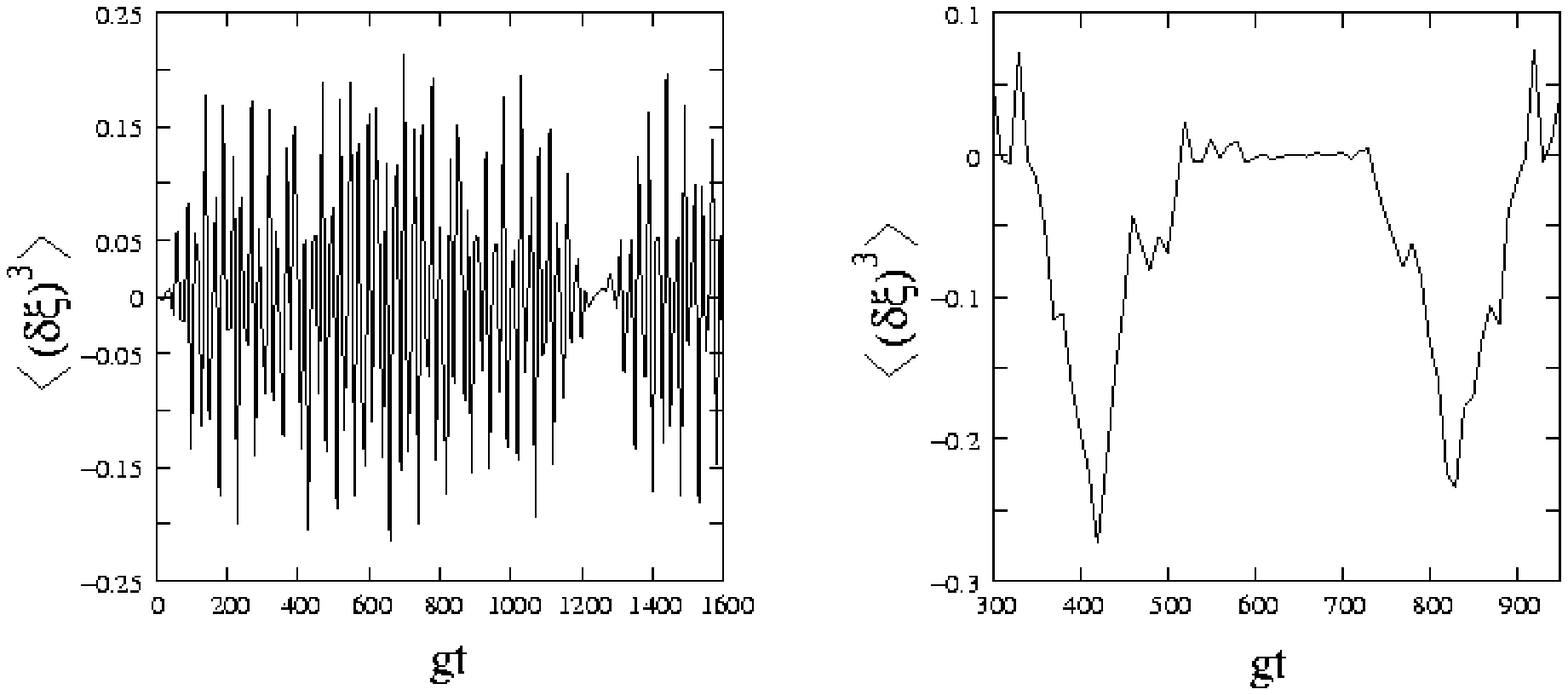}
\caption{$\aver{(\delta \xi)^3}$ versus $gt$ for an initial
state (a) $\ket{\alpha\,;\, 0}$ (b) $\ket{(\alpha,3)\,;\, 0}$ with
$\nu = 1$  and $\gamma/g = 10^{-2}$.}
\label{deltaxi3m0and5nu1}
\end{center}
\end{figure}
%\begin{figure}[htpb]
%\begin{center}
%\includegraphics[width=3in]
%{deltaxi3m0nu5.eps}
%\caption{$\aver{(\delta \xi)^3}$ versus $gt$ for an initial
%state $\ket{\alpha\,;\, 0}$ with
%$\nu = 5$  and $\gamma/g = 10^{-2}$.}
%\label{deltaxi3m0nu5}
%\end{center}
%\end{figure}

Turning to the higher moments of 
$\xi$ and $\eta$, we note that all the odd 
moments of $\xi$ vanish identically for all $t$ 
if the initial state is a direct product of Fock states (see Appendix D).
Analogous to the single-mode case,
for small values of $\nu$,  
the higher moments  of $\xi$ 
show distinct signatures 
at fractional 
revivals only if $m$ is sufficiently 
large. This is evident from Fig. \ref{deltaxi3m0and5nu1}, where we have 
plotted $\aver{(\xi-\aver{\xi})^3}\equiv \aver{(\delta \xi)^3}$   
versus time for initial states 
$\ket{\alpha\,;\, 0}$ and  $\ket{(\alpha\,,\,3)\,;\, 0}$, 
respectively, with
$\nu = 1$  and $\gamma/g = 10^{-2}$.} In the latter case, the 
signatures are manifested in the pronounced 
dips in $\aver{(\delta\xi)^3}$ at $\frac{1}{3}
T_{\rm rev}$ and $\frac{2}{3} 
T_{\rm rev}$.	
%\begin{figure}[htpb]
%\begin{center}
%\includegraphics[width=5.5in]
%{deltaxi3m0and3nu1.eps}
%\caption{$\aver{(\delta \xi)^3}$ versus $gt$ for an initial 
%state (a) $\ket{\alpha\,;\, 0}$ (b) $\ket{(\alpha\,3);\, 0}$ with
%$\nu = 1$  and $\gamma/g = 10^{-2}$.}
%\label{deltaxi3m0and5nu1}
%\end{center}
%\end{figure}
\begin{figure}[htpb]
\begin{center}
\includegraphics[width=3in]
{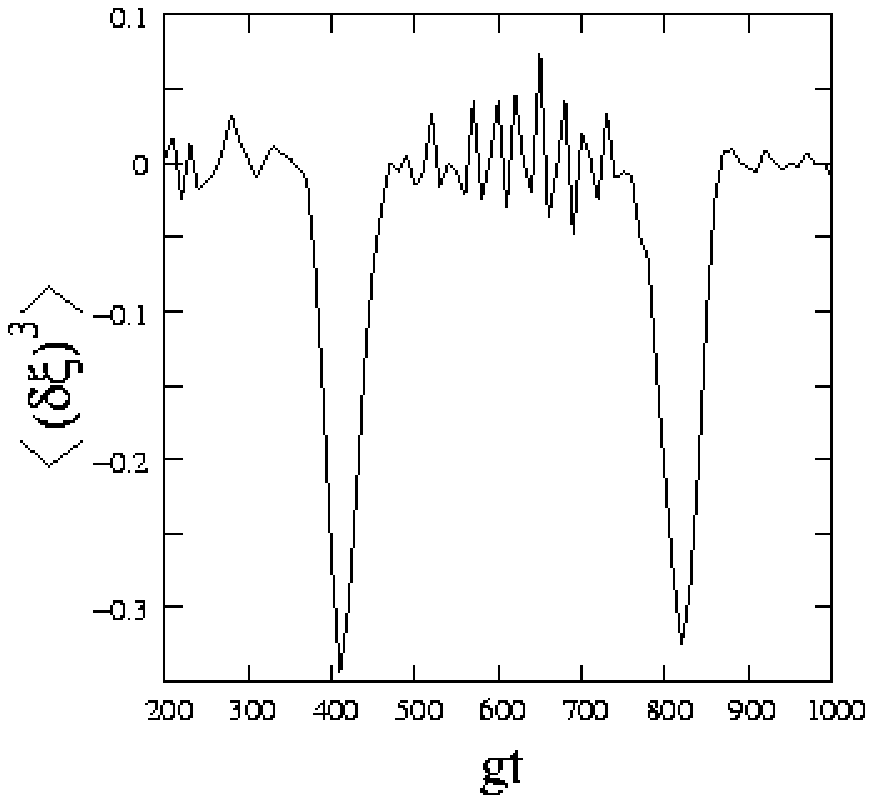}
\caption{$\aver{(\delta \xi)^3}$ versus $gt$ for an initial 
state $\ket{\alpha\,;\, 0}$ with
$\nu = 5$.}
\label{deltaxi3m0nu5}
\end{center}
\end{figure}
However, for larger values of $\nu$, 
such signatures appear even in the 
case of an initial CS ($m = 0$), as is evident from Fig. 
\ref{deltaxi3m0nu5}.
 In contrast to this,  the 
variances and higher 
moments of the subsystem variables $x_a$, $x_b$, $p_a$ and $p_b$ do not 
display these signatures. For instance, the plot of $\varDelta x_a$ versus 
time shown in Fig. \ref{deltaxam0nu5} does not display  a 
distinctive signature at 
$\frac{1}{2}T_{\rm rev}$ alone: apart from oscillations 
in the vicinity of this instant, 
we see marked dips around $\frac{1}{3} T_{\rm rev}$ 
and $\frac{2}{3} T_{\rm 
rev}$ as well. 

\section{Concluding remarks}
In the light of our discussion, it is clear that the quadrature 
variables $\xi$ and $\eta$ corresponding to the full system are the 
appropriate 
observables whose dynamics signals the occurrence of the analogs of  
revival phenomena, while the 
expectation values of sub-system variables would seem to be
less suitable  choices for 
highlighting collapse and revival 
phenomena. On heuristic grounds, we may, however,  
expect them to be better candidates 
for  investigating non-periodic 
temporal behavior --- in particular, time  
evolution that is actually indicative of exponential instability. 
We address this issue in the next chapter, in which we 
analyze the dynamics of the sub-system 
variable $\aver{a^{\dagger}a}$ 
for a wide range of values of the parameter 
$\nu$ and the ratio $\gamma/g$.

%% file: timeseries.tex
\chapter{Ergodicity properties of entangled two-mode 
states - II}\label{timeseries}
\section{Properties of the coupled-system Hamiltonian}
\label{timeseries}

In this chapter, we continue our investigation of the dynamics of 
a single-mode electromagnetic field propagating in a nonlinear medium, 
with the model 
Hamiltonian of Eq. (\ref{coupledhamil1}). We write down $H$ again for
ready reference, restoring the over-all factor of $\hbar$ for the
moment, as in Eq. (\ref{2modehamiltonian}):   
\begin{equation}
H = \hbar \,[ \omega \,a^\dagger a + \omega_0 \,b^\dagger b +  \gamma\,
b^{\dagger 2} b^2 +  g \,(a^\dagger b + b^\dagger a)].
\label{coupledhamil2}
\end{equation}
We recall that the operators $a$ and $a^{\dagger}$ 
pertain to the 
sub-system represented by the field, 
while $b$ and $b^{\dagger}$ are the ladder  
operators for the other sub-system, the anharmonic  
oscillator modeling an atom of the medium. We 
have found in Chapter 5
that sub-system variables do not adequately 
capture the effects of revival 
phenomena. On the other hand, this would make them 
potentially good candidates for examining  the 
ergodicity properties of the system. The mean photon number 
$\aver{\sf N} = \aver{a^{\dagger}a}$, which is also essentially 
the average energy of the field mode,   
turns out to be an ideal candidate for this purpose. 
Our reasons for this choice are as follows. 

In the single-mode model with Hamiltonian 
$\hbar\,\chi\,{\sf N}({\sf N} -1)$ considered in Chapters 2,
3 and 4,  
it is evident that the expectation value of  
${\sf N}$, as well as all its higher moments, are
constants of the motion for any initial 
state. In contrast to this, 
${\sf N}$ does not 
commute with the Hamiltonian (\ref{coupledhamil2}) of the 
coupled system. As a consequence of the 
entanglement between the atom and field modes, 
the mean photon number is 
no longer a constant of the motion for 
any non-zero value of the coupling 
constant $g$. However, as already stated, 
the total number operator 
${\sf N}_{\rm tot} 
= a^{\dagger}a + b^{\dagger}b$ commutes with 
$H$ for all values of the parameters in $H$. Its expectation value is
therefore always a constant of the motion. In the absence of 
the anharmonicity parameter $\gamma$, the coupled Hamiltonian is 
essentially linear in each of the sub-system variables, and can be
diagonalized in terms of linear combinations of the original ladder 
operators. In physical terms, this leads to fairly simple dynamics, 
essentially entailing a simple periodic exchange of energy 
between the two sub-systems or modes. (This mechanism 
becomes even more 
transparent if we set $\omega = \omega_0\,$.) When $\gamma$ is
non-zero, the dynamics is more complicated, but the existence 
of the operator ${\sf N}_{\rm tot}$ ensures that the system as 
a whole is ``well-behaved'', with a discrete spectrum (labeled, as
 we have seen, by the quantum numbers $N$ and $s$, where 
$N = 0,1,\ldots$ and $s = 0,1,\ldots , N$). The 
{\it time-dependence} of 
$\aver{\sf N}$ 
is thus a direct consequence 
of the {\it coupling} of the field mode to another degree of 
freedom; and the 
{\it deviation} of 
$\aver{\sf N}$ from {\it periodic} temporal variation is a 
probe of the effects of the {\it nonlinearity} 
in the latter. It turns out that a remarkable diversity of temporal 
behavior is exhibited by $\aver{\sf N}$, 
in different ranges of the parameters in $H$.  

In this context, it is very helpful to understand the dynamical 
behavior of   
the completely {\it classical} counterpart of the Hamiltonian 
of Eq. (\ref{coupledhamil2}). This analysis is also needed in 
order to facilitate subsequent comparison of the quantum  
dynamics with the classical one. We may obtain the classical 
Hamiltonian as follows. Let the linear harmonic 
oscillator associated with $a$ and $a^{\dagger}$ have a mass $m$,  
position $x$ and momentum $p_x\,$. Similarly, let the oscillator 
associated with $b$ and $b^{\dagger}$ have a mass $M$, position $y$
and momentum $p_y\,$. Putting in all constant factors (including  
$\hbar$), we have 
\begin{equation}
a = \sqrt{\frac{m\omega}{2\hbar}}\,x + \frac{ip_x}
{\sqrt{2m\hbar\omega}}\,,\quad
b = \sqrt{\frac{M\omega_0}{2\hbar}}\,y + \frac{ip_y}
{\sqrt{2M\hbar\omega_0}}\,.
\label{effectivecoords}
\end{equation}
The hermitian conjugates $a^{\dagger}$ and $b^{\dagger}$ are 
of course obtained by replacing $i$ with $-i$ in the
above. Substituting the foregoing in Eq. (\ref{coupledhamil2}), 
we get $H$ in terms of $x,\,p_x,\,y$ and $p_y\,$. 
Passing to the classical limit in which 
$\hbar \rightarrow 0$, we find   
that the only consistent way to obtain a non-trivial, finite 
expression for the classical Hamiltonian in this limit is to 
let $\gamma$ also tend to zero simultaneously, such that 
the ratio $\gamma/\hbar$ tends to a finite value $\lambda$, say. When
this is done, the classical Hamiltonian obtained is 
\begin{eqnarray}
H_{\rm cl} &=& 
\frac{p_{x}^{2}}{2m} + \frac{1}{2}m\omega^2 x^2 + 
\frac{p_{y}^{2}}{2M} + \frac{1}{2}M\omega_0^2 y^2 + 
\frac{\lambda}{\omega_0^2}
\left(\frac{p_{y}^{2}}{2M} 
+ \frac{1}{2}M\omega_0^2 y^2\right)^2\nonumber \\
&+& \frac{g}{\sqrt{\omega \omega_0}}\left(
\sqrt{mM}\,\omega\omega_0\, xy + \frac{p_x \,p_y}{\sqrt{mM}}\right),
\label{classicalcoupledhamil1}
\end{eqnarray}
with the canonical Poisson bracket relations $\{x\,,\,p_x\} = 
\{y\,,\,p_y\} = 1$. As it stands, it is not immediately obvious
that the
corresponding motion is bounded for all initial conditions, owing 
to the presence of the cross terms $xy$ and $p_x\,p_y$ that can 
assume either sign. However, 
the counterpart of ${\sf N}_{\rm tot} = a^{\dagger}a + 
b^{\dagger}b$, namely, 
\begin{equation}
N_{\rm cl} = 
\frac{1}{\omega}\left( \frac{p_{x}^{2}}{2m} +
 \frac{1}{2}m\omega^2 x^2 \right) 
+ \frac{1}{\omega_0}\left(
\frac{p_{y}^{2}}{2M} + \frac{1}{2}M\omega_0^2 y^2 \right),  
\label{Nclassical}
\end{equation}
is a second, analytic, constant of the motion $\left(\{N_{\rm cl}\,,\,
H_{\rm cl}\} = 0\right)$. 
This ensures, of course, that the 
two-freedom system is Liouville-Arnold integrable. 
Further, since $N_{\rm cl} = {\rm constant}$ is evidently a
 bounded hypersurface (an ellipsoid) 
in the four-dimensional phase space,
it is clear that motion under $H_{\rm cl}$ is indeed bounded 
for any set of initial conditions. All four Lyapunov exponents
vanish, and the classical motion is always regular (and restricted  
to a $2$-torus) for each set of initial conditions. 
The dynamics can be analyzed in further detail,  
including the identification of 
the location and nature of 
the corresponding critical or equilibrium points. 
Several features emerge 
that are interesting in their own right, 
but we do not go into these here, as 
they are not directly 
germane to our present purposes. We note, though, that 
the equation of motion of the classical counterpart of ${\sf N}~\,\,
~(=~a^{\dagger}a)$, on evaluating the Poisson bracket concerned, is  
\begin{equation}
\frac{d}{dt} \left[
\frac{1}{\omega}\left( \frac{p_{x}^{2}}{2m} +
 \frac{1}{2}m\omega^2 x^2 \right) \right] = 
g\left(\sqrt{\frac{m\omega}{M\omega_0}}\, x\,p_y -
\sqrt{\frac{M\omega_0}{m\omega}} \, y\,p_x \right).
\label{eomclassicalN}
\end{equation} 
The equation of motion of the combination on the right-hand side 
of Eq. (\ref{eomclassicalN}) involves other functions of the 
dynamical variables, and so on. The resulting set of 
equations does not
close. This is brought out more clearly in the quantum mechanical
problem, to which we now revert.  
  
Returning to the quantum mechanical Hamiltonian 
(\ref{coupledhamil2}), some additional insight into its precise nature
is provided by re-writing it in terms of angular momentum operators 
\cite{agar1}. The latter are defined in the standard manner in terms of
the two mutually commuting 
sets of boson operators, according to 
\begin{equation}
J_{+} = a^{\dagger} b\,,\,\,J_{-} = ab^{\dagger}\,,\,\,
J_z = \frac{1}{2}(a^{\dagger}a - b^{\dagger}b).
\label{angmomoperators}
\end{equation}
The Hamiltonian in Eq. (\ref{coupledhamil2}) becomes 
\begin{equation}
H = \hbar\left[\frac{1}{2}(\omega + \omega_0 -\gamma){\sf N}_{\rm tot} 
+ (\omega -\omega_0 + \gamma) J_z 
+ \frac{1}{4}\gamma ({\sf N}_{\rm tot} - 2J_z)^2 
+ 2g J_x\right].
\label{coupledhamil3}
\end{equation}
We also have the well known relation 
$J^2 = \frac{1}{2} {\sf N}_{\rm tot}
\left(\frac{1}{2}{\sf N}_{\rm tot} +
1\right)$. Therefore, 
as $J^2$ is a function of ${\sf N}_{\rm tot}\,$, it is evident that 
${\sf N}_{\rm tot}$ commutes not only with $H$, but also with every
component $J_i$ of the angular momentum. However, 
$[J_z\,,\,J_x] \neq 0$, of course, and this suffices to 
preclude exact diagonalization of $H$. It is clear that we have 
a problem similar to that of an Ising spin in a 
transverse magnetic field. 
The quantity whose time 
evolution we shall track is the expectation value of 
${\sf N} = a^{\dagger}a = J_z + \frac{1}{2}{\sf N}_{\rm tot}\,$. 
Using the standard algebra of 
angular momentum generators, 
the equation of motion of this operator works out to  
\begin{equation}
\frac{d {\sf N}}{dt} = 
\frac{d J_z}{dt} = 
(i\hbar)^{-1}\,[{\sf N}\,,\,H] 
= -i g\,(a^{\dagger}b - ab^{\dagger}) = 2g\,J_y\,.
\label{eomforphotonnumber}
\end{equation}
But we also find  
\begin{equation}
\frac{d J_y}{dt} = 
(\omega -\omega_0 -\gamma \,{\sf N}_{\rm tot})\,J_x 
- 2g\,J_z + \gamma \,[J_z\,,\,J_x]_{+}
\label{eomjy}
\end{equation}
and 
\begin{equation}
\frac{d J_x}{dt} = (\gamma\,{\sf N}_{\rm tot} - \omega + \omega_0 
- \gamma)\,J_y -\gamma\,[J_y\,,\,J_z]_{+}\,,
\label{eomjx}
\end{equation}
where $[\cdots\,,\,\cdots]_{+}$ stands for the 
anti-commutator of the
operators concerned. When differentiated with respect to $t$, these
terms lead to trilinear combinations of the angular momentum
operators, and so on. It is easily 
checked that the system of equations does not
close.  

Having obtained some idea of the dynamics we can expect of the
operator ${\sf N}$, we now turn to a numerical study. We shall regard 
the mean photon number 
$\aver{\sf N}$ (equivalently, the mean energy of the field mode) 
as an effective classical dynamical variable, and
analyze its time series over long intervals of time. We carry out this 
analysis for a wide range of values of the 
ratio $\gamma/g$, which is a measure of the relative strengths of 
the nonlinearity and the interaction present in the coupled 
system. As in Chapter 5, we shall take the atomic oscillator to be 
initially in the ground state, and the field to be in a CS or a PACS.  

In order to set the framework, in the next section 
we briefly outline the relevant   
features of time-series analysis \cite{abar}, as this topic is quite
distinct from those involved in the preceding parts of this thesis.   

\section{Time-series analysis in brief}

We consider a typical situation, in which 
the time-series data of just one observable (signal), or that of a 
certain sub-set of observables 
corresponding to a given physical system,  
are available for 
analysis. In general, there will be 
other dynamical 
variables   
whose time series are not 
readily accessible, but which also control
the full dynamics of the system.
The dynamical equations are typically a  set of 
nonlinear coupled 
evolution equations for the complete set 
of dynamical variables.
In time-series analysis one attempts to  
ascertain,  from the available data, the {\it effective} number of 
variables that are actually 
relevant in determining 
the temporal behavior of the system,  
consequently establishing the 
minimum number of effective dimensions  
$d_{\rm emb}$ (the {\it embedding dimension}) of the 
``phase space''  of these relevant variables. 
Evidently, these 
variables have to be generated from the 
values of the observed variables 
at a given 
time and those corresponding to their 
time-delayed copies. 
Hence this space is only 
formally equivalent to the actual phase space of the system, 
and the dynamical variables generated from the data 
are merely effective variables. However, if the analysis is 
carried out appropriately,  
they will (in most cases) 
capture the salient features of the dynamics in the 
actual phase space of the system --- in particular, of 
dynamical chaos, if any.        

The embedding theorem \cite{takens} indicates 
how $d_{\rm emb}$ can 
be obtained. To begin with, one assumes different 
integer values $d$ 
for the dimensionality of the effective phase space, and 
generates the $d$ dynamical variables from 
the time-series data available. 
Each of these variables is regarded as a 
component of a $d$-dimensional vector.
The embedding dimension will turn out to be one 
of the values assumed for $d$, obtained by the procedure 
described below.  
                                                                               
Generating the set of effective dynamical variables 
at any instant involves the identification of 
a {\it time delay} $\tau$ such that,  
from  the available time-series of the signal, 
data points separated by time intervals $\geq \tau$ 
are sufficiently independent of each other 
to be treated as the independent dynamical variables. 
We assume that the time series is obtained by sampling 
the variable(s) concerned at time steps $\delta t$. 
Clearly, if two data points $s(n)$ and $s(n+T)$ at discrete times 
$n$ and $(n+T)$ (in units of $\delta t$) 
are sufficiently independent of each other,  
the information deducible regarding 
the measured value of $s(n)$ 
through the measurement of $s(n+T)$ 
must tend to zero, for sufficiently 
large $T$. More precisely, if $p\big(s(n)\big)$ 
and $p\big(s(n+T)\big)$ are the 
individual probability densities
for obtaining the values $s(n)$ and $s(n+T)$ at times $n$ and $(n+ T)$, 
respectively, and $p\big(s(n)\,,\,s(n+T)\big)$ is the
corresponding joint probability density, 
the average mutual information 
\begin{equation}
I(T) = \sum_{s(n)\,,\,s(n+T)}\,p\big(s(n),s(n+T)\big)\,
\log_{2}\,
\Big[\frac{p\big(s(n)\,,\,s(n+T)\big)}{p\big(s(n)\big)\,
p\big(s(n+T)\big)}\Big]
\label{mutualinformation}
\end{equation}
should tend to zero for sufficiently large $T$. 
In practice, a frequently-used 
prescription \cite{fraser} is 
to take $\tau$ to be that value of $T$ 
at which the first minimum in 
$I(T)$ occurs. Thus, from a 
time series 
$s(0),\, s(1),\, s(2),\ldots$ of the 
signal measured at discrete times $t = 0,\, 
1,\,2,\,\ldots \,$, one constructs the 
phase space vector $\mathbf{s}_0$ 
at $t = 0$, with 
components $s(0),\, 
s(\tau),\, s(2 \tau),\,\ldots ,\,s((d-1) \tau)$. 
After the first time step, the vector 
$\mathbf{s}_0$ evolves to $\mathbf{s}_1\,$, 
 with components  $s(1),\, s(1 + \tau),\, 
s(1 + 2 \tau),\,\ldots\,,s(1 + 
(d - 1) \tau)$, and so on.    
This procedure is used to 
reconstruct sets of time-delayed vectors in the 
$d$-dimensional phase space. Each vector defines a point in the space. 
Taken in the correct sequence, these points 
describe the time evolution of the underlying dynamical system.   

This procedure is carried out for different values of $d$. The next 
task is to identify the value of $d$ that corresponds to 
$d_{\rm emb}\,$. 
For this purpose, we evaluate 
the correlation integral  
\begin{equation}
C(r)= \lim_{n\rightarrow\infty}\,\frac{1}{n^2}\sum_{i,j=0}^{n-1}
\theta\,(r-|\mathbf{s}_i-\mathbf{s}_j|)
=\int_{0}^{\,r}\!d^d r\,'\, c (\mathbf{r}\,')\,.
\label{correlnintegral}
\end{equation}
$C(r)$ is an estimate of the average correlation 
between the various 
points in a phase space of a given dimensionality $d$ that 
contains a large number of points $n$ 
generated from the time series data. The integrand 
$c(\mathbf{r})$ itself is the standard correlation function.  
It has been argued \cite{grass} that, for sufficiently small $r$, 
$C(r)$ scales like a power  
$r^{\zeta}$ of $r$\,; moreover, once $d$ reaches the 
correct minimum embedding dimension $d_{\rm emb}\,$, 
the exponent $\zeta$ will not change with 
an increase in $d$ beyond 
$d_{\rm emb}$. This determines the  
exact value of $d_{\rm emb}\,$.
In practice, $d_{\rm emb}$
could be much smaller than the dimension of the actual phase 
space of the
system considered. However, in a generic situation, the dynamics of 
the effective variables alone, in the phase space of dimensionality 
$d_{\rm emb}\,$,  should suffice to extract 
quantitative information 
about the dynamical chaos in the signal.

A  specific objective of our study is to determine whether 
a sub-system variable ( an appropriately chosen expectation value) can
exhibit exponential sensitivity and 
instability, i.~e., ``chaotic'' behavior. For 
this purpose we need to calculate the maximal Lyapunov 
exponent in the reconstructed phase space of dimensionality  
$d_{\rm emb}\,$. 
There are, in general, $d_{\rm emb}$ Lyapunov exponents 
corresponding to the dynamics in the reconstructed phase space. 
For our purposes, it 
suffices to estimate the largest Lyapunov exponent $\lambda_{\rm max}$ of 
this set. If this quantity 
turns out to be a positive number, we have 
chaotic behavior. We use a robust algorithm 
\cite{rosen} developed for 
the estimation of the maximal Lyapunov exponent from  
data sets represented by time series (see also \cite{kantz}).
We start with the set of ``distances'' 
$\{d_j(0)\}$,  where $d_j(0)$  
is the separation 
between the $j^{\rm th}$ pair of nearest 
neighbors in the phase space. This 
evolves under the dynamics to the set of distances  
$\{d_j(k)\}$ after $k$ time steps $\delta t$, i.~e., at time $t =
k\,\delta t$.  
The quantity $\aver{\ln\,d_{j}(k)}$, where the angular brackets denote the 
average over all values of $j$, is plotted against the time, 
$k\,\delta t$. The result is, in general, a curve that has a short 
initial transient region, a long, clear-cut linear region, and a 
subsequent saturation region. The slope of the linear region, 
identified by a least squares fit, 
is precisely the largest Lyapunov exponent sought, 
$ \lambda_{\rm max}$. Once again, 
it can be established that, if the procedure 
for phase-space reconstruction is 
implemented carefully and the minimum embedding dimension
$d_{\rm emb}$ obtained correctly,
any further increase in the dimensionality of the 
reconstructed phase space
should not alter the inferences made regarding the 
exponential instability, if any, of the system.

With this summary of the relevant aspects of time series analysis, 
 we proceed to  
carry out a detailed time 
series 
analysis of the values of the mean photon number obtained at regular 
intervals of time, as the 
single-mode field propagates in the nonlinear medium. The corresponding 
power spectra of the signal have been obtained using a 
standard package (Labview). 

\section{Ergodicity properties of the mean photon 
number}

\begin{figure}[htpb]
\begin{center}
\includegraphics[width=5.7in]
{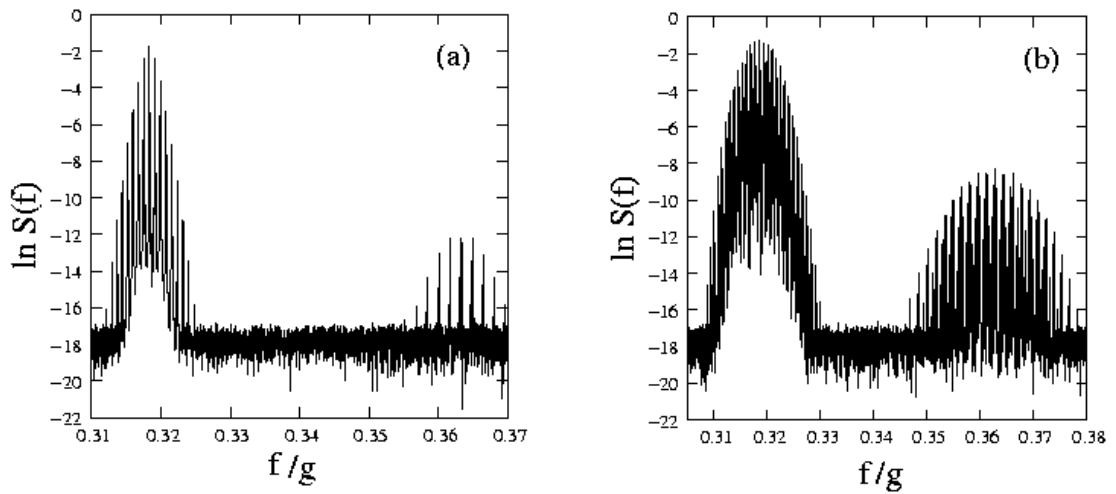}
\caption{Power spectrum of 
the mean photon number for the initial states 
(a) $\ket{\alpha\,;\,0}$ and (b) $\ket{(\alpha,5)\,;\,0}$
with $\gamma/g=10^{-2}$ and $\nu=1$.}
\label{spectrumm0qbyg0.01nu1}
\end{center}
\end{figure}

As already stated, we take the atomic oscillator in its ground
state and the field to be either in a CS or a PACS, initially. 
Again, as in Chapter 5, 
we set $\omega = \omega_0 = 1$ for illustrative purposes.  
We first consider the case of weak nonlinearity, 
i.~e., $\gamma/g \ll 1$, 
with $\nu = 1$. We recall from Chapter 5 that, if $\gamma/g \ll 1$ and 
$\nu$ is sufficiently small, then 
the state exhibits revivals (or approximate revivals) if  
the field mode is initially a CS or a PACS.   
Distinctive signatures of such revivals show up both in  
the entropy of entanglement and in the 
dynamics of appropriately chosen 
quadrature variables. At revivals (or near revivals), the 
expectation values of all observables 
return to their initial values (or to the neighborhood
of those values). In a classical phase space spanned by  
these expectation values, the trajectories may
therefore be expected to be  
periodic or ergodic (as in the case of quasi-periodicity), 
but not chaotic, in any case. 
However, while  this 
indicates that the system as a whole 
(comprising, in our case,  the field and atom modes in 
interaction with each other) is a non-chaotic 
dynamical system, there is no {\it \`a priori} reason to 
assume that this remains true for the 
field or atom sub-systems taken individually ---  
particularly because the 
sub-system variables 
do not necessarily capture signatures 
of the revival phenomena, as pointed 
out in the preceding chapter. 
Keeping these aspects in mind, we have analyzed
the time series data of the mean photon 
number ${\sf N}$, 
and examined the plots of the 
corresponding power spectrum $S(f)$ 
(which is essentially the 
Fourier transform of the autocorrelation function of the mean photon 
number, $f$ denoting the frequency), in order to 
make deductions regarding the nature of the dynamics. 
The frequency ratio $f/g$ is clearly the counterpart of 
the dimensionless time variable $gt$  
in terms of which the dynamics of the 
field mode has been investigated in Chapter 5. 
As a representative case 
of weak nonlinearity, we have chosen (as before) 
the parameter values $\gamma = 1$ and $g = 100$.    
The time step we have used in the numerical work 
related to this case is $\delta t = 10^{-2}$.  
Figure \ref{spectrumm0qbyg0.01nu1} (a), which is a 
plot of the logarithm of the power spectrum 
versus $f/g$ when the initial state is $\ket{\alpha\,;\,0}$,  
corresponds to quasi-periodic behavior. 
With increasing departure from coherence 
of the initial field mode, the number 
of frequencies seen in the power spectrum increases, 
as is evident from 
Fig. \ref{spectrumm0qbyg0.01nu1} (b), which 
 corresponds to an initial state $\ket{(\alpha,5)\,;\,0}$.  
Further, for a given initial state, 
the number of characteristic frequencies in the power 
spectrum increases with an increase in $\nu$.
\begin{figure}[htpb]
\begin{center}
\includegraphics[width=4.0in]
{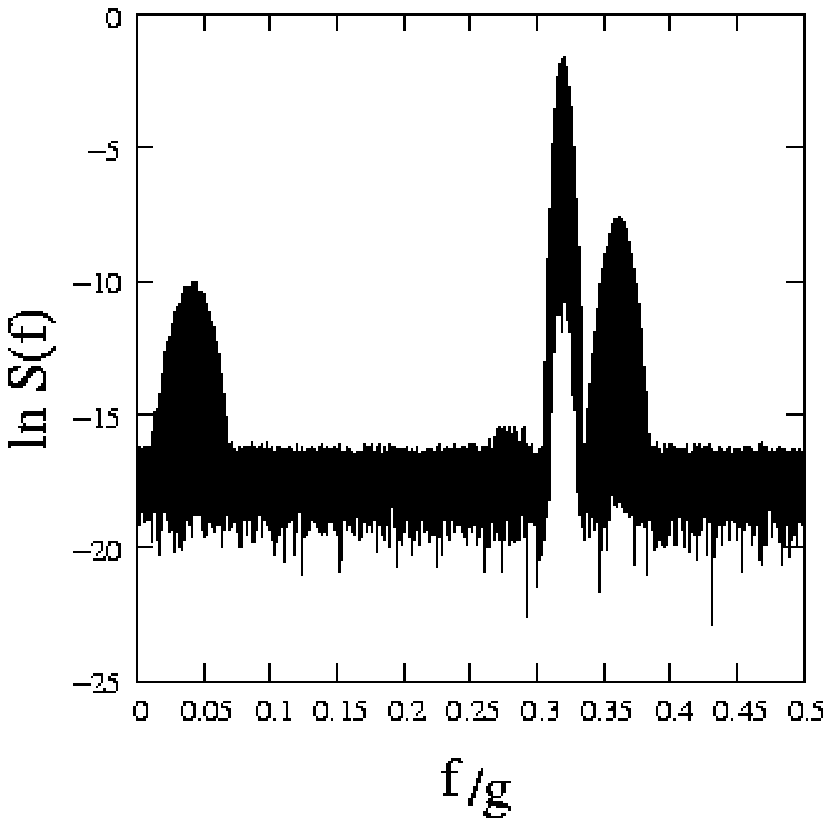}
\caption{Power spectrum of 
$\aver{\sf N}$ for an initial state 
$\ket{(\alpha,5)\,;\,0}$ with $\gamma/g=10^{-2}$ and $\nu=5$.}
\label{spectrumm5qbyg0.01nu5}
\end{center}
\end{figure}
This is evident when we compare 
Figs. \ref{spectrumm0qbyg0.01nu1} (b) and 
Fig. \ref{spectrumm5qbyg0.01nu5}, which depicts 
the power spectrum for the same situation, but with $\nu = 5$. 
The embedding dimension turns 
out to be $d_{\rm emb} = 3$ in all these cases, 
and a detailed time-series analysis 
confirms that the maximum Lyapunov exponent is zero.

In contrast to the case of weak nonlinearity, 
the nature of the dynamics of the sub-system changes 
drastically when $\gamma/g \gtrsim 1$. It    
ranges from quasi-periodicity to 
fully-developed chaos, depending on the precise 
nature of the initial state. As representative values 
for this nonlinearity-dominated regime, we  have set 
$\gamma = 5\,,\,\,g = 1$. We  
use a time step $\delta t = 10^{-1}$ in this 
instance, so that $g\,\delta t = 10^{-1}$, as this suffices to 
pick up the oscillations in the variables concerned.
 We first examine the 
case corresponding to $\nu = 1$. For an initial field 
mode which is a CS, both the time series 
and the power spectrum 
confirm that the sub-system dynamics is not chaotic. 
In contrast to this,  
an initial field state that is a PACS leads to 
chaotic behavior, for sufficiently large values of $m$. 
The time series analysis yields a value 
$d_{\rm emb} = 5$ in this case. 
\begin{figure}[htpb]
\begin{center}
\includegraphics[width=3.5in]
{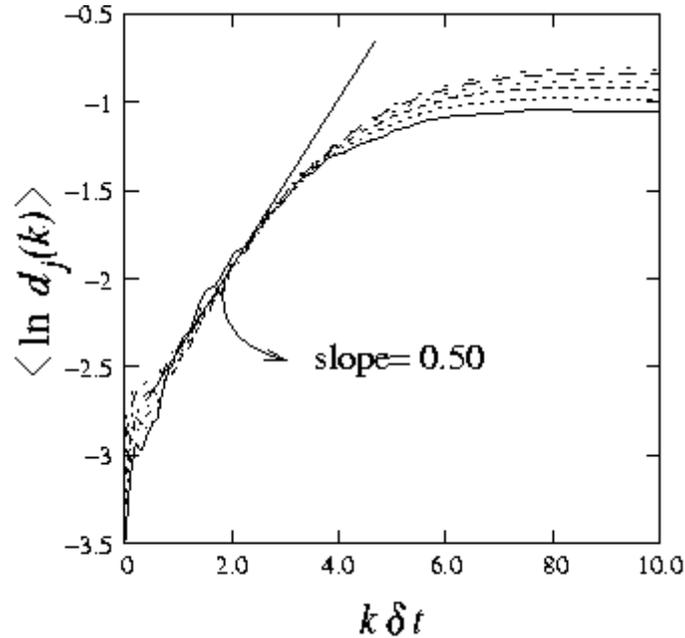}
\caption{$\aver{\ln\,d_j(k)}$ versus $t = k\,\delta t$ 
for an initial state 
$\ket{(\alpha,5)\,;\,0}$ with $\gamma/g=5$ and $\nu=1$.
The solid line corresponds to $d_{\rm emb} = 5$. 
The dotted lines correspond to values of $d_{\rm emb}$  
from $6$ to $10$.}
\label{lyapm5nu1}
\end{center}
\end{figure}
Figure \ref{lyapm5nu1} shows the variation of 
$\aver{\ln{d_j(k)}}$ with $t$ for an initial state 
$\ket{(\alpha,5)\,;\,0}$.  It is evident that the slope is positive (the 
actual value being $\approx 0.5$), 
indicating that the mean energy of the 
field mode varies chaotically in time. 

In each case, we have increased the dimensions of the phase space 
beyond the 
embedding dimension, and verified that the value of the maximal Lyapunov 
exponent is a constant independent of this increase. Further, we have 
verified in each case that, if $\aver{b^{\dagger}b}$ is chosen as 
the signal for which the time series data is computed, 
 the behavior is similar to that inferred from the dynamics of 
$\aver{a^{\dagger}a}$, in such a manner that $\aver{a^{\dagger}a} + 
\aver{b^{\dagger}b}$ is a constant, as required. This lends support to 
the conclusion that round-off or truncation errors 
may be ruled out as the source of the 
computed chaotic behavior.  

\begin{figure}[htpb]
\begin{center}
\includegraphics[width=3.5in]
{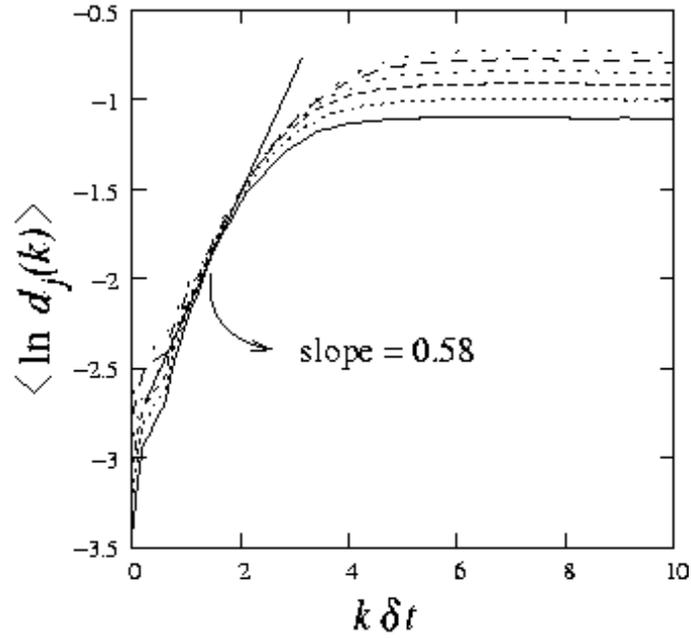}
\caption{$\aver{\ln\,d_j(k)}$ versus $t$ for an initial state 
$\ket{(\alpha,1)\,;\,0}$ with $\gamma/g=5$ and $\nu=5$.
The solid line corresponds to embedding dimension equal to $5$. 
The dotted lines correspond to values of $d_{\rm emb}$ from $6$ to $10$.}
\label{lyapm1nu5}
\end{center}
\end{figure}
\begin{figure}[htpb]
\begin{center}
\includegraphics[width=3.5in]
{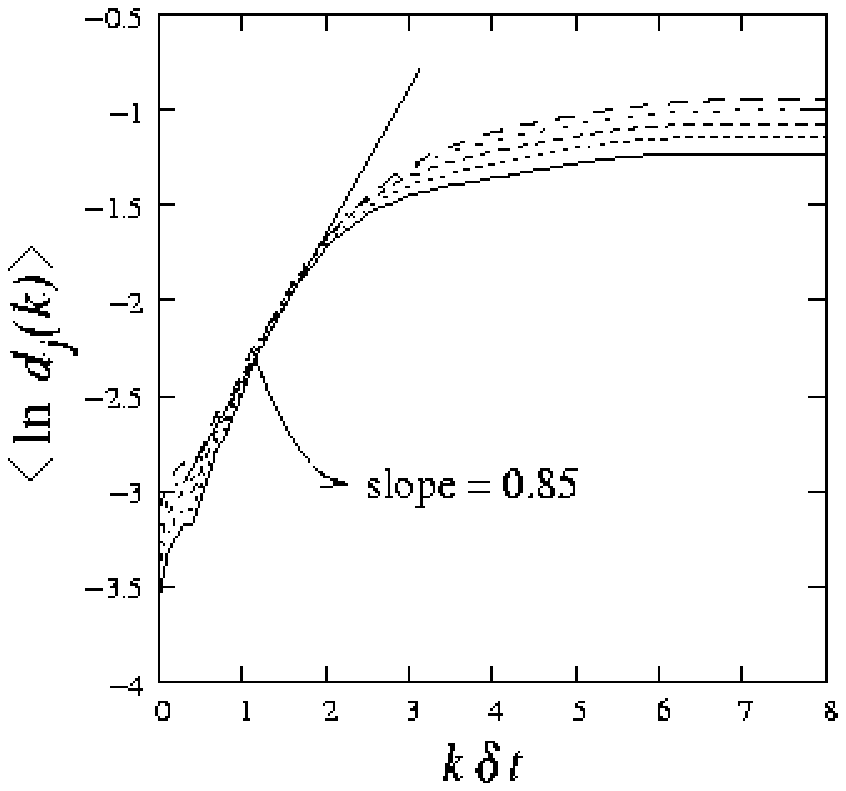}
\caption{$\aver{\ln\,d_j(k)}$ versus $t$ for an initial state 
$\ket{(\alpha,5)\,;\,0}$ with $\gamma/g=5$ and $\nu=5$.
The solid line corresponds to $d_{\rm emb } = 6$. 
The dotted lines correspond to values of $d_{\rm emb}$ from $7$ to $10$.}
\label{lyapm5nu5}
\end{center}
\end{figure}
\begin{figure}[htpb]
\begin{center}
\includegraphics[width=3.5in]
{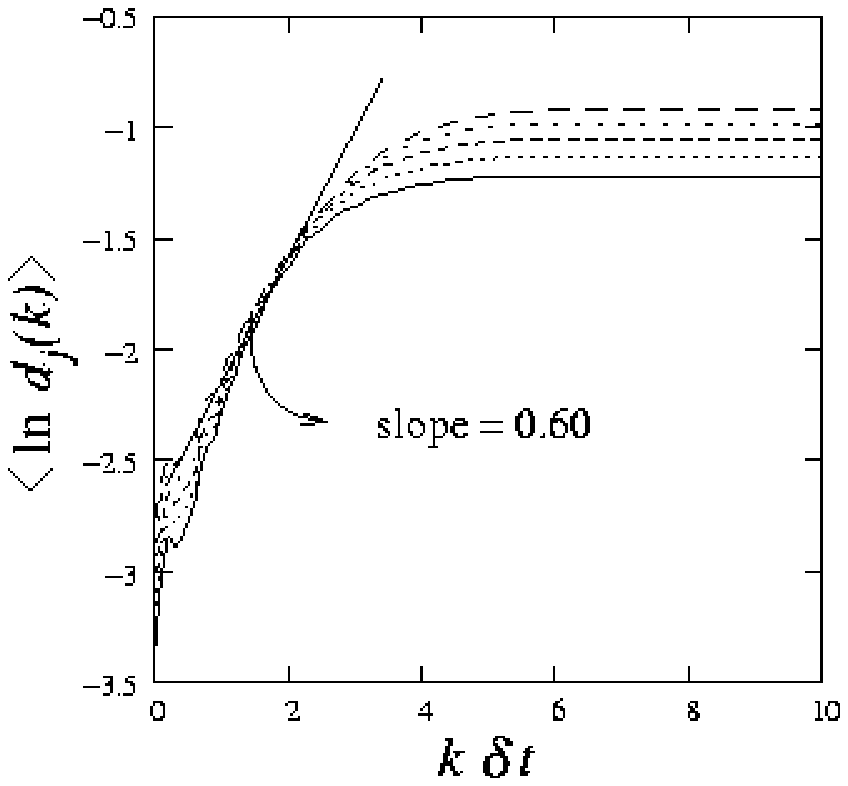}
\caption{$\aver{\ln\,d_j(k)}$ versus $t$ for an initial state 
$\ket{\alpha\,;\,0}$ with $\gamma/g=5$ and $\nu=10$.
The solid line corresponds to embedding dimension equal to $5$. 
The dotted lines correspond to values of $d_{\rm emb}$ from $6$ to $10$.}
\label{lyapm0nu10}
\end{center}
\end{figure}
For a given value of $\gamma/g$, an increase in either $\nu$ or  
$m$ in the initial PACS results in an 
increase in the value of $\lambda_{\rm max}$. This is evident 
from Figs. \ref{lyapm1nu5} and \ref{lyapm5nu5},   
where plots of $\aver{\ln{d_j(k)}}$ versus $t$ are 
shown for initial 
states $\ket{(\alpha,1)\,;\,0}$ and $\ket{(\alpha,5)\,;\,0}$ 
for $\nu = 5$. We find that 
$\lambda_{\rm max}\approx 0.58\,\, {\rm and}\,\, 0.85$, 
respectively, in these two 
cases. In contrast, there is no chaotic behavior 
for these values of the parameters 
if the initial state of the field is coherent.
However, with a further increase in $\nu$, this situation 
changes, and $\aver{\sf N}$ varies chaotically  
even for an initial coherent field mode. 
This is demonstrated in Fig. \ref{lyapm0nu10} 
showing $\aver{\ln{d_j(k)}}$ versus $t$ 
for an initial state 
$\ket{\alpha\,;\,0}$ with $\nu = 
10$. The corresponding value of $\lambda_{\rm max}$ is 
also quite large, $\approx 0.60$.      

To summarize: a time series and power spectrum analysis 
shows that, for small values of the 
ratio of the anharmonicity parameter to the strength of the inter-mode 
coupling, 
the dynamics of the expectation value of the mean photon number (or
mean energy) of the field mode ranges from periodic to ergodic, but
not chaotic, essentially 
independent of the nature of the initial state. 
On the other hand, for sufficiently large 
values of this ratio, the 
ergodicity properties of the sub-system representing the field mode 
depend significantly on the extent of 
departure from coherence of the initial field mode.  
In particular, $\lambda_{\rm max}$ is larger in the case of an 
initial field state that is a 
photon-added coherent state, compared to its value  
for initial coherent state. 
Table 6.1 provides a bird's eye view of these conclusions. 
\begin{figure}[htpb] 
\begin{center} \includegraphics[width=5.8in] 
{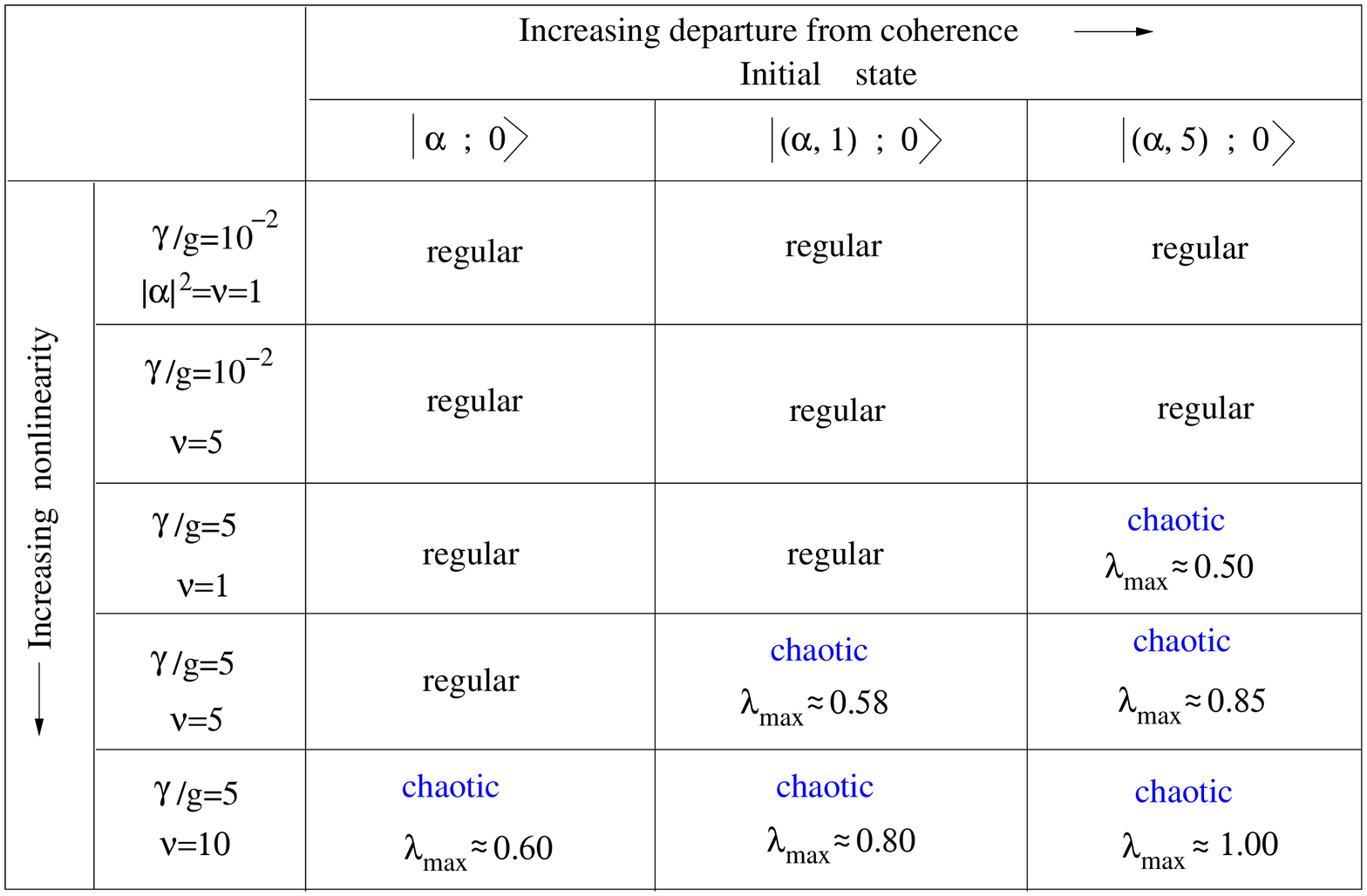}
\end{center}
%\begin{center}
Table 6.1: Qualitative dynamical behavior of the 
mean photon number of a 
single-mode electromagnetic 
field interacting with a nonlinear medium. ``Regular'' 
$\Rightarrow \lambda_{\rm max} = 0$. 
%\end{center}
\end{figure}

We have already shown that the classical counterpart of the 
Hamiltonian under study (Eq. (\ref{classicalcoupledhamil1})) 
corresponds to bounded motion 
in phase space, and that the system is integrable, 
with all its Lyapunov exponents equal to zero.
We have also seen that the quantum Hamiltonian 
(Eq.(\ref{coupledhamil2})) commutes with the total number operator
${\sf N}_{\rm tot}\,$, and has a discrete energy
spectrum for all values of the parameters in $H$. What is the 
interpretation, then, of a positive value for 
$\lambda_{\rm max}$ as deduced from the time series for 
the sub-system variable $\aver{a^{\dagger}a}$ 
(equivalently, $\aver{b^{\dagger} b}$)? 

This is best understood by appreciating the role of recurrence 
theorems and ergodicity in both classical and quantum dynamics. 
In classical physics, ergodicity itself is intimately 
linked with the Poincar\'e 
recurrence theorem. This theorem essentially 
states that any phase space configuration of a 
system enclosed in a finite 
volume will be repeated to an arbitrary degree 
of accuracy after a 
finite, though 
perhaps very long, time. Quantum analogs of this 
classical recurrence 
theorem have been established and applied 
to diverse situations. Exact revivals or 
near-revivals of a quantum state 
occur \cite{bocch} in systems with discrete 
energy eigenvalues. Further, 
it has been argued \cite{hogg} that under 
any time-periodic Hamiltonian, a 
non-resonant, bounded quantum system 
will return to its original form 
several times during its dynamics, 
and this has been illustrated using 
computer experiments. Further, the link 
between wave packet revivals, 
and geometric phases in quantum physics and classical
recurrences on the circle map has been 
examined \cite{sesh1}. Inspired by
the seminal work of Fermi, Pasta and Ulam \cite{fermi} 
on the recurrences of the initial 
state of a chain of nonlinearly coupled 
oscillators, several investigations have 
been carried out over the years 
on a wide 
variety of physical systems, 
in order to understand the conditions under which 
recurrences occur. In particular, 
it has been argued \cite{ermos} that 
wave packet 
recurrences (exact revivals and near-revivals) 
and the Fermi-Pasta-Ulam 
recurrences have the same origin: 
if the initial quantum 
state can be represented as a superposition of eigenstates, and the 
spectrum is non-equispaced, recurrences are bound to occur.

On the other hand, high-resolution 
tracking of quantum and classical 
evolutions of a system will reveal differences in 
the dynamics of the expectation values of 
corresponding observables in these two cases. 
As indicated in Chapter 1, the 
origin of this 
dichotomy can be traced back to the inadequacy 
of the naive Ehrenfest 
theorem (which does not generally take into account 
the non-commutativity 
between 
$x$ and $p$, as seen from Eq.(\ref{ehrenfestxp})) in 
retrieving the classical regime of the quantum system. An 
outcome of this feature is that {\it the Lyapunov exponents  that 
characterize the dynamical behavior of classical and quantum 
expectation values of the same observable can indeed be very different 
from each other} \cite{balle}. Two interesting observations 
\cite{habib} have been 
made in this context: (a) In isolated quantum systems with a discrete 
energy spectrum, using unitarity and the Schwarz 
inequality, it can be established that the Lyapunov exponents would 
vanish, if computed from time-series data collected over a 
sufficiently long 
time (which, in 
practice, could even turn out to be considerably longer 
than the characteristic time 
scales in the 
problem), indicative of  
non-chaotic behavior. (b) However, once 
measurement upon the system 
is included through appropriate interaction 
with an external system, the 
corresponding Lyapunov exponents need not vanish.
This would suggest that, in principle, the dynamics of a {\it
  sub}-system, as deduced from time-series data 
of certain variables, may show exponential 
instability, even if the 
system as a whole does not. The presence of hyperbolic 
(sets of) critical points in the corresponding 
classical version of the dynamical system 
plays a significant role in this regard.  
The ``chaotic'' 
behavior displayed by the mean photon number 
$\aver{a^{\dagger} a}$ in the system we have considered 
may be understood in the light of these observations. 
Ultimately, the exponential instability associated with a positive 
Lyapunov exponent is an indication of the manner in which 
an initial wave packet spreads as it evolves in time.

%% file: conclude.tex
\chapter{Conclusion}
\label{conclusion}

In the preceding chapters, we have described our results in detail,
and also summarized them at appropriate places throughout the text. It
remains to place the work in a broader perspective, 
and to list interesting open problems and avenues for
future work. 

In general terms, the present work may be viewed as addressing 
an aspect of the broad theme of understanding the manner 
in which the notions of ergodicity and varying degrees of randomness,
that are well-defined and notable features of classical dynamics, get
modified in quantum dynamics. In specific terms, we have studied
several aspects of wave packet dynamics that shed light on
non-classical effects such as revivals and fractional revivals. As
emphasized in the preceding chapters, such effects are striking
manifestations of quantum mechanical interference phenomena. They show
up in distinctive fashion in the expectation values of observables and
their higher moments, in different kinds of squeezing, and in other
indicators such as entropies and quasi-probability distributions, as
elaborated upon in the foregoing. We have worked in the
framework of two specific, rather simple, model Hamiltonians pertinent
to a nonlinear medium. However, these models capture many of the
essential features of nonlinearity, anharmonicity and inter-mode
coupling. In this sense, the results arrived at 
enable us to draw conclusions that may be expected to remain valid in
their broad features, if not in detail, in other modified situations
as well. 

Our results also suggest a number of avenues for
further exploration. 
We list some of these here, roughly in the order of 
progression of the text of this
thesis.  

We have examined the revival phenomena exhibited by a single-mode
coherent state or a photon-added coherent state as it propagates
through a Kerr-like medium. It would be of interest to investigate  
their  counterparts for other initial
states, such as photon-{\it subtracted} coherent states \cite{dodo1} 
and various squeezed states 
of light.  

The Wigner function is only one particular way of arriving at a
non-classicality indicator in the situations considered in this
thesis. Others, such as the Sudarshan-Glauber 
$P$-representation \cite{mandel} and the Husimi-Kano function
\cite{wolfgang}, are also worth
exploring, as is the extension of these considerations 
to the interacting two-mode case. 

The interacting two-mode case we have considered involves a bipartite
Hamiltonian. What would happen if this were not the case --- for
instance, if we considered three sub-systems coupled to each other?
Further, if such a system were classically non-integrable, what sort
of dynamical behavior would the quantum mechanical counterpart exhibit?  

The sub-system entropies we have considered 
(the von Neumann entropy and the linear entropy) are just two of several
possible measures of entanglement, and they seem to exhibit 
essentially similar behavior, at least in the cases we have
examined. What about other measures of entanglement, and what is the 
most appropriate or optimal measure for the purposes at hand? 

Finally, as perhaps the most interesting general question that 
arises in the present context: what are the possibilities and
implications for quantum computation using continuous variables
(e.~g., Gaussian wave packets), exploiting revival phenomena? 
It is evident that there exists a rich vein of interesting problems for
further research in this area.

%% file: appendixA.tex
\chapter{Calculation of $\aver{a^{\dagger r} a^{r+s}}$ 
for an initial PACS $\ket{\alpha,m}$ propagating in a Kerr-like medium} 
The Hamiltonian
governing the dynamics of the wave packet $\ket{\psi}$ is given by 
$H=\chi\, {\sf N}({\sf N}-1)$ 
where ${\sf N}=a^{\dagger} a$. 
\label{generalform}
The initial state $\ket{\psi(0)}$ is the PACS $\ket{\alpha,m}$.
The state at any $t\geq0$ is given by 
\begin{equation}
\ket{\psi(t)}=e^{-i\chi {\sf N}({\sf N}-1)t}\ket{\alpha,m}.
\end{equation}
Introducing complete sets of Fock states 
$\sum_{l=0}^{\infty}\ket{l}\bra{l}\,(=1)$ at appropriate places, we have 
\begin{eqnarray}
&\expect{\psi(t)}{a^{\dagger r} a^{r+s}}{\psi(t)}&
=\sum_{l=0}^{\infty}\sum_{n=0}^{\infty}\sum_{u=0}^{\infty}
\sum_{v=0}^{\infty}\inner{\alpha,m}{l}\,
\bra{l}e^{i\chi {\sf N}({\sf N}-1)t}\ket{n}\nonumber\\
&&\times\bra{n}a^{\dagger r} a^{r+s}\ket{u}\,
\bra{u}e^{-i\chi {\sf N}({\sf N}-1)t}
\ket{v}\,\inner{v}{\alpha,m}.
\label{appendixeq2}
\end{eqnarray}
Since our initial state is the PACS $\ket{\alpha,m}$, 
it is convenient to use the notation
\begin{equation}
\aver{a^{\dagger r}a^{r+s}}_m=
\expect{\psi(t)}{a^{\dagger r}a^{r+s}}{\psi(t)}.
\end{equation}
When $m=0$ (corresponding to an initial CS), we will write the 
expectation value as simply $\aver{a^{\dagger r}\,a^{r+s}}$. On using
\begin{equation}
a^p\ket{u}=\Big[\frac{u!}{(u-p)!}\Big]^{1/2}\,\ket{u-p}\nonumber
\end{equation}
and the orthogonality property of the basis states,
Eq. (\ref{appendixeq2}) simplifies to 
\begin{eqnarray}
\aver{a^{\dagger r} a^{r+s}}_m
&=&
\sum_{l=0}^{\infty}
\Big[\frac{(l+s)!}{l!}\Big]^{1/2}\frac{l!}{(l-r)!}
\,e^{-i\chi [(l+s)(l+s-1)-l(l-1)]t}\nonumber\\
&\times&\inner{\alpha,m}{l}\inner{l+s}{\alpha,m}.
\label{generalm1}
\end{eqnarray}
Here, and in the rest of this Appendix, it is understood that contributions 
from terms of the form $1/(-n)!$, where $n$ is a positive integer, vanish. 
(Note that the gamma function 
$\Gamma(1-n)$ diverges for $n=1,2,\cdots.$)
Expanding $\ket{\alpha,m}$ in the Fock basis, we find 
\begin{equation}
\inner{\alpha,m}{l}=\frac{e^{-\nu/2}\,(\alpha^*)^{l-m}}{\sqrt{m!L_m(-\nu)}}
\frac{\sqrt{l!}}{(l-m)!}\,,
\end{equation}
where $\nu = |\alpha|^2$ and $L_m$ is the Laguerre polynomial of order $m$.
We use the definitions \cite{gradshteyn}
\begin{equation}
L_m(\mu)=\sum_{n=0}^{m}\binom{m}{n}\,\frac{(-\mu)^n}{n!}
=\frac{e^{\mu}}{m!}\frac{d^m}{d\mu^m}(\mu^me^{-\mu})
\label{laguerre}
\end{equation}
and 
\begin{equation}
L_m^s(\mu)=\sum_{n=0}^{m}\binom{m+s}{n+s}\frac{(-\mu)^n}{n!}
=\frac{e^{\mu}\mu^{-s}}{m!}\frac{d^m}
{d\mu^m}(\mu^{m+s}e^{-\mu})
\label{assoclaguerre}
\end{equation}
for the Laguerre and associated Laguerre polynomials. Equation 
(\ref{generalm1}) reduces to  
\begin{eqnarray}
\aver{a^{\dagger r} a^{r+s}}_m=
\frac{e^{-\nu}}{m!L_m(-\nu)}\sum_{l=0}^{\infty}
\frac{l!}{(l-r)!}\frac{(l+s)!}{(l+s-m)!}\,
\frac{\alpha^s\,\nu^{l-m}}{(l-m)!}\,
e^{-i\chi s(s-1)\,t}\,e^{-2i\chi ls\,t}.\nonumber\\
\end{eqnarray}
But
\begin{equation}
\frac{(l+s)!}{(l+s-m)!}\,\,\nu^{l-m}=\nu^{-s}\frac{d^m}{d\nu^m}\,
(\nu^{l+s}),
\end{equation}
so that
\begin{eqnarray}
\aver{a^{\dagger r} a^{r+s}}_m=\frac{\alpha^s \,
e^{-\nu}e^{-i\chi\,s(s-1)t}}
{\nu^ s\, m!L_m(-\nu)}
\sum_{l=0}^{\infty}\frac{e^{-2i\chi \,lst}}{(l-m)!}\,\frac{d^m}{d\nu^m}
\,\Big(\nu^{r+s}\,\frac{d^{\,r}\nu^l}{d\nu^r}\Big).
\label{eqn1}
\end{eqnarray}
This can be re-expressed as
\begin{equation}
\aver{a^{\dagger r}a^{r+s}}_m=\frac{\alpha^s \,e^{-\nu}\,
e^{-i\chi\,s(s-1)t}}
{\nu^ s \,m!L_m(-\nu)}
\sum_{n=0}^{m}\binom{m}{n}\Big(\frac{d^n\nu^{r+s}}{d\nu^n}\Big)
\frac{d^{m-n+r}}{d\nu^{m-n+r}}
\sum_{l=0}^{\infty}\frac{e^{-2i\chi \,lst}\,\nu^l}{(l-m)!}\,.
\end{equation}
Simplifying, we find (after some straightforward algebra) 
\begin{eqnarray}
\aver{a^{\dagger r}a^{r+s}}_m=\frac{\alpha^s e^{-\nu}e^{-i\chi(s-1+2m)st}}
{\nu^ s L_m(-\nu)m!}
\frac{d^m}{d\nu^m}\Big(\nu^{r+s}\frac{d^r}{d\nu^r}\,[\nu^m\,
\exp\,(\nu \,e^{-2is\chi t})]\Big).
\end{eqnarray}
This can be re-cast in the form 
\begin{eqnarray}
\aver{a^{\dagger r}a^{r+s}}_m&=&\frac{\alpha^s \,
e^{-\nu}e^{-i\chi(s-1+2m)st}}
{\nu^ s \,L_m(-\nu)}
\sum_{n=n_{\rm min}}^{r}\binom{r}{n}\frac{e^{-2is\chi nt}}{(m-r+n)!}
\nonumber\\
&\times&\frac{d^m}{d\nu^m}\,[\nu^{m+s+n}
\,\exp\,(\nu e^{-2is\chi t})],
\end{eqnarray}
where $n_{\rm min}= \max\,(0,r-m)$. 
Using the definition in Eq. (\ref{assoclaguerre})
for the associated Laguerre polynomial $L_m^s$
we get, finally, 
\begin{eqnarray}
\aver{a^{\dagger r}\,a^{r+s}}_m
&=&\alpha^s \,\exp\,\big[-\nu-i\chi( s-1+2m)st
+\nu \,e^{-2is\chi t}\big]\nonumber\\
&\times&
\sum_{n=n_{min}}^{r}\,\binom{r}{n}\,
\frac{m!\,(\nu \,e^{-2is\chi t})^n}{(m-r+n)!}
\,\frac{L_m^{s+n}(-\nu e^{-2is\chi t})}{L_m(-\nu)}\,.
\label{generalmappendix} 
\end{eqnarray}
This form of the expression for $\aver{a^{\dagger r}\,a^{r+s}}_m$ at 
an arbitrary time 
$t$, corresponding to the initial state $\ket{\alpha,m}$, is used for 
numerical  calculations.

Setting $m=0$ in Eq. (\ref{generalmappendix}), we obtain
\begin{equation}
\aver{a^{\dagger r}\,a^{r+s}} = \alpha^{s}\,\nu^{r}
e^{-\nu\,(1-\cos \,2 s \chi t)}\,
\exp \left[
-i \chi \big(s(s-1) + 2rs\big)\,t - i\nu \,\sin\,2s \chi t\right]
\label{nthmomentappendix}
\end{equation}
as the corresponding expression for an initial coherent state 
$\ket{\alpha}$.

%% file: appendixB.tex
\chapter{Analytic expressions for $\aver{x(t)}$ and $\aver{p(t)}$
for an initial PACS}

Setting $r=0$ and $s=1$ in Eq. (\ref{generalmappendix}), we get 
\begin{equation}
\aver{a}_m=\frac{\alpha \,e^{-\nu}}{L_m(-\nu)}\,
e^{-2im\chi t}\,
\exp\,\big(\nu e^{-2 i\chi t}\big)\, L^1_m(-\nu \,e^{-2 i\chi t}).
\label{expectat}
\end{equation}
The expectation values $\aver{x(t)}_m$ and $\aver{p(t)}_m$  
(i.e.,  the mean values of $x$ and $p$ at time $t$ in the case of the 
initial PACS $\ket{\alpha,m}$) 
can now be calculated using Eq. 
(\ref{expectat}) and its complex conjugate, which yields 
$\aver{a^{\dagger}}_m$.

Recall that 
\begin{equation}
\aver{x(t)}_m=\sqrt{2}\,\, {\rm Re}\,\aver{a}_m\qquad {\rm and} \qquad
\aver{p(t)}_m=\sqrt{2}\,\, {\rm Im}\,\aver{a}_m,
\label{eqn3}
\end{equation}
where we emphasize that $\aver{a}_m$ 
is computed at time $t$, and the suffix $m$ is meant to denote 
 the fact that the 
initial state considered is $\ket{\alpha,m}$. 
We find
\begin{eqnarray}
\aver{x(t)}_m&=&\frac{\sqrt{2}\,
e^{-\nu(1-\cos 2\chi t)}}{L_m(-\nu)}
\,{\rm Re}\,\Big\{\alpha \,\exp\,[-i(2m\chi t+\nu \sin 2\chi t)] \,
L^1_m(-\nu \,e^{-2 i\chi t})
\Big\}\nonumber\\
&&\label{xtmappendix}\\
{\rm and}\quad&&\nonumber\\
\aver{p(t)}_m&=&\frac{\sqrt{2}\,e^{-\nu(1-\cos 2\chi t)}}{L_m(-\nu)}
\,{\rm Im}\,\Big\{\alpha \,\exp\,[-i(2m\chi t+\nu \sin 2\chi t)]\,
 L^1_m(-\nu \,e^{-2 i\chi t})
\Big\}.\nonumber\\
&&
\label{ptmappendix}
\end{eqnarray}
Setting $m=0$ in the above equations, we obtain the corresponding 
expressions for the case of an initial CS, as written down in 
Eqs. (\ref{xt}) and (\ref{pt}).

Returning to the case of an initial PACS $\ket{\alpha,m}$, we see from 
Eqs. (\ref{xtmappendix}) and (\ref{ptmappendix}) 
that the initial values of the expectation values concerned are given by
\begin{equation}
\aver{x(0)}_{m} =
\frac{L_{m}^{1}(-\nu)}{L_{m}(-\nu)}\,x_{0}\,,\,\,
\aver{p(0)}_{m} =
\frac{L_{m}^{1}(-\nu)}{L_{m}(-\nu)}\,p_{0}\,,
\label{xpinitialappendix}
\end{equation}
where $x_0+ip_0=\alpha\sqrt{2}$.
The difference 
\begin{equation}
\aver{x(t)}_m-i\aver{p(t)}_m=\sqrt{2}\aver{a^\dagger}_m
\end{equation}
deserves special mention. 
We find
\begin{eqnarray}
\aver{x(t)}_m-i\aver{p(t)}_m=\sqrt{2}\frac{\alpha^* e^{-\nu}}{L_m(-\nu)}
e^{2im\chi t}\,\exp\,\big(\nu e^{2 i\chi t}\big)  
\,L^1_m(-\nu e^{2 i\chi t}).
\end{eqnarray}
This can be rewritten in the compact form
\begin{equation}
\aver{x(t)}_m -i \aver{p(t)}_m = (x_{0} -
ip_{0})\,\frac{e^{-\nu(1-\zeta)}\,\zeta^{m}
\,L_{m}^{1}(-\nu\,\zeta)}{L_{m}(-\nu)},
\label{xmpmsoln}
\end{equation}
where  $\zeta(t)=\exp\,(2i\chi t)$.

It is convenient to define the quantities
\begin{eqnarray}
X_{m}(t)=\aver{x(t)} _{m}\,
\exp\,[\nu\,(1-\cos 2\chi t)]
\,\, {\rm and}\,\,  
P_{m}(t)=\aver{p(t)} _{m}\,\exp\,[\nu\,(1-\cos 2\chi t)].\nonumber\\
\end{eqnarray}
Using the expressions for $\aver{x(t)} _{m}$ and $\aver{p(t)} _{m}$ 
in Eqs. (\ref{xtmappendix}) and (\ref{ptmappendix}),
we get
\begin{eqnarray}
X_m(t)&=&\frac{1}{L_m(-\nu)}
\,{\rm Re}\,\Big\{(x_0+ip_0) \exp\,[-i(2m\chi t+\nu \sin 2\chi t)]\,
 L^1_m(-\nu e^{-2 i\chi t})
\Big\},\nonumber\\
P_m(t)&=&\frac{1}{L_m(-\nu)}
\,{\rm Im}\,\Big\{(x_0+ip_0) \exp\,[-i(2m\chi t+\nu \sin 2\chi t)]\,
 L^1_m(-\nu e^{-2 i\chi t})
\Big\}.\quad\qquad
\end{eqnarray}
These can be written after some straightforward algebra in the concise 
form
\begin{eqnarray}
\left.\begin{array}{ll}
X_{m}(t)= x_{0}\,{\rm Re}\,z_{m}(t)+p_{0}\,{\rm Im}\,z_{m}(t),\\
P_{m}(t)= p_{0}\,{\rm Re}\,z_{m}(t)-x_{0}\,{\rm Im}\,z_{m}(t),
\end{array}\right\}
\label{mxtptappendix}
\end{eqnarray}
where
\begin{equation}
z_{m}(t)=
\frac{L_{m}^{1}(-\nu\,e^{2i\chi t})}{L_{m}(-\nu)}
\,\exp\,[i\,(2 m \chi t+\nu\,\sin\,2\chi t)]\,.
\label{zmappendix}
\end{equation}
For ready reference, we have written out Eqs. (\ref{mxtptappendix}) and 
(\ref{zmappendix}) once again in Chapter 3 (Eqs. (\ref{mxtpt}) and (\ref{zm}),
respectively). In deriving the expressions above, we have made frequent 
use 
of the fact that  $e^z$ and $L_m^s(z)$ are real analytic functions 
of $z\,$: 
that is, they satisfy $f(z^*)=f^*(z)$.

%% file: appendixC.tex
\chapter{Calculation of $D_q^{(m)}(t)$ 
for an initial PACS $\ket{\alpha,m}$}
\label{Dqmoft}

Setting $r=0$ and $s=q$ (a non-negative integer)
in Eq. (\ref{generalmappendix}) of Appendix A and using 
the series 
expansion for the associated 
Laguerre polynomial, we get
\begin{eqnarray}
\aver{a^q}_m&=&\frac{e^{-\nu(1-\cos 2q\chi t)}}
{L_m(-\nu)}\,\nu^q\,\sum_{n=0}^{m}
\frac{(m+q)!\,\nu^n}{(m-n)!\,(n+q)!\,n!}\nonumber\\
&\times&\exp \,
\big\{\negthickspace-\!i\,[(q-1+2m+2n)\,\chi qt+
\nu\sin 2\chi qt-q\theta]\big\},
\end{eqnarray}
where $\theta$ is the argument of $\alpha$, that is, 
$\alpha=\nu^{1/2}\,e^{i\theta}$.
Similarly, when $r=q$ and $s=0$, Eq. (\ref{generalmappendix}) gives 
\begin{eqnarray}
\aver{a^{\dagger q}\,a^q}_m=\frac{1}{L_m(-\nu)}\sum_{n=n_{\rm min}}^{q}
\binom{q}{n}
\frac{m!}{(m-q+n)!}\,\nu^{n}\,L_m^{n}(-\nu),
\end{eqnarray}
where $n_{\rm min}$ is now given by $\max\,(0,q-m)$.
The quantity $D_q^{(m)}(t)$ 
for an initial state $\ket{\alpha,m}$ can now be calculated using the 
definition (see Chapter 4, Eq. (\ref{Dqtwo}))
\begin{equation}
D_q^{(m)}(t) 
=\frac{2\,\left[{\rm Re}\,\aver{a^{2q}}_m 
-2 \,\big({\rm Re} \,\aver{a^q}_m \big)^2+ 
\aver{a^{\dagger q} \,a^{q}}_m\right]}
{\aver{F_q({\sf N})}_m}\,.
\label{appendixc3}
\end{equation}
Here $F_q({\sf N})=[a^q,\,a^{\dagger q}]$. After some algebra, we find that
the operator $F_q({\sf N})$ is 
a polynomial of order $(q-1)$ in the number operator
${\sf N}=a^{\dagger} a$, given by
\begin{eqnarray}
F_q({\sf N})
=q!\,\bigg[1+\sum_{n=1}^{q-1}\binom{q}{n}
\frac{1}{n!}\,\Big\{{\sf N}({\sf N}-1)\cdots
\big({\sf N}-(n-1)\big)\Big\}\bigg].
\label{fqnappendix}
\end{eqnarray}
Since ${\sf N}$ commutes with the Hamiltonian $H = \chi \,{\sf N}
({\sf N} - 1)$, it follows that $\aver{F_q({\sf N})}$ is also a 
constant of the motion, and its value remains the same as its initial
value computed in the state $\ket{\alpha, m}$.

Substituting the values for ${\rm Re}\, \aver{a^q}_m$, 
${\rm Re}\, \aver{a^{2q}}_m$ 
and $\aver{a^{\dagger q}\,a^q}_m$ in  Eq. (\ref{appendixc3}), we get   
\begin{eqnarray}
\frac{1}{2}L_m(-\nu)\aver{F_q({\sf N})}_mD_q^{(m)}(t) &=&
e^{-\nu(1-\cos4\chi qt)}
\sum_{n=0}^{m}\binom{m+2q}{n+2q}
\,\frac{\nu^{n+q}}{n!} \nonumber\\
&\times& \cos\,\Big[2(2m+2n + 2q -1)\chi qt+\nu\sin4\chi qt
-2q\theta\Big]\nonumber\\
&-& \frac{2e^{-2\nu(1-\cos2\chi qt)}}{L_m(-\nu)}
\biggl\{\sum_{n=0}^{m}\binom{m+q}{n+q}\,\frac{\nu^{n+q/2}}{n!}\nonumber\\
&\times&\cos\,\Big[(q-1+2m+2n)\chi qt+\nu\sin\,2\chi qt-q\theta\Big]
\biggr\}^2\nonumber\\
&+&\sum_{n=n_{\rm min}}^{q}\binom{q}{n}\frac{m!}{(m-q+n)!}\,\,
\nu^n\,L_m^n(-\nu),
\end{eqnarray}
where $n_{\rm min}={\rm Max}\,(0\,,\,q-m)$.
When $m=0$
(corresponding to an initial CS), $D_q^{(0)}$ simplifies to 
\begin{eqnarray}
D_q^{(0)}(t)&=&\frac{2\nu^q}{\aver{F_q({\sf N})}}
\bigg\{1+e^{-2\nu\sin^{2}2q\chi t}
\cos\,\big(2\chi q(2q-1) t+\nu\sin 4\chi qt-2q\theta\big)\nonumber\\
&-&2e^{-4\nu\sin^2q\chi t} \cos^{2}\,\big(\chi q(q-1) t+
\nu\sin 2\chi qt-q\theta\big)\bigg\}.\nonumber\\
\label{Dqtcsappendix}
\end{eqnarray}
As mentioned in the text, $D_q^{(0)}(t)$ has been written as simply 
$D_q(t)$ in Sec. 4.2, in the case of an initial CS.

%% file: appendixD.tex
\chapter{Density matrix for a field 
mode interacting with a nonlinear medium}  
The Hamiltonian which describes the interaction of a single-mode field 
interacting with the nonlinear medium is given by \cite{agar1}
\begin{equation}
H =  \omega \,a^{\dagger} a + \omega_0 \,b^\dagger b +  \gamma\, 
b^{\dagger 2} b^2 +  g \,(a^\dagger b + b^\dagger a).
\label{couplehamilappendix}
\end{equation}
The total number operator ${\sf N}_{\rm tot} 
= (a^{\dagger}a + b^{\dagger}b)$ 
commutes with $H$. We use $N$ to denote its 
eigenvalues. 
The eigenstates $\ket{\psi_{Ns}}$ of $H$ are then   
labeled by integers $N$ and $s$,
where $N = 0,\,1,\,\ldots \,{\it ad \ inf.}$ and 
$s = 0,\,1,\,\ldots ,N$  
for a given  value of $N$, as in Chapter 5, Sec. 5.2.
These can be expanded in terms of the basis states 
$\ket{N-n}_a\otimes\ket{n}_b\equiv
\ket{N-n;n}$ as
\begin{equation}
\ket{\psi_{Ns}} = \sum_{n = 0}^{N}\, d_n^{Ns}\,\ket{N-n\,;\, n},
\label{psins}
\end{equation}
with  expansion coefficients
\begin{equation}
d_n^{Ns} = \langle{N-n\,;\, n}\,|\,\psi_{Ns}\rangle.
\label{dnns}
\end{equation}
The Hamiltonian in Eq. (\ref{couplehamilappendix})
 can be represented in the basis $\ket{N-n;\,n}$
as a real, symmetric, tridiagonal matrix. The  eigenvalues 
$\lambda_{Ns}$ and eigenstates $\ket{\psi_{Ns}}$ of $H$ 
can be found numerically using appropriate matrix algebra routines 
\cite{press,gnu}. 

The unitary time evolution operator 
is given by
\begin{equation}
U(t)=\sum_{N=0}^{\infty}\sum_{s=0}^{N}\,\exp\,(-i\lambda_{Ns}t)\ket{\psi_{Ns}}
\bra{\psi_{Ns}}. 
\end{equation}
Hence, an initial state $\ket{\psi(0)}$
evolves in time  to the state
\begin{equation}
\ket{\psi(t)}=U(t) \ket{\psi(0)}=  \sum_{N=0}^{\infty}\sum_{s=0}^{N}
\,\exp\,(-i\lambda_{Ns}t)\inner{\psi_{Ns}}{\psi(0)}\ket{\psi_{Ns}}.
\end{equation}
The corresponding time-dependent density matrix $\rho(t)$ is 
\begin{equation}
\rho(t)=\sum_{N=0}^{\infty}\sum_{s=0}^{N}\sum_{N'=0}^{\infty}\sum_{s'=0}^{N'}
\,\exp\,[-i(\lambda_{Ns}-\lambda_{N's'})\,t\,]
\inner{\psi_{Ns}}{\psi(0)}\inner{\psi(0)}{\psi_{N's'}}\ket{\psi_{Ns}}\bra{\psi_{N's'}}.
\label{rhotgeneral}
\end{equation}

If the atomic oscillator is initially in the ground state 
$\ket{0}_b$ 
and the field is in the Fock 
state $\ket{N}_a$, we have
$\ket{\psi(0)}=\ket{N; 0}$, and  
\begin{eqnarray}
\rho(t) = \sum_{s = 0}^{N}\,\sum_{s' = 
0}^{N}\,\exp{[-i(\lambda_{Ns} - \lambda_{Ns'})t]} 
\,\langle{\psi_{Ns}}|N\,;\, 0 \rangle
 \,\langle{N\,;\, 0}|{\psi_{Ns'}}\rangle\,
\ket{\psi_{Ns}}\bra{\psi_{Ns'}}.
\label{}
\end{eqnarray}
(This expression is explicitly $N$-dependent, as expected.) 
Substituting for $d_n^{Ns}$ from Eq. (\ref{dnns}), $\rho(t)$  
can be re-expressed in the form
\begin{eqnarray}
\rho(t) &=& \sum_{s = 0}^{N}\,\sum_{s' = 
0}^{N}\,\exp\,[-i(\lambda_{Ns} - \lambda_{Ns'})t]
\,\,d_0^{Ns}\,\,d_0^{Ns'}\,
\ket{\psi_{Ns}}\bra{\psi_{Ns'}}.
\label{rhotN}
\end{eqnarray} 
Hence
\begin{eqnarray}
\expect{\psi_{Ml}}{\rho(t)}{\psi_{M'l'}}=
\,\exp\,[-i(\lambda_{Ml}-\lambda_{M'l'})\,t\,]\,\,
d_0^{Ml}\,\,d_0^{M'l'}\,\delta_{MN}\delta_{NM'}.
\end{eqnarray} 
It is evident that $\rho(t)$ is an 
$(N+1)$-dimensional diagonal matrix in this case.

For an initial state $\ket{(\alpha,m); 0}$ we find, using
Eq. (\ref{rhotgeneral}),
\begin{eqnarray}
\rho(t)&=&\frac{e^{-\nu}}{m!L_m(-\nu)}\sum_{N=m}^{\infty}\sum_{s=0}^{N}
\sum_{N'=m}^{\infty}\sum_{s'=0}^{N'}
\frac{\sqrt{N!\,N'!}\,\,(\alpha)^{N-m}\,\,(\alpha^*)^{N'-m}}{(N-m)!\,(N'-m)!}\nonumber\\
&\times&\,\exp\,[-i(\lambda_{Ns}-\lambda_{N's'})\,t\,]\,
d_0^{Ns}\,\,d_0^{N's'}\,\ket{\psi_{Ns}}\bra{\psi_{N's'}},
\end{eqnarray}
where we have used Eq. (\ref{photonadded}) of Chapter 3, and expanded 
$\ket{\alpha, m}$ in the Fock basis.
The corresponding matrix elements of the density matrix are given by
\begin{eqnarray}
\expect{\psi_{Ml}}{\rho(t)}{\psi_{M'l'}}&=&\frac{e^{-\nu}}
{m!L_m(-\nu)}\frac{\sqrt{M!\,M'!}\,\,(\alpha)^{M-m}\,\,(\alpha^*)^{M'-m}}
{(M-m)!\,(M'-m)!}\nonumber\\ 
&\times&
\,\exp\,[-i(\lambda_{Ml}-\lambda_{M'l'})\,t\,]\,d_0^{Ml}\,\,d_0^{M'l'}.
\end{eqnarray}
Here, and in the rest of this Appendix, it is understood that 
contributions 
from terms of the form $1/(-n)!$, where $n$ is a positive integer, vanish.

The expectation values and higher moments of the quadrature 
variables $\xi(t)$ and $\eta(t)$, defined in Eq. (\ref{xiandeta}) of 
Chapter 5, can now be 
obtained numerically, using the 
above 
expressions for the density matrix and the
matrix elements of the operators 
$a$ and $b$ in the basis $\ket{\psi_{Ns}}$. The latter are given by
\begin{equation}
\expect{\psi_{Ns}}{a}{\psi_{N's'}}=
\sum_{n=0}^{N'} (N'-n)^{1/2}\,\,d_{n}^{Ns}\,d_{n}^{N's'}\,\delta_{N,N'-1}
\label{matrixa}
\end{equation}
and
\begin{equation}
\expect{\psi_{Ns}}{b}{\psi_{N's'}}=
\sum_{n=1}^{N'} n^{1/2}\,\,d_{n-1}^{Ns}\,d_{n}^{N's'}\,\delta_{N,N'-1}\,,
\label{matrixb}
\end{equation}
respectively. As these are purely off-diagonal, and $\rho(t)$ is 
diagonal for an initial state that is a direct product of Fock states, 
it follows that all the odd moments of $\xi$ and $\eta$ vanish identically
for all $t$, as asserted in the text. This is no longer 
true for the other classes of initial states considered.
 
The time-dependent reduced density matrices $\rho_k(t)$ ($k = a,b)$ 
are given by
\begin{eqnarray}
\rho_a(t)&=&{\rm Tr}_b\,[\rho(t)]=
\sum_{n=0}^{\infty}
\,_b\!\expect{n}{\rho(t)}{n}_{b}\,,\nonumber\\
\rho_b(t)&=&{\rm Tr}_a\,[\rho(t)]=
\sum_{n=0}^{\infty}\,_a\!\expect{n}{\rho(t)}
{n}_{a}.
\end{eqnarray}
Corresponding to an initial state $\ket{N\,;\, 0}$, 
these reduced density matrices $\rho_k(t)$ take the form
\begin{eqnarray}
\rho_a(t) &=& \sum_{n=0}^{N}\sum_{s = 0}^{N}\,\sum_{s' = 
0}^{N}\,\exp\,[-i(\lambda_{Ns} - \lambda_{Ns'})\,t\,]\nonumber \\
&&\times \quad d_0^{Ns}\,\,d_0^{Ns'}\,\,d_n^{Ns}\,\,d_n^{Ns'}\,
\ket{(N-n)}_{a} \,_a\!\bra{(N-n)}
\label{}
\end{eqnarray}
and
\begin{eqnarray}
\rho_b(t) &=& \sum_{n=0}^{N}\sum_{s = 0}^{N}\,\sum_{s' = 
0}^{N}\,\exp\,[-i(\lambda_{Ns} - \lambda_{Ns'})\,t\,]\nonumber \\
&&\times \quad  d_0^{Ns}\,\,d_0^{Ns'}\,\,d_{N-n}^{Ns}\,\,d_{N-n}^{Ns'}\,
\ket{(N-n)}_b\,_b\!\bra{(N-n)}.
\label{}
\end{eqnarray}
Hence we have, in the Fock basis, 
\begin{eqnarray}
_a\!\expect{n}{\rho_{a}(t)}{n'}_a &=& \sum_{s = 0}^{N}\,\sum_{s' = 
0}^{N}\,\exp\,[-i(\lambda_{Ns} - \lambda_{Ns'})\,t\,]\,
\,d_0^{Ns}\,\,d_0^{Ns'}\,\,d_{N-n}^{Ns}\,\,d_{N-n'}^{Ns'}\,\delta_{nn'}
\nonumber\\
\label{subrhoatfock}
\end{eqnarray}
and
\begin{eqnarray}
_b\!\expect{n}{\rho_b(t)}{n'}_b &=& \sum_{s = 0}^{N}\,\sum_{s' = 
0}^{N}\,\exp\,[-i(\lambda_{Ns} - \lambda_{Ns'})\,t\,]\,
\,d_0^{Ns}\,\,d_0^{Ns'}\,\,d_{n}^{Ns}\,\,d_{n'}^{Ns'}\,\delta_{nn'}\,
.\nonumber\\
\label{subrhobtfock}
\end{eqnarray}
As before, these are explicitly $N$-dependent finite-dimensional 
matrices. 

For an initial 
state $\ket{(\alpha,m)\,;\, 0}$  
the expressions for $\rho_k(t)$ are given by
\begin{eqnarray}
\rho_a(t)&=&\frac{e^{-\nu}}{m!L_m(-\nu)}
\sum_{n=0}^{\infty}\sum_{N=N_{\rm min}}^{\infty}\sum_{s=0}^{N}
\sum_{N'=N_{\rm min}}^{\infty}
\sum_{s'=0}^{N'}
\frac{\sqrt{N!\,N'!}\,\,(\alpha)^{N-m}\,\,(\alpha^*)^{N'-m}}
{(N-m)!\,(N'-m)!}\nonumber\\
&\times&\,\exp\,[-i(\lambda_{Ns}-\lambda_{N's'})\,t\,]\,
d_0^{Ns}\,\,d_0^{N's'}\,\,d_n^{Ns}\,\,d_n^{N's'}\,\,\ket{(N-n)}_a\,
_a\!\bra{(N'-n)}\nonumber\\
\end{eqnarray}
and
\begin{eqnarray}
\rho_b(t)&=&\frac{e^{-\nu}}{m!L_m(-\nu)}\sum_{n=0}^{\infty}
\sum_{N=N_{\rm min}}^{\infty}\sum_{s=0}^{N}\sum_{N'=N_{\rm min}}^{\infty}
\sum_{s'=0}^{N'}
\frac{\sqrt{N!\,N'!}\,\,(\alpha)^{N-m}\,\,(\alpha^*)^{N'-m}}
{(N-m)!\,(N'-m)!}\nonumber\\
&\times&
\exp\,[-i(\lambda_{Ns}-\lambda_{N's'})\,t\,]\,
d_0^{Ns}\,\,d_0^{N's'}\,d_{N-n}^{Ns}\,\,d_{N'-n}^{N's'}\,\,
\ket{(N-n)}_b\,
_b\!\bra{(N'-n)},\nonumber\\
\end{eqnarray}
where $N_{\rm min}=\max\,(n,m)$. These are infinite-dimensional 
matrices. 
The corresponding matrix elements of $\rho_k(t)$  
in the Fock basis are given by 
\begin{eqnarray}
_a\!\expect{l}{\rho_a(t)}{l'}_a&=&\frac{e^{-\nu}}{m!L_m(-\nu)}
\sum_{n_{\rm min}}^{\infty}\sum_{s=0}^{n+l}
\sum_{s'=0}^{n+l'}
\frac{\sqrt{(n+l)!\,(n+l')!}\,\,(\alpha)^{n+l-m}\,\,(\alpha^*)^{n+l'-m}}
{(n+l-m)!
\,(n+l'-m)!}\nonumber\\
&\times&
\,\exp\,[-i(\lambda_{(n+l)s}-\lambda_{(n+l')s'})\,t\,]\,
\,d_0^{(n+l)s}\,\,d_0^{(n+l')s'}\,\,
d_n^{(n+l)s}\,\,d_n^{(n+l')s'}\nonumber\\
\label{subrhoatphoton}
\end{eqnarray}
and
\begin{eqnarray}
_b\!\expect{l}{\rho_b(t)}{l'}_b&=&\frac{e^{-\nu}}{m!L_m(-\nu)}
\sum_{n_{\rm min}}^{\infty}\sum_{s=0}^{n+l}
\sum_{s'=0}^{n+l'}
\frac{\sqrt{(n+l)!\,(n+l')!}\,\,(\alpha)^{n+l-m}\,\,(\alpha^*)^{n+l'-m}}
{(n+l-m)!
\,(n+l'-m)!}\nonumber\\
&\times&
\,\exp\,[-i(\lambda_{Ns}-\lambda_{N's'})\,t\,]\,\,
d_0^{(n+l)s}\,\,d_0^{(n+l')s'}\,\,d_{l}^{(n+l)s}\,\,d_{l'}^{(n+l')s'},\nonumber\\
\label{subrhobtphoton}
\end{eqnarray}
where $n_{\rm min}=\max\,(m-l,m-l')$.

All the reduced density matrices (with elements 
given by Eqs. (\ref{subrhoatfock}), 
(\ref{subrhobtfock}), (\ref{subrhoatphoton}) 
and (\ref{subrhobtphoton})) 
are hermitian. They are diagonalized numerically, and their eigenvalues 
are used to compute the entropies $S_k(t)$ and $\delta_k(t)$ given 
by Eqs. (\ref{svnelne}). In the case of infinite-dimensional 
matrices, convergence in numerical computation 
is provided by the factorials 
in the denominators of the summands in the expressions derived above for 
the matrix  elements. We use double precision arithmetic with an accuracy 
of $1$ part in $10^{6}$.  As mentioned in the text, we use the equality of
$\rho_a(t)$ and $\rho_b(t)$ as one of the checks on the numerical 
computations. Other checks include the condition $\rho^2(t)=\rho(t)$ 
for the density matrix of the total system, as we are only 
dealing with pure states.

%% file: thesis.bbl
\begin{thebibliography}{10}
\addcontentsline{toc}{section}{Bibliography}

\bibitem{haake}
F.~Haake.
\newblock {\em Quantum Signatures of Chaos}.
\newblock Springer-Verlag,\ \ 1991.

\bibitem{berry1}
M.~V. Berry and M.~Tabor.
\newblock {\em Level clustering in the regular spectrum}.
\newblock Proc. R. Soc. Lond. A\ \ {\bf 356}, 375--394\ \ (1977).

\bibitem{berry2}
M.~V. Berry and M.~Tabor.
\newblock {\em Closed orbits and the regular bound spectrum}.
\newblock Proc. R. Soc. Lond. A\ \ {\bf 349}, 101--123\ \ (1976).

\bibitem{berry4}
M.~V. Berry.
\newblock {\em Semiclassical mechanics of regular and irregular motion}.
\newblock In {\it Chaotic Behavior of Deterministic Systems}, Editors G. Iooss,
  R. H. G. Helleman, and R. Stora, Les Houches Session XXXVI (1981),
  North-Holland,\ \ 1983.

\bibitem{peres}
A.~Peres.
\newblock {\em Stability of quantum motion in chaotic and regular systems}.
\newblock Phys. Rev. A\ \ {\bf 30}, 1610--1615\ \ (1984).

\bibitem{aver}
I.~Sh. Averbukh and N.~F. Perelman.
\newblock {\em The dynamics of wave packets of highly-excited atoms and
  molecules}.
\newblock Sov. Phys. Usp.\ \ {\bf 34}, 572--591\ \ (1991).

\bibitem{bluhm}
R.~Bluhm, V.~A. Kostelecky and J.~Porter.
\newblock {\em The evolution and revival structure of localized quantum wave
  packets}.
\newblock Am. J. Phys.\ \ {\bf 64}, 944--953\ \ (1996).

\bibitem{robi}
R.~W. Robinett.
\newblock {\em Quantum wave packet revivals}.
\newblock Phys. Rep.\ \ {\bf 392}, 1--119\ \ (2004).

\bibitem{dodo1}
V.~V. Dodonov.
\newblock {\em `Non-classical' states in quantum optics: a `squeezed' review of
  the first 75 years}.
\newblock J. Opt. B: Quant. Semiclass. Opt.\ \ {\bf 4}, R1--R33\ \ (2002).

\bibitem{keller}
J.~B. Keller and W.~Streifer.
\newblock {\em Complex rays with an application to Gaussian beams}.
\newblock J. Opt. Soc. Am.\ \ {\bf 61}, 40--43\ \ (1971).

\bibitem{deschamps}
G.~A. Deschamps.
\newblock {\em Ray techniques in electromagnetics}.
\newblock Proc. IEEE\ \ {\bf 60}, 1022--1035\ \ (1972).

\bibitem{littlejohn}
R.~G. Littlejohn.
\newblock {\em The semiclassical evolution of wave packets}.
\newblock Phys. Rep.\ \ {\bf 138}, 193--291\ \ (1986).

\bibitem{tara}
K.~Tara, G.S Agarwal and S.~Chaturvedi.
\newblock {\em Production of Schr\"odinger macroscopic quantum-superposition
  states in a Kerr medium}.
\newblock Phys. Rev. A\ \ {\bf 47}, 5024--5029\ \ (1993).

\bibitem{nauenberg}
M.~Nauenberg, C.~Stroud and J.~Yeazell.
\newblock {\em The classical limit of an atom}.
\newblock Sci. Am.\ \ {\bf 270}, 44--49\ \ (1994).

\bibitem{berry3}
M.~V. Berry and S.~Klein.
\newblock {\em Integer, fractional, and fractal Talbot effects}.
\newblock J. Mod. Opt.\ \ {\bf 43}, 2139--2164\ \ (1996).

\bibitem{banaszek}
K.~Banaszek, K.~W\'{o}dkiewicz and W.~P. Schleich.
\newblock {\em Fractional Talbot effects in phase space: A compact summation
  formula}.
\newblock Opt. Exp.\ \ {\bf 2}, 169--172\ \ (1998).

\bibitem{styer}
D.~F. Styer.
\newblock {\em The motion of wave packets through their expectation values and
  uncertainties}.
\newblock Am. J. Phys.\ \ {\bf 58}, 742--744\ \ (1990).

\bibitem{sudh1}
C.~Sudheesh, S.~Lakshmibala and V.~Balakrishnan.
\newblock {\em Manifestations of wave packet revivals in the moments of
  observables}.
\newblock Phys. Lett. A\ \ {\bf 329}, 14--21\ \ (2004).

\bibitem{aver1}
I.~Sh. Averbukh and N.~F. Perelman.
\newblock {\em Fractional revivals: Universality in the long-term evolution of
  quantum wave packets beyond the correspondence principle dynamics}.
\newblock Phys. Lett. A\ \ {\bf 139}, 449--453\ \ (1989).

\bibitem{park1}
J.~Parker and C.~R.~Stroud Jr.
\newblock {\em Coherence and decay of Rydberg wave Packets}.
\newblock Phys. Rev. Lett.\ \ {\bf 56}, 716--719\ \ (1986).

\bibitem{yeaz1}
J.~A. Yeazell, M.~Mallalieu and C.~R.~Stroud Jr.
\newblock {\em Observation of the collapse and revival of a Rydberg electronic
  wave packet}.
\newblock Phys. Rev. Lett.\ \ {\bf 64}, 2007--2010\ \ (1990).

\bibitem{yeaz2}
J.~A. Yeazell and C.~R.~Stroud Jr.
\newblock {\em Observation of fractional revivals in the evolution of a Rydberg
  atomic wave packet}.
\newblock Phys. Rev. A\ \ {\bf 43}, 5153--5156\ \ (1991).

\bibitem{tanas}
R.~Tanas.
\newblock {\em Nonclassical states of light propagating in Kerr media}.
\newblock In {\it Theory of Non-classical States of Light}, Editors V. V.
  Dodonov and V. I. Man'Ko, Taylor \& Francis,\ \ 2003.

\bibitem{cerf}
N.~J. Cerf, A.~Ipe and X.~Rottenberg.
\newblock {\em Cloning of continuous quantum variables}.
\newblock Phys. Rev. Lett.\ \ {\bf 85}, 1754--1757\ \ (2000).

\bibitem{shapiro}
E.~A. Shapiro, M.~Spanner and M.~Y. Ivanov.
\newblock {\em Quantum logic approach to wave packet control}.
\newblock Phys. Rev. Lett.\ \ {\bf 91}, 237901\ \ (2003).

\bibitem{vanenk}
S.~J. van Enk.
\newblock {\em Entanglement capabilities in infinite dimensions:
  Multidimensional entangled coherent states}.
\newblock Phys. Rev. Lett.\ \ {\bf 91}, 017902\ \ (2003).

\bibitem{zavatta}
A.~Zavatta, S.~Viciani and M.~Bellini.
\newblock {\em Quantum-to-classical transition with single-photon-added
  coherent states of light}.
\newblock Science\ \ {\bf 306}, 660--662\ \ (2004).

\bibitem{sudh2}
C.~Sudheesh, S.~Lakshmibala and V.~Balakrishnan.
\newblock {\em Wave packet dynamics of photon-added coherent states}.
\newblock Europhys. Lett.\ \ {\bf 71}, 744--750\ \ (2005).

\bibitem{sudh3}
C.~Sudheesh, S.~Lakshmibala and V.~Balakrishnan.
\newblock {\em Squeezing and higher-order squeezing of photon-added coherent
  states propagating in a Kerr-like medium}.
\newblock J. Opt. B: Quant. Semiclass. Opt.\ \ {\bf 7}, S728--S735\ \ (2005).

\bibitem{ken}
A.~Kenfack and K.~Zyezkowski.
\newblock {\em Negativity of the Wigner function as an indicator of
  non-classicality}.
\newblock J. Opt. B: Quant. Semiclass. Opt.\ \ {\bf 6}, 396--404\ \ (2004).

\bibitem{lvovsky}
A.~I. Lvovsky, H.~Hansen, T.~Aichele, O.~Benson, J.~Mlynek and S.~Schiller.
\newblock {\em Quantum state reconstruction of the single-photon Fock state}.
\newblock Phys. Rev. Lett.\ \ {\bf 87}, 050402\ \ (2001).

\bibitem{mandel}
L.~Mandel and E.~Wolf.
\newblock {\em Optical Coherence and Quantum Optics}.
\newblock Cambridge University Press,\ \ 1995.

\bibitem{sanz}
L.~Sanz, R.~M. Angelo and K.~Furuya.
\newblock {\em Entanglement dynamics in a two-mode nonlinear bosonic
  Hamiltonian}.
\newblock J. Phys. A\ \ {\bf 36}, 9737--9754\ \ (2003).

\bibitem{agar1}
G.~S. Agarwal and R.~R. Puri.
\newblock {\em Collapse and revival phenomenon in the evolution of a resonant
  field in a Kerr-like medium}.
\newblock Phys. Rev. A\ \ {\bf 39}, 2969--2977\ \ (1989).

\bibitem{abar}
H.~D.~I. Abarbanel.
\newblock {\em Analysis of Observed Chaotic Data}.
\newblock Springer,\ \ 1995.

\bibitem{grass}
P.~Grassberger and I.~Procaccia.
\newblock {\em Characterization of strange attractors}.
\newblock Phys. Rev. Lett.\ \ {\bf 50}, 346--349\ \ (1983).

\bibitem{fraser}
A.~M. Fraser and H.~L. Swinney.
\newblock {\em Independent coordinates for strange attractors from mutual
  information}.
\newblock Phys. Rev. A\ \ {\bf 33}, 1134--1140\ \ (1986).

\bibitem{sudh4}
C.~Sudheesh, S.~Lakshmibala and V.~Balakrishnan.
\newblock {\em Ergodicity properties of entangled two-mode states}.
\newblock (Submitted for publication).

\bibitem{robi1}
R.~W. Robinett.
\newblock {\em Visualizing the collapse and revival of wavepackets in the
  infinite square well using expectation values}.
\newblock Am. J. Phys.\ \ {\bf 68}, 410--420\ \ (2000).

\bibitem{don}
M.~A. Doncheski and R.~W. Robinett.
\newblock {\em Expectation value analysis of wave packet solutions for the
  quantum bouncer: Short-term classical and long-term revival behaviors}.
\newblock Am. J. Phys.\ \ {\bf 69}, 1084--1090\ \ (2001).

\bibitem{agar2}
G.~S. Agarwal and K.~Tara.
\newblock {\em Nonclassical properties of states generated by the excitations
  on a coherent state}.
\newblock Phys. Rev. A\ \ {\bf 43}, 492--497\ \ (1991).

\bibitem{kita}
M.~Kitagawa and Y.~Yamamoto.
\newblock {\em Number-phase minimum-uncertainty state with reduced number
  uncertainty in a Kerr nonlinear interferometer}.
\newblock Phys. Rev. A\ \ {\bf 34}, 3974--3988\ \ (1986).

\bibitem{grei}
M.~Greiner, O.~Mandel, T.~W. H\"ansch and I.~Bloch.
\newblock {\em Collapse and revival of the matter wave field of a Bose-Einstein
  condensate}.
\newblock Nature\ \ {\bf 419}, 51--54\ \ (2002).

\bibitem{siva}
S.~Sivakumar.
\newblock {\em Photon-added coherent states as nonlinear coherent states}.
\newblock J. Phys. A\ \ {\bf 32}, 3441--3447\ \ (1999).

\bibitem{schleich}
W.~Schleich, R.~J. Horowicz and S.~Varro.
\newblock {\em Bifurcation in the phase probability distribution of a highly
  squeezed state}.
\newblock Phys. Rev. A\ \ {\bf 40}, 7405--7408\ \ (1989).

\bibitem{wall1}
D.~F. Walls and P.~Zoller.
\newblock {\em Reduced quantum fluctuations in resonance fluorescence}.
\newblock Phys. Rev. Lett.\ \ {\bf 47}, 709--711\ \ (1981).

\bibitem{wall2}
D.~F. Walls.
\newblock {\em Squeezed states of light}.
\newblock Nature\ \ {\bf 306}, 141--146\ \ (1983).

\bibitem{loudon}
R.~Loudon and P.~L. Knight.
\newblock {\em Squeezed light}.
\newblock J. Mod. Opt.\ \ {\bf 34}, 709--759\ \ (1987).

\bibitem{hong1}
C.~K. Hong and L.~Mandel.
\newblock {\em Generation of higher-order squeezing of quantum electromagnetic
  fields}.
\newblock Phys. Rev. A\ \ {\bf 32}, 974--982\ \ (1985).

\bibitem{hill}
M.~Hillery.
\newblock {\em Amplitude-squared squeezing of the electromagnetic field}.
\newblock Phys. Rev. A\ \ {\bf 36}, 3796--3802\ \ (1987).

\bibitem{zhang}
Z.~Zhang, L.~Xu, J.~Chai and F.~Li.
\newblock {\em A new kind of higher-order squeezing of radiation field}.
\newblock Phys. Lett. A\ \ {\bf 150}, 27--30\ \ (1990).

\bibitem{dodo2}
V.~V. Dodonov, M.~A. Marchiolli, Y.~A. Korennoy, V.~I. Man'ko and Y.~A.
  Moukhin.
\newblock {\em Dynamical squeezing of photon-added coherent states}.
\newblock Phys. Rev. A\ \ {\bf 58}, 4087--4094\ \ (1998).

\bibitem{yurke}
B.~Yurke and D.~Stoler.
\newblock {\em Generating quantum mechanical superpositions of macroscopically
  distinguishable states via amplitude dispersion}.
\newblock Phys. Rev. Lett\ \ {\bf 57}, 13--16\ \ (1986).

\bibitem{gerry}
C.~Gerry and S.~Rodrigues.
\newblock {\em Higher-order squeezing from an anharmonic oscillator}.
\newblock Phys. Rev. A\ \ {\bf 35}, 4440--4442\ \ (1987).

\bibitem{du}
S.-D. Du and C.-D. Gong.
\newblock {\em Higher-order squeezing for the quantized light field:
  $K$th-power amplitude squeezing}.
\newblock Phys. Rev. A\ \ {\bf 48}, 2198--2212\ \ (1993).

\bibitem{perina}
J.~Perina.
\newblock {\em Quantum Statistics of Linear and Nonlinear Optical Phenomena}.
\newblock Reidel,\ \ 1984.

\bibitem{brune}
M.~Brune, S.~Haroche, J.~M. Raimond, L.~Davidovich and N.~Zagury.
\newblock {\em Manipulation of photons in a cavity by dispersive atom-field
  coupling: Quantum-nondemolition measurements and generation of
  ``Schr\"odinger cat'' states}.
\newblock Phys. Rev. A\ \ {\bf 45}, 5193--5214\ \ (1992).

\bibitem{gerry2}
C.~C. Gerry and P.~L. Knight.
\newblock {\em Introductory Quantum Optics}.
\newblock Cambridge University Press,\ \ 2005.

\bibitem{breit}
G.~Breitenbach and S.~Schiller.
\newblock {\em Homodyne tomography of classical and non-classical light}.
\newblock J. Mod. Opt.\ \ {\bf 44}, 2207--2225\ \ (1997).

\bibitem{sudh5}
C.~Sudheesh, S.~Lakshmibala and V.~Balakrishnan.
\newblock {\em Entanglement dynamics in the propagation of a single-mode field
  through a nonlinear medium}.
\newblock Topical Conference on Atomic, Molecular and Optical Physics, Indian
  Association for the Cultivation of Science, Kolkata, December 2005.

\bibitem{takens}
F.~Takens.
\newblock {\em Detecting strange attractors in turbulence}.
\newblock In {\it Dynamical Systems and Turbulence}, Editors D. Rand and L. S.
  Young, Springer,\ \ 1981.

\bibitem{rosen}
M.~T. Rosenstein, J.~J. Collins and C.~J.~D. Luca.
\newblock {\em A practical method for calculating largest Lyapunov exponents
  from small data sets}.
\newblock Physica D\ \ {\bf 65}, 117--134\ \ (1993).

\bibitem{kantz}
H.~Kantz.
\newblock {\em A robust method to estimate the maximal Lyapunov exponent of a
  time series}.
\newblock Phys. Lett. A\ \ {\bf 185}, 77--87\ \ (1994).

\bibitem{bocch}
P.~Bocchieri and A.~Loinger.
\newblock {\em Quantum recurrence theorem}.
\newblock Phys. Rev.\ \ {\bf 107}, 337--338\ \ (1957).

\bibitem{hogg}
T.~Hogg and B.~A. Huberman.
\newblock {\em Recurrence phenomena in quantum dynamics}.
\newblock Phys. Rev. Lett.\ \ {\bf 48}, 711--714\ \ (1982).

\bibitem{sesh1}
S.~Seshadri, S.~Lakshmibala and V.~Balakrishnan.
\newblock {\em Quantum revivals, geometric phases and circle map recurrences}.
\newblock Phys. Lett. A\ \ {\bf 256}, 15--19\ \ (1999).

\bibitem{fermi}
E.~Fermi, J.~Pasta and S.~Ulam.
\newblock {\em Studies in nonlinear problems}.
\newblock In {\it Collected Papers of Enrico Fermi}, Vol. II, Editor E.
  Segr\`e, University of Chicago Press,\ \ 1965.

\bibitem{ermos}
V.~A. Ermoshin, M.~Erdmann and V.~Engel.
\newblock {\em Fermi-Pasta-Ulam recurrences, normal modes and wave-packet
  revivals}.
\newblock Chem. Phys. Lett.\ \ {\bf 356}, 29--35\ \ (2002).

\bibitem{balle}
L.~E. Ballentine, Y.~Yang and J.~P. Zibin.
\newblock {\em Inadequacy of Ehrenfest's theorem to characterize the classical
  regime}.
\newblock Phys. Rev. A\ \ {\bf 50}, 2854--2859\ \ (1994).

\bibitem{habib}
S.~Habib, K.~Jacobs and K.~Shizume.
\newblock {\em The quantum emergence of chaos}.
\newblock arXiv: quant-ph/0412159; Phys. Rev. Lett. (to appear)\ \ (2005).

\bibitem{wolfgang}
W.~P. Schleich.
\newblock {\em Quantum Optics in Phase Space}.
\newblock Wiley-VCH,\ \ 2001.

\bibitem{gradshteyn}
I.~S. Gradshteyn and I.~M. Ryzhik.
\newblock {\em Table of Integrals, Series, and Products}.
\newblock Academic Press,\ \ 1980.

\bibitem{press}
W.~H. Press, S.~A. Teukolsky, W.~T. Vetterling and B.~P. Flannery.
\newblock {\em Numerical Recipes in C}.
\newblock Cambridge University Press, pp. 475-481,\ \ 2001.

\bibitem{gnu}
M.~Galassi, J.~Davies, J.~Theiler, B.~Gough, G.~Jungman, M.~Booth and F.~Rossi.
\newblock {\em GNU Scientific Library}.
\newblock GNU Software,\ \ 2005.

\end{thebibliography}
